\documentclass[aps,prd,preprintnumbers,longbibliography,onecolumn,nofootinbib]{revtex4-2}

\usepackage{multirow}
\usepackage{siunitx}

\usepackage{tikz}
\usetikzlibrary{decorations.markings, arrows, positioning, calc}

\usepackage{graphicx}
\usepackage{latexsym}
\usepackage{mathtools}

\usepackage{amsmath,amssymb,amsbsy,amsfonts}
\usepackage{array}
\usepackage{bm}
\usepackage{bbold}

\usepackage{graphics}

\usepackage{mathrsfs}
\usepackage{xcolor}
\usepackage{cancel}
\usepackage[normalem]{ulem}

\usepackage[unicode, pdfprintscaling=None, colorlinks]{hyperref}
\hypersetup{
    colorlinks,
    allcolors=blue!50!black,
}
\usepackage{inconsolata}
\usepackage{multirow}
\usepackage{makecell}
\usepackage{tabu}
\usepackage{tabularx}
\newcolumntype{Y}{>{\centering\arraybackslash}X}
\usepackage{diagbox}

\usepackage[capitalise]{cleveref}

\usepackage{relsize}
\usepackage{microtype}

\usepackage{xfrac}
\usepackage{orcidlink}
\usepackage{ulem} 
\usepackage{amssymb}
\usepackage{subcaption}
\usepackage{caption}
\usepackage{stackengine}

\usepackage{bbm}
\usepackage{rotating}
\usetikzlibrary{arrows.meta}
\usepackage{quantikz}

\tikzset{
    gateO/.style={
        draw,
        circle,
        minimum width=0.8em,
        inner sep=1pt,
        append after command={
            \pgfextra {
                \fill (\tikzlastnode) circle[radius=2pt];
            }
        }
    }
}

\DeclareExpandableDocumentCommand{\gateO}{O{}{m}}{|[gateO,#1]| {#2} \qw}

\usepackage{adjustbox}
\usepackage{placeins}
\usepackage{float}

\begin{document}
\begin{figure}
  \vskip -1.cm
  \leftline{\includegraphics[width=0.15\textwidth]{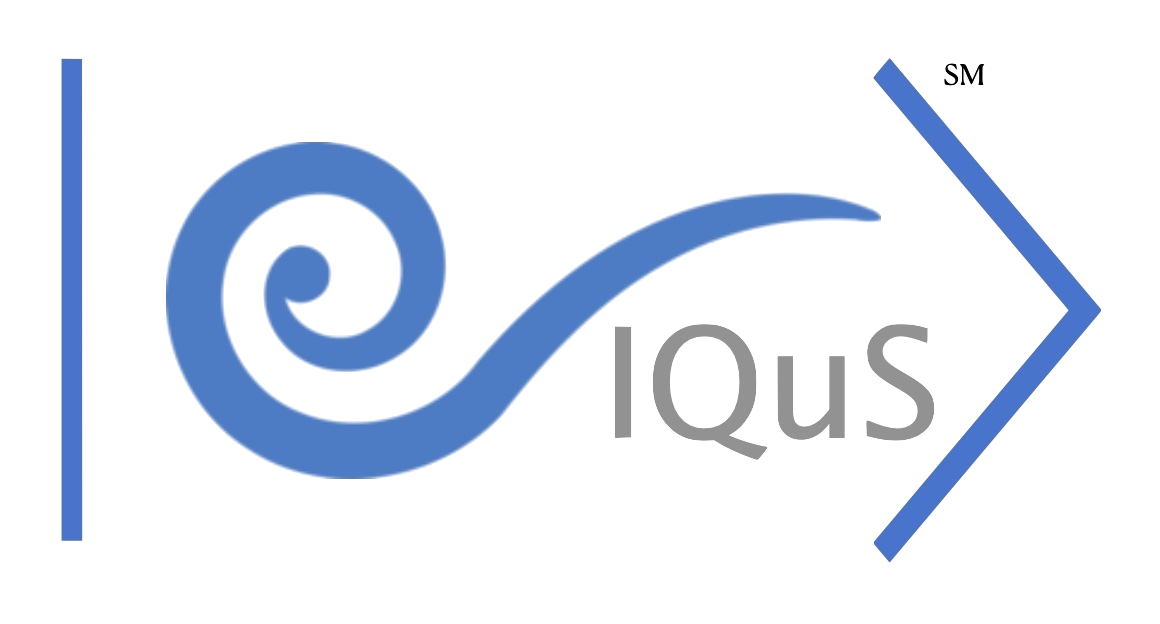}}
\end{figure}

\title{Scalable Quantum Simulations of Scattering in Scalar Field Theory on 120 Qubits}
\author{Nikita A. Zemlevskiy\,\orcidlink{0000-0002-0794-2389}}
\email{zemlni@uw.edu}
\affiliation{InQubator for Quantum Simulation (IQuS), Department of Physics, University of Washington, Seattle, WA 98195, USA.}

\preprint{IQuS@UW-21-090}
\date{\today}

\begin{abstract}
Simulations of collisions of fundamental particles on a quantum computer are expected to have an exponential advantage over classical methods and promise to enhance searches for new physics. Furthermore, scattering in scalar field theory has been shown to be BQP-complete, making it a representative problem for which quantum computation is efficient. As a step toward large-scale quantum simulations of collision processes, scattering of wavepackets in one-dimensional scalar field theory is simulated using 120 qubits of IBM's Heron superconducting quantum computer {\tt ibm\_fez}. Variational circuits compressing vacuum preparation, wavepacket initialization, and time evolution are determined using classical resources. By leveraging physical properties of states in the theory, such as symmetries and locality, the variational quantum algorithm constructs scalable circuits that can be used to simulate arbitrarily-large system sizes. A new strategy is introduced to mitigate errors in quantum simulations, which enables the extraction of meaningful results from circuits with up to 4924 two-qubit gates and two-qubit gate depths of 103. The effect of interactions is clearly seen, and is found to be in agreement with classical Matrix Product State simulations. The developments that will be necessary to simulate high-energy inelastic collisions on a quantum computer are discussed.
\end{abstract}

\maketitle
\newpage{}
\tableofcontents
\newpage{}
\section{Introduction}
The rapid growth of the computational capabilities of quantum devices over the last several years has fueled the pursuit of quantum advantage. Despite the steady improvement of quantum hardware, both in qubit counts and error rates, its inherent noise is a key challenge for the realization of general-purpose quantum computation, and is expected to persist until the development of fault-tolerant devices. In the Noisy Intermediate-Scale Quantum (NISQ) regime \cite{Preskill:2018jim}, the implementation of many quantum algorithms with provable speedups over their classical counterparts is out of reach, due to their assumption of fault tolerance enabled by error correction \cite{Jordan_2024}. Driven by the chase for practical quantum advantage, known as quantum utility, sophisticated error mitigation techniques have been developed \cite{Temme:2016vkz,Cai:2022rnq,Liao:2023eug}. These methods are used to account for the effects of hardware noise without real-time error correction, enabling the extraction of expectation values from noisy quantum states.

With these error mitigation techniques and quantum devices available today, digital calculations are becoming possible beyond the capacity of brute-force or approximate classical methods \cite{Arute:2019zxq,Yu:2022ivm,Shtanko:2023tjn,Kim:2023bwr,Farrell:2023fgd,Baumer:2023vrf,Chen:2023tfg,Liao:2023eug,Bluvstein:2023zmt,Pelofske:2023nwg,Farrell:2024fit,Chowdhury:2023vbk,Ciavarella:2024fzw,Shinjo:2024vci,Bao:2024jls,Hayata:2024fnh,Hayata:2024smx}. Because an achievement of quantum utility requires a computation done at scale, the development of scalable methods is critical to fully use the power of state-of-the-art quantum computers. Variational algorithms have recently emerged as a promising avenue for such computations \cite{McClean:2015vup,Cerezo:2020jpv,Biamonte:2021ntr}. These hybrid quantum-classical techniques use classical resources to determine the circuits that are executed on quantum computers. Variational algorithms have been developed for a variety of tasks, famously for ground-state preparation using the Variational Quantum Eigensolver (VQE) \cite{Peruzzo:2013bzg}, but also for combinatorial optimization \cite{Farhi:2014ych}, quantum machine learning \cite{Biamonte:2016ugo}, linear algebra problems \cite{Bravo-Prieto:2019kld}, quantum simulation \cite{Yuan:2018jdl}, and other applications. The power of these algorithms stems from their ability to significantly compress circuits that would otherwise be too deep for current hardware capabilities. 

Quantum simulation has been identified as a uniquely-positioned computational task capable of demonstrating scientifically-useful quantum advantage \cite{Daley:2022eja}. Unlike problems that assume error correction, quantum simulations have a built-in tolerance of errors, coming from the nature of computing expectation values in a systematically-improvable manner. As a result of this, error mitigation is an ideal method to deal with noise in the setting of quantum simulation on near-term devices. Furthermore, quantum simulation is able to take advantage of physical properties of target systems to increase the scale of the computations, while maintaining control over approximations and uncertainties. For these reasons, quantum simulation is a well-suited problem to exercise quantum machines available today.

Quantum simulation holds the potential for predictive power in a wide class of fundamental physics problems that are out of reach for classical methods. In particular, quantum simulations of high-energy inelastic scattering of relativistic particles, such as those that take place in extreme cosmological events or particle colliders, promise to shed light onto the structure and properties of the Standard Model and what may lie beyond \cite{Bauer:2023qgm,DiMeglio:2023nsa}. Standard techniques in quantum field theory predict the outputs of scattering events in regimes where perturbation theory is valid. Although finite-volume lattice methods \cite{Luscher:1985dn,Luscher:1986pf} are able to compute non-perturbative quantities for specific processes (for example, see Refs.~\cite{Beane:2008dv,Beane:2010em,Luu:2011ep,Briceno:2013hya,Hansen:2014wea}), studying arbitrary scattering events is challenging. Even in the perturbative regime, conventional analytical methods only give static quantities, or results that are valid at infinitely-late times. While this approximation is valid for many quantities of interest, the dynamics of collision events are generally inaccessible. Numerical lattice techniques suffer from an exponential increase of resources due to the sign problem when attempting to simulate real-time dynamics. Furthermore, classical simulations of highly-inelastic scattering using methods based on truncations of entanglement, such as Matrix Product States (MPS), are expected to fail because of the large amount of entanglement that is believed to be generated in these events \cite{Milsted:2020jmf,Vanhecke:2021noi,PhysRevResearch.3.013078,Rigobello:2021fxw,Belyansky:2023rgh, Papaefstathiou:2024zsu}. As such, the dynamics of scattering of fundamental particles is a prime example of a physically-relevant problem that is challenging for classical computation, making it a promising target for quantum simulation.  

Much work has been done toward the goal of full-scale quantum simulations of scattering in theories and regimes where analytic results are scarce and classical computations are intractable. Proposals of wavepacket creation techniques \cite{Gustafson:2019mpk,Milsted:2020jmf,Surace:2020ycc,huang2023quantum,Tutunnikov:2023aol,Kane:2023jdo,Davoudi:2024wyv,Su:2024uuc} and strategies to extract scattering observables from dynamics \cite{Li:2023kex,Farrell:2024mgu} have complemented demonstrations of quantum simulations of scattering in various models \cite{Dreher:2022scr,Turco:2023rmx,Kreshchuk:2023btr,Chai:2023qpq,Briceno:2023xcm,Sharma:2023bqu,Turro:2023dhg,Wang:2024scd,Wu:2024rod,Bennewitz:2024ixi,Turro:2024ksf,Yusf:2024igb}. One of the simplest relativistic field theories in which scattering may be studied is the scalar field theory in one dimension.\footnote{While it has been proven that the scalar field theory is trivial in five spacetime dimensions, and believed to be trivial in four, the existence of a non-trivial continuum limit has been rigorously established in the case of one and two spatial dimensions \cite{Brydges:1983di}. The continuum limit can be recovered from lattice simulations by studying systems with successively finer lattices and carefully tuning the system parameters as functions of the lattice spacing.} This is the theory of coupled harmonic oscillators at each point in space with a quartic potential that governs interactions between particles. Scalar and pseudoscalar particles play key roles in many areas, including nuclear physics \cite{Weinberg:1978kz,Holland:2019zju}, condensed matter physics \cite{Shankar:1993pf,Schafer:2006yf}, high energy physics \cite{Higgs:1964pj,Higgs:1964ia,Caprini:2005zr}, as well as cosmology \cite{PhysRevD.23.347} and physics beyond the Standard Model \cite{Peccei:1977hh,Peccei:2006as,Rinaldi:2021jbg}. Scalar fields have also been used as a venue to study phase transitions and spontaneous symmetry breaking \cite{Rychkov:2014eea,Rychkov:2015vap,Thompson:2023kxz}. Moreover, scattering in scalar field theory has been shown to be BQP-complete \cite{Jordan_2018}, meaning all ``quantum-efficient" problems may be solved by mapping them to quantum simulations of scattering with various initial conditions. As such, this ``simple" theory is a good sandbox for demonstrations of scalable quantum simulation methods. Following the seminal work of Jordan, Lee, and Preskill (JLP) \cite{Jordan:2011ci,Jordan:2012xnu,Jordan_2018}, there have been numerous efforts targeting aspects of quantum simulations of scattering in scalar field theory \cite{Klco:2018zqz,Yeter-Aydeniz:2018mix,Klco:2019yrb,Klco:2019xro,Barata:2020jtq,Klco:2020aud,Kurkcuoglu:2021dnw, Macridin:2021uwn, Liu:2021otn,Illa:2022jqb,Li:2022ped,Ozzello:2023dzn,Hardy:2024ric}. 

In this work, variational algorithms on the basis of brickwall \cite{Haghshenas:2021nsg,Leone:2022aux,Filippov:2022exc,Okada:2022fiy,Tepaske:2022uad,Barthel:2023ine,Miao:2023her,Tepaske:2023mfc,Robertson:2023jlp,Zhang:2024kuf,Wright:2024yxm} circuits\footnote{``Brickwall", ``brickwork", and ``hardware-efficient" are commonly used to refer to dense variational ansatze using native single- and two-qubit gates. ``Brickwall" is used to describe this class of circuits in this work.} and other ansatze are used to simulate the scattering of two particle wavepackets on IBM's quantum computers. While the application of variational algorithms to quantum simulations has been proposed and executed on devices in the past \cite{Ibrahim:2022liy,Causer:2023wpp,Wang:2024pap,Zhuang:2024mdh,Sachdeva:2024kob}, this work serves as the first example of a quantum simulation using physics-informed brickwall circuits at scale for state preparation and time evolution. Furthermore, it is the first quantum simulation of wavepacket scattering in an interacting quantum field theory. These proof-of-concept simulations are a demonstration of the techniques that will be foundational for simulations of high-energy, high-inelasticity scattering events in more complicated theories, where rich physical phenomena are expected. The simulations proceed in several steps: 
\begin{enumerate}
    \item Vacuum preparation
    \item Particle creation
    \item Time evolution
    \item Measurement
\end{enumerate}
By incorporating the symmetries and physical structure of the scalar field theory into the circuit design, the simulation is carried out in a scalable way, unlocking the full capabilities of state-of-the-art devices. Throughout this work, ``scalable" will be used to indicate scalability for states whose structure does not change with increasing system size. Recent work has determined upper bounds of $O(10^{12})$ fault-tolerant operations required for simulations of scattering in one-dimensional scalar field theory \cite{Hardy:2024ric}. The variational techniques to minimize circuit depth developed in this work enable approximate simulations of scattering processes with feasible gate counts for devices available today. While previous quantum simulations have taken a hybrid approach, (e.g., state preparation was done variationally, but time evolution was implemented using an algorithm with rigorous performance guarantees), this work establishes the application of physics-informed variational algorithms to all parts of the simulation. The variational methods developed in this work are expected to have broad relevance to simulating physical models of interest.

This paper is organized as follows. An overview of lattice scalar field theory in one dimension and its mapping to qubits is given in Sec.~\ref{sec:lattice_scalar_field_theory}. The algorithm for implementing scalable variational circuits (SVCs) for a class of quantum simulation problems is outlined in Sec.~\ref{sec:scalable_variational_circuits}. This algorithm may be seen as a general framework capable of compressing physical simulation circuits at scale. Sec.~\ref{sec:circuits} describes how knowledge of the physics of the theory is used to determine SVCs both for state preparation and time evolution. The errors stemming from these approximate circuit compression methods are quantified. In Sec.~\ref{sec:results}, SVCs are used to simulate scattering in scalar field theory using 120 qubits of IBM's superconducting quantum computer {\tt ibm\_fez}. Both the free and interacting cases are studied. A new error mitigation strategy on the basis of simulations of vacuum evolution is introduced. With this technique, the results are found to be in qualitative agreement with MPS circuit simulations, and the effect of the interaction strength on the nature of the collisions is clearly seen. Improvements and comments on the extensibility of the developed methods are presented in Sec.~\ref{sec:discussion}. 

\section{Lattice scalar field theory}
\label{sec:lattice_scalar_field_theory}
The Hamiltonian of a real scalar field $\phi$ in one dimension with periodic boundary conditions (PBCs) defined on a lattice of $L$ spatial sites is written as
\begin{align}
    H &= \sum_{j=0}^{L-1} \frac{1}{2} \Pi_j^2 + \frac{1}{2}m^2 \phi_j^2 + \frac{1}{2} (\phi_{j+1} - \phi_j)^2 + \frac{\lambda}{4!} \phi_j^4 \label{eq:lattice_h} \\
    &\equiv H_\Pi + H_\phi + H_\text{kin} + H_\text{int}. \nonumber
\end{align}
Here $m$ is the bare mass, $\lambda$ is the coupling controlling the strength of $\phi^4$ interactions, $\Pi_j$ is the conjugate momentum operator obeying the canonical commutation relation $[\phi_i, \Pi_j] = i\delta_{ij}$, and the nearest-neighbor finite-difference operator $\nabla \phi = (\phi_{j+1} - \phi_j)$ is used in the lattice representation of the kinetic term. The lattice spacing is set to $a=1$ in this work. 

With PBCs, the eigenstates of $H$ are labeled by their spatial momenta $k=2\pi n/L$, where $n$ is an integer such that $k \in (-\pi,\pi]$. The vector of fields $\vec{\phi} = (\phi_0, \phi_1, \dots, \phi_{L-1})$ is related to the vector of spatial momentum eigenstates $\vec{\chi}$ by a spatial Fourier transform: $\vec{\chi} = V \vec{\phi}$, where $V = \frac{1}{\sqrt{L}}e^{i \vec{k}\cdot\vec{j}}$ is the Fourier transform matrix. In the free theory $(\lambda=0)$, the energy of a single-particle eigenstate with spatial momentum $k$ is 
\begin{equation}
    E_k^2 = m^2 + 4\sin^2\left(\frac{k}{2}\right).
    \label{eq:dispersion}
\end{equation}

The spectrum of $H$ in the free theory with PBCs can be constructed by introducing creation and annihilation operators
\begin{equation}
    a_k = \sum_j e^{-ikj} \left(\sqrt{\frac{E_k}{2}} \phi_j + i \sqrt{\frac{1}{2E_k}} \Pi_j\right).\label{eq:ap}
\end{equation}
This can be inverted to find expressions for $\phi$ and $\Pi$ in terms of $a_k$ and $a_k^\dagger$. Written in terms of the creation and annihilation operators, $H$ takes the following form:
\begin{equation}
    H = \frac{1}{L}\sum_k E_k \left(a_k^\dagger a_k + \frac{1}{2}\right).
\end{equation}

The vacuum of the free theory can be written down analytically. In the $a_k$ Fock basis, the ground state of the system can be written as a tensor product of the $L$ harmonic oscillators $\vec{\chi}$, each in its ground state (denoted by the subscript 0):
\begin{equation}
    \ket{\psi_\text{vac}} = \ket{\psi_{\chi_0}}_0 \otimes \ket{\psi_{\chi_1}}_0 \otimes \dots \otimes \ket{\psi_{\chi_{L-1}}}_0.
    \label{eq:vacuum_ket}
\end{equation}
The eigenfunctions of the harmonic oscillator can be used to construct wavefunctions in the free scalar field theory. The momentum-space representation of the $n^\text{th}$ harmonic oscillator eigenfunction is given by 
\begin{align}
    \langle \chi | n\rangle = \frac{1}{\sqrt{2^n n!}} \left( \frac{E_n}{\pi} \right)^\frac{1}{4} e^{-\frac{1}{2}E_n \chi^2}\text{H}_n\!\left(\sqrt{E_n}\chi\right),
    \label{eq:sho_eigenfunctions}
\end{align}
where $\text{H}_n$ is the $n^\text{th}$ Hermite polynomial. Since $\text{H}_0=1$, the ground-state wavefunction of the scalar field theory in this basis can be written as a product of Gaussians: 
\begin{equation}
    \langle \vec{\chi} | \psi_\text{vac} \rangle = \frac{(\text{det} E)^{1/4}}{\pi^{L/4}} e^{-\frac{1}{2}\vec{\chi}^T E\vec{\chi}},
    \label{eq:vacuum_wavefunction}
\end{equation} 
where $E$ is the matrix consisting of the energies in Eq.~\eqref{eq:dispersion} on the diagonal. The position-space wavefunction can be found using the Fourier transform $\vec{\chi} = V\vec{\phi}$.

Similar to the vacuum, wavepackets in the free theory can be specified analytically. Using the vacuum state of Eq.~\eqref{eq:vacuum_ket}, the operator $a_k^\dagger$ defined in Eq.~\eqref{eq:ap} creates a single-particle excitation with momentum $k$:
\begin{equation}
    |k\rangle = a_k^\dagger|\psi_\text{vac}\rangle = |\psi_{\chi_0}\rangle_0 \otimes \dots\otimes |\psi_{\chi_k}\rangle_1 \otimes\dots\otimes |\psi_{\chi_{L-1}}\rangle_0.
\end{equation}
A general single-particle state is a superposition of these excitations:
\begin{equation}
    |\psi_\text{wp}(p)\rangle = \mathcal{N}\sum_k g(k,p) |k\rangle,
    \label{eq:single_particle_superposition}
\end{equation}
where $g(k,p)$ is an envelope in $k$-space that specifies the form of the state, and $\mathcal{N}$ is a normalization constant. A localized wavepacket state can be created by choosing a Gaussian profile $g(k,p;\sigma_p)=\frac{1}{\sqrt{2\pi}\sigma_p}e^{-\frac{1}{2}(k-p)^2/\sigma_p^2}$. The wavefunction of a Gaussian wavepacket with momentum $p$ and spread $\sigma_p$ is found using Eq.~\eqref{eq:sho_eigenfunctions} to be
\begin{equation}
    \langle \vec{\chi}|\psi_\text{wp}(p;\sigma_p)\rangle = \mathcal{N} e^{-\frac{1}{2}\vec{\chi}^T E \vec{\chi}}\sum_k g(k,p;\sigma_p)\,\text{H}_1\!\left(\sqrt{E_k}\chi_k\right).
    \label{eq:wp_wavefunction}
\end{equation}
This specifies a wavepacket centered at $j=0$; it can be moved elsewhere using a spatial translation operator. As before, the position-space wavefunction is found by the Fourier transform $\vec{\chi} = V\vec{\phi}$. Multiparticle states can be constructed in a similar fashion. 

Time-evolved observables in the free theory can be computed analytically by expanding the state in the appropriate basis of multiparticle excitations. For example, the time evolution of a single-wavepacket state can be written as $|\psi_\text{wp}\rangle(t) = \mathcal{N}\sum_k e^{-i t E_k} g(k,p) a_k^\dagger|\psi_\text{vac}\rangle$. By rewriting $\phi^2_j$ in terms of $a_k$ and $a_k^\dagger$, $\langle \phi^2_j\rangle(t)$ can be obtained using Wick contractions. In the interacting $(\lambda \neq 0)$ theory, $E_k,a_k^\dagger,$ and $|\psi_\text{vac}\rangle$ are all functions of $\lambda$. In cases where $\frac{\lambda}{4!}$ is small, observables can be obtained by treating the $\phi^4$ term as a perturbation.

In the context of quantum simulation, the interacting theory can be studied provided that the required states can be prepared. JLP introduced a procedure to create interacting particle wavepackets from their free counterparts \cite{Jordan:2011ci,Jordan:2012xnu}. By the adiabatic theorem, if the $\phi^4$ interactions are turned on in a manner that is slow enough for states to adjust, then eigenstates in the free theory will flow into the corresponding eigenstates in the interacting theory. However, because wavepackets are superpositions of eigenstates, they will propagate and broaden during this adiabatic evolution. To counter this, the adiabatic evolution turning on $\lambda$ can be broken up into several steps, and the unwanted evolution can be ``undone" at each step by evolving backwards. Let $s$ parameterize the interactions in the Hamiltonian
\begin{align}
    H(s) = H_\phi + H_\Pi + H_\text{kin} + s H_\text{int},\quad s \in [0,1].
\end{align}
The adiabatic turn-on is then implemented in $N$ steps by interleaving forward time evolution with increasing $s$ at each step with reverse time evolution at fixed $s$:
\begin{align}
    U_\text{ad} = \prod_{n=0}^{N-1} T\left\{\exp\left(-i\frac{t_\text{ad}}{2}\int_{\frac{(2n+1)}{2N}}^{\frac{(n+1)}{N}}H(s)ds\right)\right\} \exp\left(it_\text{ad}H\!\left(\frac{2n+1}{2N} \right)\right) T\left\{\exp\left(-i\frac{t_\text{ad}}{2}\int_{\frac{n}{N}}^{\frac{(2n+1)}{2N}}H(s)ds\right)\right\},
    \label{eq:u_adiabatic}
\end{align}
where $T$ is the time-ordering operator. With a proper choice of $N$ and $t_\text{ad}$ depending on the parameters $m$ and $\lambda$ \cite{ Jordan:2011ci,Jordan:2012xnu,Li:2024lrl}, it can be shown that unwanted evolution can be sufficiently suppressed.  

\subsection{Qubit representation}
\label{sec:qubit_representation}
At each lattice site $j$, the field takes continuous values $\phi_j \in (-\infty, \infty)$. Since the Hilbert space at each $j$ is infinite, the field must be digitized to be represented on a quantum computer. The JLP digitization is used to map this system onto qubits \cite{Jordan:2011ci,Jordan:2012xnu}. Given a register of $n_q$ qubits, a maximum magnitude $\phi_\text{max}$ is chosen,\footnote{There exists an optimal choice of $\phi_\text{max}$ for a given $n_q$ as a result of the Nyquist-Shannon Sampling Theorem \cite{shannon_collected_1993,10.5555/3179430.3179434,Macridin:2018oli,Macridin:2018gdw}. The optimal $\phi_\text{max}$ also depends on $m$ and $\lambda$, since these parameters control the support of the eigenstates of the theory (see App.~\ref{sec:digitization_effects}). The optimal values of $\phi_\text{max}$ are determined numerically for small $n_q$ in Ref.~\cite{Klco:2018zqz}, and analytically in Refs.~\cite{Bauer:2021gek,Kane:2022ejm}.} and the field at each point in space is evenly discretized in $2^{n_q}$ steps of size $\delta_\phi$ in the range $[-\phi_\text{max}, \phi_\text{max}]$. At each site $j$, the digitized field can take on the values
\begin{equation}
    \phi_j = -\phi_\text{max} + \ell \delta_\phi, \quad \delta_\phi = \frac{2 \phi_\text{max}}{2^{n_q} - 1}, \quad \ell \in [0, 2^{n_q}-1].
    \label{eq:phi_digitized}
\end{equation}
Twisted boundary conditions in field space \cite{Klco:2018zqz} are used to preserve the symmetry in the digitizations of $\phi$- and $\Pi$-space. The allowed conjugate momenta $k_\phi$\footnote{The conjugate momenta $k_\phi$, which are used to construct the $\Pi_j$ operator, should not be confused with the spatial momenta $k$, which label eigenstates of the Hamiltonian.} are distributed symmetrically around 0:
\begin{equation}
    k_{\phi,j}= -\frac{\pi}{\delta_\phi} + \left(\ell + \frac{1}{2}\right)\frac{2\pi}{2^{n_q} \delta_\phi}, \quad \ell \in [0, 2^{n_q}-1].
    \label{eq:pi_digitized}
\end{equation} 
The allowed state of the field at each site is represented by qubits using a binary encoding of the $2^{n_q}$ allowed values from Eqs.~\eqref{eq:phi_digitized} and \eqref{eq:pi_digitized}:
\begin{equation}
    |\phi_j = -\phi_\text{max} + \ell \delta_\phi\rangle = |\ell\rangle,
\end{equation} 
(e.g., $|\phi_j = -\phi_\text{max} + \delta_\phi\rangle = |1\rangle = |01\rangle$ for $n_q=2$).

This digitization is particularly convenient because the form of the operators in the Hamiltonian is simple when expressed in terms of their action on qubits:
\begin{equation}
    \phi_j = -\frac{\phi_\text{max}}{2^{n_q}-1}\sum_{\ell=0}^{n_q-1} 2^\ell Z_{n_q j + \ell},
    \label{eq:phi_qubits}
\end{equation}
and in $\Pi$-space:
\begin{equation}
    \Pi_j = -\frac{\pi}{2^{n_q}\delta_\phi} \sum_{\ell=0}^{n_q-1} 2^\ell Z_{n_q j + \ell}.
    \label{eq:pi_qubits}
\end{equation}

\begin{figure*}
{\includegraphics[width=0.8\linewidth]{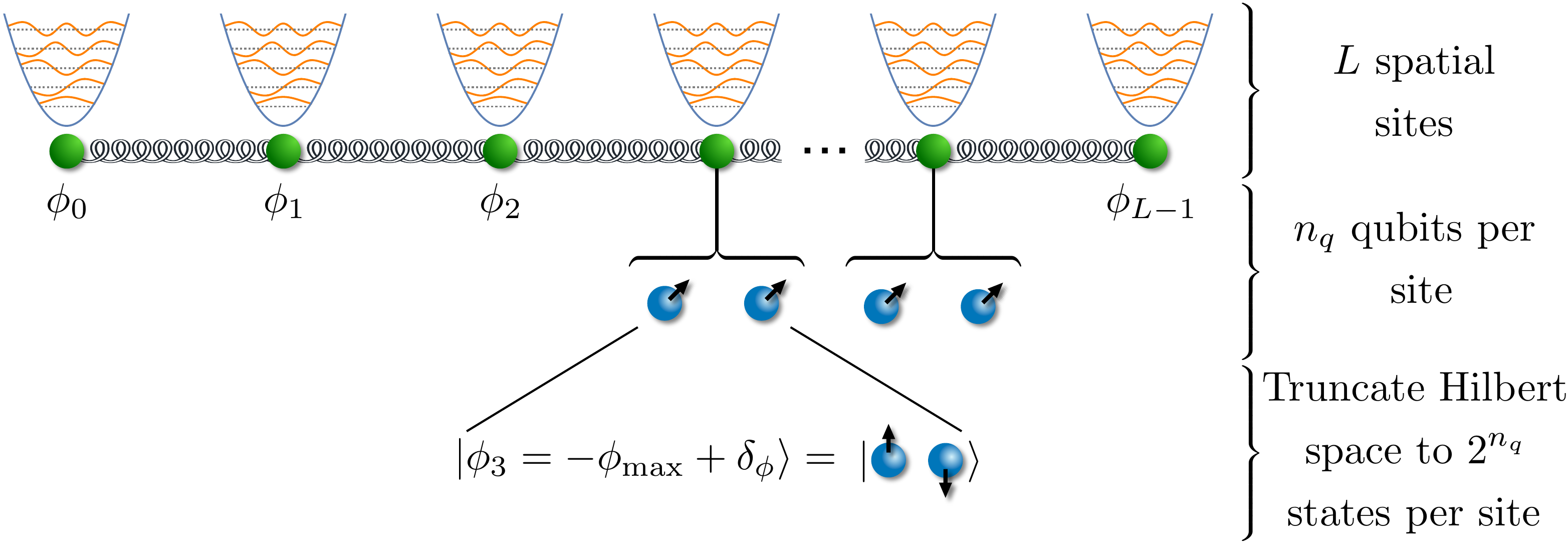}}
\caption{Mapping the one-dimensional lattice scalar field theory to qubits. For a system of $L$ spatial sites, $n_q$ qubits are used to represent the state of the field at each spatial site. The infinite-dimensional bosonic Hilbert space of $\phi$ at each site is truncated to $2^{n_q}$ values. A maximum magnitude of the field $\phi_\text{max}$ is chosen, and the values of the field in increments of $\delta_\phi$ are assigned to the possible qubit states.}
\label{fig:map_to_qubits}
\end{figure*} 

The representation of the lattice scalar field by qubits is shown in Fig.~\ref{fig:map_to_qubits}. This construction efficiently recovers the low-energy physics in a way that is a systematically extendable to include higher-energy contributions. In Ref.~\cite{Klco:2018zqz} it was shown that, aside from device noise and errors stemming from approximations to the time evolution operator, the error in local observables decays double-exponentially with the amount of qubits used in the site-wise digitization.\footnote{That is, for a local observable, the percent error $\epsilon$ in the low-energy eigenvalues of the digitized observable compared to their undigitized counterparts scales as $\epsilon \sim 2^{-2^{n_q}}$. In fact, it was found in Ref.~\cite{Klco:2018zqz} that $n_q=4$ provides sufficient precision in local observables, so that other sources of error dominate. Furthermore, it was also found that the $\phi$ basis is advantageous over other representations, such as the basis of harmonic oscillator eigenstates, and strategies associated with improved actions that are often used in lattice QCD. This advantage stems from the simplicity and efficiency of the implementations of operators in the theory (Eqs.~\eqref{eq:phi_qubits} and \eqref{eq:pi_qubits}).}

The JLP prescription gives an efficient implementation of the operators using quantum gates \cite{Jordan:2011ci,Jordan:2012xnu}. Operators in the Hamiltonian of Eq.~\eqref{eq:lattice_h} can be broken into two non-commuting groups: those that contain $\phi$, and those that contain $\Pi$. The operators made up of powers and tensor products of $\phi$ are diagonal in the $\phi$ basis, and are straightforward to implement. Aside from factors of the identity, $\phi^2_j$ and $(\phi_{j+1}-\phi_j)^2$ contain only $ZZ$ operators and can be implemented using $O(n_q^2)$ two-qubit gates. Similarly, $\phi^4_j$ can be implemented with $O(n_q^4)$ two-qubit gates, as it has additional $ZZZZ$ terms for $n_q \geq 4$. $\Pi^2_j$ is translated into gates by using a site-wise quantum Fourier transform (QFT) to switch into the conjugate momentum basis where $\Pi$ is diagonal. Using this, $\Pi^2_j$ can be implemented with $O(n_q^2)$ two-qubit gates as well.\footnote{Note that for a device with nearest-neighbor (NN) two-qubit gate connectivity, as is the case for most superconducting quantum computers, there is an extra two-qubit gate overhead to convert circuits from arbitrary connectivity to NN. To reduce the two-qubit gate depth, efficient methods for the decomposition of specific circuit elements to NN connectivity have been developed \cite{10.5555/2011827.2011828,Park:2023goh,Farrell:2024fit}.}

Conventionally, the time evolution operator is implemented in quantum simulations using first-order Trotterization: 
\begin{align}
    e^{-itH} = \left( U_\text{trot}^{(1)}\left(\frac{t}{n}\right)\right)^n + O\left(\left(\frac{t}{n}\right)^2\right), \quad \quad \quad U_\text{trot}^{(1)}(t) = e^{-it(H_\phi + H_\text{kin} + H_\text{int})}e^{-itH_\Pi},
\end{align}
where the terms in the exponential are organized into commuting groups. Higher-order product formulas can be recursively constructed by symmetric combinations of $U_\text{trot}$ via the formula: $U_\text{trot}^{(2n)}(t) = U_\text{trot}^{(n)}(t/2)U_\text{trot}^{(n)}(-t/2)^\dagger$ \cite{suzuki_1976}. These formulas have more favorable error scaling with the step size $t/n$, at the cost of a larger gate depth. In this work, $U_\text{trot}^{(2)}$ is used as a point of comparison against variational methods. After two-qubit gate cancellations and reordering terms, one second-order Trotter step can be implemented with a two-qubit gate depth of 20 for $n_q=2$.

\section{Scalable variational circuits}
\label{sec:scalable_variational_circuits}
The preparation of an arbitrary state on a quantum computer requires a number of gates that grows exponentially with system size \cite{Knill:1995kz}. Fortunately, states in physical theories have properties, such as symmetries, that can be utilized to simplify the state preparation task. This was studied in the context of using the locality of correlations to systematically truncate exact scalar field theory state preparation circuits in Refs.~\cite{Klco:2019yrb,Klco:2020aud}. As was recently demonstrated for state preparation in Refs.~\cite{Farrell:2023fgd,Farrell:2024fit}, the combination of these ideas with variational methods proves to be a powerful way to create compressed scalable circuits, while not sacrificing accuracy.

There always exists a unitary that takes one particular state $|\psi_\text{ansatz}\rangle$ to another target state $|\psi_\text{targ}\rangle$ in the Hilbert space. Implementing this unitary is the goal of state preparation (e.g., $|\psi_\text{ansatz}\rangle = |\psi_\text{vac}\rangle,\,|\psi_\text{targ}\rangle = |\psi_\text{wp}\rangle$) and time evolution (e.g., $|\psi_\text{ansatz}\rangle =|\psi_\text{wp}\rangle,\,|\psi_\text{targ}\rangle = e^{-itH}|\psi_\text{wp}\rangle$). While many methods exist to approximately construct these unitaries, they often produce circuits that are too deep to implement on currently-available hardware. Furthermore, these methods are often general, even though only a single or several matrix elements are of interest in a simulation. In this sense, the problem may be reduced from finding an implementation of an operator that properly transforms all states in the Hilbert space, to that of finding an operator that implements $|\psi_\text{ansatz}\rangle \rightarrow |\psi_\text{targ}\rangle$. To take advantage of this simplification and compress unitaries given by conventional methods, variational circuits can be used to approximate the action of these unitaries in a systematically-improvable way.

Suppose a quantum circuit $U(\vec{\theta})$ is parameterized by $\vec{\theta}$. The task of the variational algorithm is to choose the parameters $\vec{\theta}_\text{opt}$ that implement $|\psi_\text{ansatz}\rangle \rightarrow |\psi_\text{targ}\rangle$ as closely as possible. To measure the quality of the prepared state $|\psi_\text{ansatz}(\vec{\theta})\rangle = U(\vec{\theta})|\psi_\text{ansatz}\rangle$ against $|\psi_\text{targ}\rangle$, the local infidelity is used: 
\begin{align}
    I_d\left(\rho_\text{targ}, \rho_\text{ansatz}(\vec{\theta})\right) &= 1-\left(\text{tr} \sqrt{\sqrt{\rho_\text{targ}} \rho_\text{ansatz}(\vec{\theta}) \sqrt{\rho_\text{targ}}}\right)^2.
    \label{eq:local_infidelity}
\end{align}
Here, $\rho_\text{targ}$ and $\rho_\text{ansatz}$ are the density matrices of the target and prepared states, respectively, and the subscript $d$ indicates that reduced states spanning $d$ spatial sites are used.\footnote{For pure states $\rho_\text{targ}$ and $\rho_\text{ansatz}$ (i.e., $d=L$), this expression reduces to $I_L=1-|\langle \psi_\text{targ}|\psi_\text{ansatz}\rangle|^2$.} Since the global fidelity $|\langle\psi_\text{targ}|\psi_\text{ansatz}\rangle|^2$ will become arbitrarily small with increasing system size, the reduced state infidelity is used instead to measure the local quality of $|\psi_\text{ansatz}\rangle$. This quantity is particularly relevant for translationally-invariant states, or translationally-invariant states with local perturbations, since these have a repeating local structure. 

The mass gap and locality of interactions appearing in physical systems with Hamiltonians such as Eq.~\eqref{eq:lattice_h} enable scaling up quantum simulations of these systems. Correlations in the ground states of such systems decay exponentially with separation beyond the correlation length (i.e., for $|i-j|>\xi,\, \langle\phi_i\phi_j\rangle \sim e^{-|i-j|/\xi}$) \cite{Hastings:2005pr}. Here, the correlation length $\xi$ is defined to be the inverse of the gap (equivalently, the inverse of the mass of the lightest particle), $\xi=1/m_\text{particle}$.\footnote{In the free theory, $m_\text{particle}\rightarrow m$ as $\delta_\phi \rightarrow 0$, while in the interacting theory, $m_\text{particle} = m_\text{particle}(m,\lambda)$. $m_\text{particle}\neq m$ in this work as a result of the digitization used $(n_q=2)$. This digitization also introduces interactions (see App.~\ref{sec:digitization_effects}).} Because of this, the finite-volume ground-state wavefunction for $L\gg\xi$ is exponentially close to its infinite-volume form. This, in turn, means that local observables measured in the ground state agree with their infinite-volume values up to corrections that are exponentially small in the system size, $\langle A \rangle_L = \langle A \rangle_\infty + \mathcal{O}(e^{-L/\xi})$. 

The exponential convergence of ground states with system size implies that the structure of the circuits that create these states also converges exponentially \cite{Farrell:2024fit}. In turn, this means that the circuits that act on these ground states also exhibit the same exponential convergence. Importantly, this applies not only to ground states, but generally to states whose structure does not change as the system size increases (provided the correlations decay as above). In particular, fixed-width wavepackets of single particles meet these criteria due to their locality. This property enables the determination of circuits that can be scaled to arbitrary system sizes. If the target state $|\psi_\text{targ}\rangle$ is known for a small system size, the equivalent state can be created in a larger system with the following algorithm to create SVCs:
\begin{enumerate}
    \item Create a parametrized circuit ansatz that implements the unitary $U(\vec{\theta})$. The ansatz must have a structure that can be scaled up to arbitrary system sizes. \label{enum:ansatz}
    \item Optimize the parameters in $U(\vec{\theta})$:\label{enum:optimization}
    \begin{enumerate}
        \item Prepare the target state $|\psi_\text{targ}\rangle$ to a desired precision using a known method.\footnote{The classical methods to determine $|\psi_\text{targ}\rangle$ for ground-state preparation, particle creation, and time evolution are given in Sec.~\ref{sec:lattice_scalar_field_theory}. Their quantum counterparts are discussed in Sec.~\ref{sec:circuits}.}
        \item Initialize the initial state $|\psi_\text{ansatz}\rangle$.
        \item Optimize the parameters $\vec{\theta}$ by minimizing the local infidelity: $\displaystyle \min_{\vec{\theta}}\, I_d\left(\rho_\text{targ}, \rho_\text{ansatz}(\vec{\theta})\right)$, with $|\psi_\text{ansatz}(\vec{\theta})\rangle = U(\vec{\theta})|\psi_\text{ansatz}\rangle$.\label{enum:infidelity}
    \end{enumerate}
    \item Repeat steps \ref{enum:ansatz}, \ref{enum:optimization} for a set of increasing system sizes $L$. Fit an exponential to the parameters $\vec{\theta}$ and extrapolate to $L$ of interest. 
\end{enumerate}
\begin{figure*}
{\includegraphics[width=0.9\linewidth]{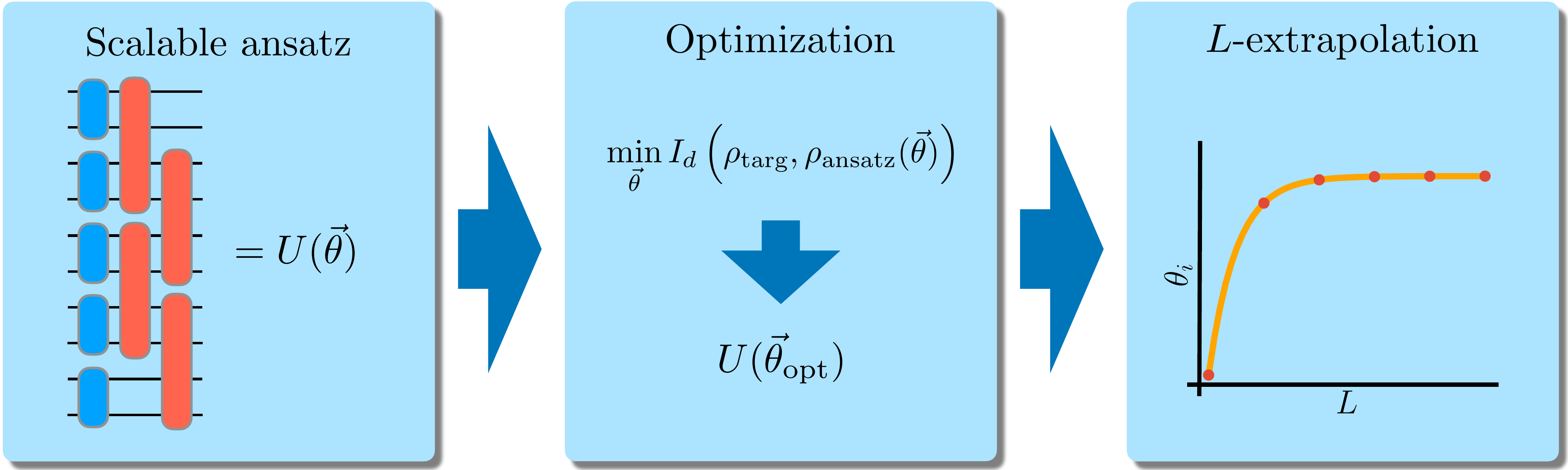}}
\caption{The algorithm to create variational circuits that can be scaled up to arbitrary system sizes \cite{Farrell:2023fgd}. A parametrized ansatz with a scalable structure is chosen by considering symmetries of the system and properties specific to the target unitary. The infidelity between the target state and the produced state is minimized to optimize the variational circuit. This process is repeated for a series of increasing system sizes. The parameters are then extrapolated to an arbitrary system size of interest. }
\label{fig:scalable_variational_circuits_procedure}
\end{figure*} 
This algorithm is shown schematically in Fig.~\ref{fig:scalable_variational_circuits_procedure}. Depending on the convergence properties of the system being studied, this algorithm can be carried out using simulations on a classical computer, or on a small register of a quantum computer.\footnote{If running on a quantum computer, methods such as those described in Ref.~\cite{Huang:2024qmc} may be used to estimate the infidelity.}

Because this algorithm creates a variational unitary by optimizing a single matrix element $\langle\psi_\text{targ}|U(\vec{\theta})|\psi_\text{ansatz}\rangle$, there is no guarantee that the implemented $U$ will behave as expected for different starting states. However, since only initial states $|\psi_\text{ansatz}\rangle$ with very specific properties are of interest in quantum simulations, this algorithm is ideal for the simulation task. If an approximation to the unitary is needed that is valid for more general states, step \ref{enum:infidelity} of the algorithm can be modified by minimizing the average of the infidelity over several states instead: 
\begin{equation}
    \min_{\vec{\theta}} \frac{1}{N}\sum_{i=0}^{N-1} I_d\left(\rho_{\text{targ},i}, \rho_{\text{ansatz},i}(\vec{\theta})\right).
\end{equation}
The set of training states states $\{|\psi_{\text{targ},i}\rangle,\,|\psi_{\text{ansatz},i}\rangle\}$ can be chosen to cover all the states of interest, or some representative subset of them \cite{Mansuroglu:2021azm,Tepaske:2022uad,Guo:2024tnb}. This process takes advantage of the simplification of the problem of approximating a general unitary in the case of quantum simulation. Creating variational operators by optimizing matrix elements makes this algorithm much more tractable both for classical and quantum computing, than if the whole operator were optimized.

The power of this method lies within the choice of the ansatz circuit. The ansatz must be carefully chosen to have enough degrees of freedom and expressivity to have a chance at replicating the action of its target unitary. Greedy algorithms such as ADAPT-VQE have recently emerged as efficient and powerful ways to build the ansatz circuit \cite{Grimsley_Economou_Barnes_Mayhall_2019,Feniou:2023gvo}. This method, in the form it is used in this work, is summarized below. 
\begin{enumerate}
    \item Begin with an initial state $|\psi_\text{ansatz}\rangle$, a target state $|\psi_\text{targ}\rangle$, and a pool of operators $\{O_i\}$ made up of commutators of terms in the Hamiltonian. These operators respect some or all of the symmetries of the system.
    \item Choose the next operator to append to the ansatz:
    \begin{enumerate}
        \item For each operator in the pool, optimize the local infidelity as a function of all parameters $\vec{\theta}$ in the ansatz, $\displaystyle \min_{\vec{\theta}}\, I_d\left(\rho_\text{targ}, \rho_\text{prep}(\vec{\theta})\right)$.\footnote{ADAPT-VQE was originally developed for general problems where $|\psi_\text{targ}\rangle$ is not known a-priori, so the optimization was implemented by minimizing the energy to find the ground state.}
        \item Append the operator that decreases the infidelity the most to the ansatz, and update all the parameters $|\psi_\text{ansatz}\rangle \rightarrow e^{-i\theta_iO_i}|\psi_\text{ansatz}(\theta_0,\theta_1,\dots,\theta_{i-1})\rangle = |\psi_\text{ansatz}(\theta_0,\theta_1,\dots,\theta_{i})\rangle.$
    \end{enumerate}
    \item Repeat the step 2 until the required convergence criteria is reached, such as a combination of circuit depth and infidelity. Use the parameters determined at the previous step as an initial guess for the next step.
\end{enumerate}
This algorithm corresponds to steps 1-2 of the SVC framework as one way to determine the ansatz and optimize the variational parameters at a fixed system size. When used with SVC, this algorithm is known as SC-ADAPT-VQE \cite{Farrell:2023fgd,Farrell:2024fit}. Incorporating symmetries directly into the ansatz makes it more optimizable, and guarantees that the resulting state obeys the symmetries of the system (provided the input state also has the symmetries). This method and its numerous variants have been shown to perform extremely well for ground-state preparation. In particular, it has been used for ground-state preparation and particle creation in the context of translationally-invariant systems \cite{Farrell:2023fgd,Feniou:2023gvo,Farrell:2024fit,VanDyke:2022ffj,Gustafson:2024bww}. Impressively, it was also shown that although greedy methods may not find the best sequence of operators, almost all other sequences perform substantially worse, which removes the need to optimize over all possible sequences. 

It is important to consider the tradeoff between ease of optimization and circuit depth when selecting a variational ansatz. The simplest shallow and expressive ansatz is the brickwall ansatz, where each layer consists of parameterized single-qubit rotations followed by entangling gates. Compared to ADAPT-VQE, this approach trades the gate depth on the quantum device for the classical computational complexity involved in optimizing a large set of parameters for a set of circuits of increasing size. Optimizing a circuit with many parameters is a computationally intensive task that is susceptible to the well-known problem of barren plateaus \cite{Cerezo:2023nqf,Larocca:2024plh}. Properties of the target state or target operator can be used to simplify the ansatz structure and reduce the number of parameters that need to be optimized. To ease the classical optimization task, new brickwall layers can be added and optimized iteratively using the same algorithm as described above.   

As such, finding the right balance between circuit depth and structure is imperative to create a trainable variational operator that mimics the target unitary with a shallow circuit. Using the physics of the system and the operator being approximated to inform the design of the ansatz proves to be a powerful way to increase the performance of this variational method \cite{Sim:2019yyv,Seki:2020nnj,Funcke:2020vkw,Anschuetz:2022wvo,Gibbs:2024ggs}.

\section{Quantum simulation of scalar field theory using scalable variational circuits}
\label{sec:circuits}
\begin{figure*}
    \includegraphics[width=0.9\linewidth]{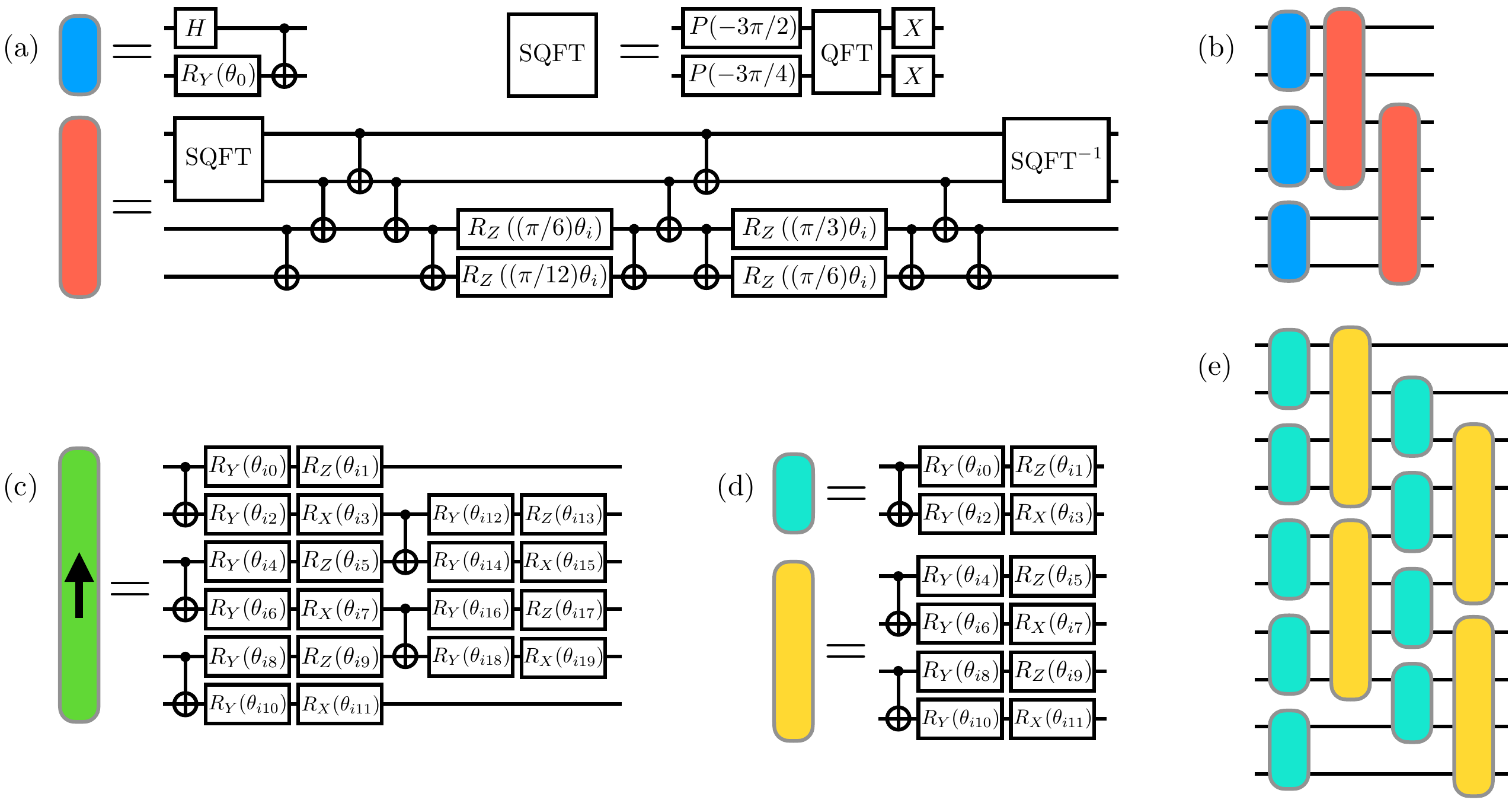}
    \caption{The circuit elements that are used for state preparation and time evolution in simulations of scattering. (a): The circuit elements used to prepare the vacuum. The input state to SC-ADAPT-VQE is a real, symmetric tensor product state over the spatial sites prepared by a single-site variational operator (blue). The circuit implementing $e^{-i\theta O_1}$ (red) is used to couple neighboring sites. The symmetric version of the QFT defined in Ref.~\cite{Klco:2018zqz}, SQFT, is used to account for the symmetric digitization of $\phi$ and $\Pi$ around 0. (b): The translationally-invariant circuit to prepare the vacuum. One layer of SC-ADAPT-VQE is Trotterized into separate terms coupling even-odd and odd-even spatial sites. This circuit has depth 25 and 3 variational parameters (one for the input state, and one for each application of $O_1$). (c): One layer of the brickwall circuit that is used to create particle wavepackets with negative momentum (indicated by the up arrow) on top of the vacuum. To create the corresponding positive-momentum particles, the circuit is flipped and site-wise SWAP gates are added. Each layer has 20 parameters and is depth 2. The building block of this circuit is the variational gate introduced in Ref.~\cite{Madden:2021dax}. (d): The circuit elements that are used to implement time evolution. Both single- (cyan) and two-site (yellow) operators are used, mimicking the terms present in the present in the Hamiltonian of Eq.~\eqref{eq:lattice_h}. The building block of these circuits is the same as (c). (e): Two layers of the translationally-invariant circuit used to implement time evolution. Each layer has depth 2 and 12 parameters. Single- and two-site operators are applied in an iterative fashion.}
    \label{fig:circuit_elements}
\end{figure*}

The scalable variational circuits procedure can be carried out on a classical computer or a small quantum computer. Throughout this work, classical computational resources are leveraged to determine parameters that are extrapolated to system sizes of interest to be run on quantum computers. This work uses $n_q=2$ qubits to represent the state of $\phi$ at each site, for the theory with bare mass $m=1/2$. The method of Ref.~\cite{Klco:2018zqz} is used to choose the cutoff $\phi_\text{max}=1.5$ to minimize field digitization effects, which are studied in App.~\ref{sec:digitization_effects}. The dynamics of free $(\lambda=0)$ and interacting $(\lambda=2)$ wavepackets is compared. For the parameters chosen, the correlation lengths $\xi_\lambda$ of the free and interacting theory are $\xi_0=2.4503$ and $\xi_2=1.5817$, respectively. These are determined, as in the previous section, by computing the inverse gap for small but increasing system sizes $L$, fitting an exponential, and extrapolating to the infinite-volume limit. The circuit elements that are used to compose the variational circuits throughout this section are shown in Fig.~\ref{fig:circuit_elements}. Details of the optimization strategies and choices are given in App.~\ref{sec:optimization}. The resulting parameters $\vec{\theta}_\text{opt}$ for all components of the variational circuit are given in App.~\ref{sec:variational_params}.

\subsection{Vacuum preparation}
\label{sec:circuits_vac_prep}
As a first step toward simulating a scattering process, the SVC framework is used to prepare the ground state of the scalar field theory. Together with a variational input state, SC-ADAPT-VQE is used to prepare the vacuum of the theory both in the free and interacting cases. Compared to other methods for preparing ground states, such as Quantum Imaginary Time Evolution \cite{Motta:2019yya} or subspace methods \cite{Yoshioka:2024lle}, SC-ADAPT-VQE gives shallow circuits with few variational parameters, making optimization easy and enabling implementation on modern-day devices.

In the free theory, $|\psi_\text{vac}\rangle$ defined in Eq.~\eqref{eq:vacuum_wavefunction} can be used as $|\psi_\text{targ}\rangle$. Approximations to the vacuum of the interacting theory can be found analytically using standard perturbative methods for small $\lambda$ as is discussed in Sec.~\ref{sec:lattice_scalar_field_theory}. Fortunately, for the chosen system parameters $m$ and $\lambda$, the variational parameters of the circuits initializing these states show convergence for small-enough system sizes where exact diagonalization (ED) is viable. Taking advantage of this, the interacting ground state determined from ED is used as $|\psi_\text{targ}\rangle$. 

It is important to choose a good initial input for the SC-ADAPT-VQE algorithm. There has been recent work using MPS or other ansatze as input to variational algorithms to improve their performance \cite{Ravi:2022cwu,Khan:2023uhz,Fomichev:2023mtt}. Time reversal symmetry (implying the reality of the ground-state wavefunction), $\phi \rightarrow -\phi$ symmetry, and translational invariance can be used to choose a judicious input state to the greedy algorithm. The circuit to initialize the input state, shown in blue in Fig.~\ref{fig:circuit_elements}a, uses one variational parameter to create an arbitrary real, symmetric state on each spatial site. This product state can be viewed as the vacuum $| \widetilde{\psi}_\text{vac}\rangle$ of the theory with the correlation matrix $K=V^\dagger E V$ truncated to only contain elements on its diagonal: 
\begin{equation}
    \langle{\vec{\phi}} | \widetilde{\psi}_\text{vac} \rangle = \mathcal{N} e^{-\frac{1}{2}\vec{\phi}^T \text{diag}(K)\vec{\phi}}. \label{eq:adapt_vqe_input_state}
\end{equation}
By choosing this state, or a state like this, the variational algorithm only has to build out the correlations between spatial sites.

The Hamiltonian of the theory is real and time reversal invariant, so it must have real eigenstates. It is found that an operator pool built out of commutators of terms in the Hamiltonian, similar to the pools used in Refs.~\cite{Farrell:2023fgd,Farrell:2024fit}, is effective. A pool with this structure ensures the reality of the prepared vacuum by applying only real unitaries, and maintains the symmetries of the system. The lowest-order operator in $\phi$ and $\Pi$ that acts over more than one site and is not included in the Hamiltonian is derived from the commutator of terms in $H_\text{kin}$ and $H_\Pi$, $O_1 = -\frac{i}{2}\sum_j[\phi_j \phi_{j+1},\Pi^2_j]$. The canonical commutation relation $[\phi_i, \Pi_j] = i\delta_{ij}$ only holds in the limit $\delta_\phi \rightarrow 0$ (i.e., only for the undigitized operators $\phi_j$, $\Pi_j$). Nevertheless, this relation can still be used at nonzero $\delta_\phi$ to simplify the operators in the SC-ADAPT-VQE pool because the requisite symmetries are maintained. This makes operators such as $O_1$ easier to implement using quantum gates. Using this, $O_1$ can be rewritten as 
\begin{align}
    O_1 = \sum_j\Pi_j \phi_{j+1}, \label{eq:pi_phi}
\end{align}
which can readily be implemented on a quantum device using the techniques described in Sec.~\ref{sec:qubit_representation}. To build out longer correlations, the separation between the sites that $O_1$ acts on can be increased: 
\begin{equation}
    O_d = \sum_j\Pi_j \phi_{j+d}. \label{eq:pi_phi_d}
\end{equation}
This operator can similarly be obtained from the commutator of $\Pi_j^2$ with a longer-range verion of $H_\text{kin}$. The pool used for SC-ADAPT-VQE consists of the operators $\{O_d\}$  for a range of increasing $d$. Note that the operators $O_d$ are imaginary in the $\phi$-basis. This is because $\phi_j$ is real and $\Pi_j$ is imaginary since it is the conjugate momentum operator. As a result, $e^{-i\theta O_d}$ is a real unitary.

The circuit implementing $e^{-i\theta O_1}$ on a pair of sites is given in red in Fig.~\ref{fig:circuit_elements}a. Because $O_1$ acts on all pairs of neighboring sites, it must be Trotterized into an operator acting on even-odd sites, and an operator acting on odd-even sites. This is shown in Fig.~\ref{fig:circuit_elements}b, where there are two layers of the $O_1$ circuit. This Trotterization breaks translational invariance, so one variational parameter is used for each Trotter layer to partially mitigate this effect. 

\begin{figure*}
    \includegraphics[width=\linewidth]{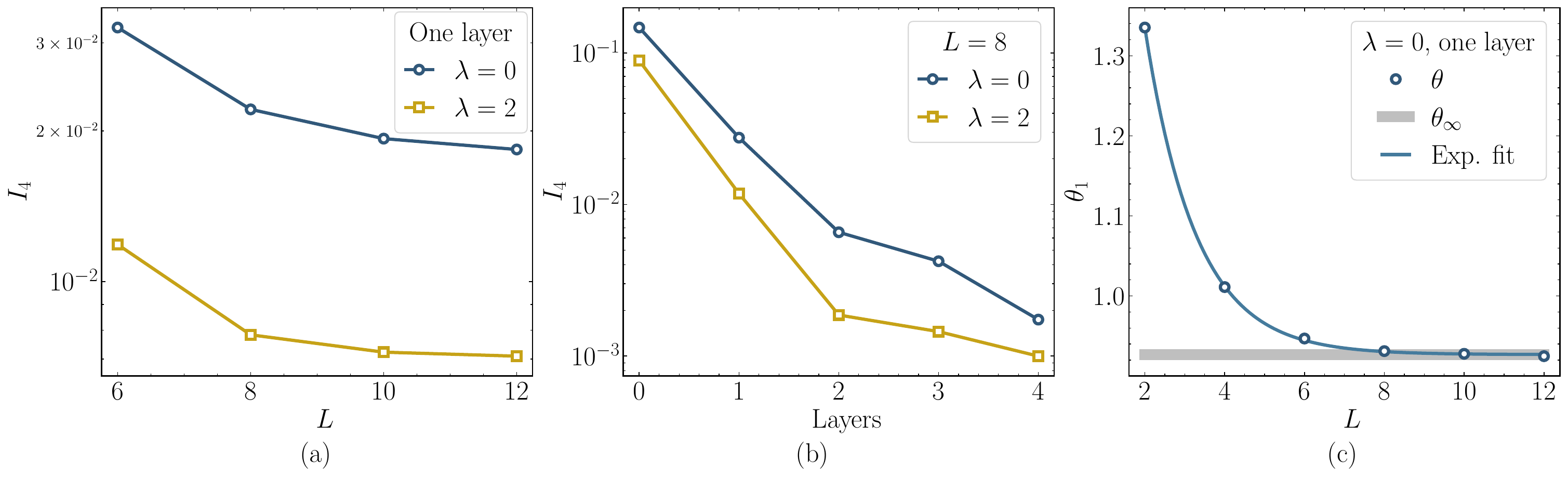}
    \caption{Convergence of the wavefunction and state preparation parameters in the vacuum preparation circuit shown in Fig.~\ref{fig:circuit_elements}b for both the free and interacting theory. (a): $I_4$ taken over a region of four contiguous lattice sites (eight qubits) as a function of system size. One layer of SC-ADAPT-VQE is used. (b): $I_4$ as a function of number of SC-ADAPT-VQE layers used in the ansatz for a $L=8$ system. Each layer uses a longer-range operator $O_d$ given in Eq.~\ref{eq:pi_phi_d} to create correlations that span more lattice sites. (c): An example of the exponential convergence of one of the angles parameterizing the $\lambda=0$ vacuum preparation circuit with one layer of SC-ADAPT-VQE, as a function of system size. The remainder of the angles are given numerically in App.~\ref{sec:variational_params}.
    }
    \label{fig:vac_prep_figs}
\end{figure*}

Figure \ref{fig:vac_prep_figs} shows the convergence of the variationally-prepared vacuum. The local infidelity over four sites, $I_4$, is used to measure the quality of prepared state, spanning roughly 2-3$\times$ the correlation length of the infinite-volume vacuum. With just one SC-ADAPT-VQE layer, $I_4$ can be seen to be non-increasing in Fig.~\ref{fig:vac_prep_figs}a, meaning that the local quality of the prepared function is not degrading with system size for both $\lambda=0$ and $\lambda=2$. The infidelity is systematically improved by applying more layers of $O_d$ with successively larger $d$. The result of longer-range operators creating correlations that span more lattice sites is seen in the plots of $I_4$ in Fig.~\ref{fig:vac_prep_figs}b. In both of these figures, it can be seen that the $\lambda=2$ wavefunction converges faster as a result of the smaller correlation length in the interacting theory. Figure \ref{fig:vac_prep_figs}c is a representative example of the convergence of the parameters with increasing system size. These plots indicate that the $L$-extrapolation is applicable and the parameters to initialize the ground state for an arbitrary system size can be determined using modest $L$, where classical simulation is possible.\footnote{If the SVC method is run using classical computation, the correlation length of ground states that can be prepared with the presented method is limited by the maximum number of qubits that can be simulated classically.}

The circuit shown in Fig.~\ref{fig:circuit_elements}b, consisting of the circuit preparing the symmetric input state and one layer of SC-ADAPT-VQE, is used to initialize the vacuum. It has a two-qubit gate depth of 25, making it shallow enough to save circuit depth for wavepacket preparation and time evolution. The extrapolated parameters $\vec{\theta}_\text{opt}$ that implement the vacuum preparation are given in Table \ref{tab:gs_prep_angles}, and numerical $I_4$ values are given in Table \ref{tab:error_budget}.

\subsection{Wavepacket creation}
\label{sec:circuits_wp_prep}
The creation of particles on top of the prepared vacuum requires initializing local excited states. JLP proposed a method of creating wavepackets in the free theory by applying unitary evolution under $a_k^\dagger$ and $a_k$ defined in Eq.~\eqref{eq:ap}, and then adiabatically turning on the interaction as described in Sec.~\ref{sec:lattice_scalar_field_theory}.\footnote{Whereas JLP specify the form of their wavepackets in position space, the method used in this work specifies the target profile in $k$-space. See App.~\ref{sec:wp_creation_details} for a comparison.} Instead of this, SVC is used to compress the wavepacket preparation circuit by taking advantage of classical computation. This step of the algorithm prepares a wavepacket whose size remains fixed as the total system size increases. While this method does not allow for the scalable preparation of wavepackets of increasing size (as will be necessary for simulations approaching the continuum), it utilizes the scalability in total system size, including vacuum regions.\footnote{In this case, the classical computing capability limits the maximum size of the wavepacket.} This is useful because a finer spacing in $k$-space increases the spatial momentum resolution of the wavepacket, building a more accurate representation of wavepackets in the continuum. Furthermore, wavepackets may only scatter between different lattice momenta in a lattice quantum field theory. Increasing the system size increases total number of lattice momenta and thus more faithfully reproduces the target continuum process.

\begin{figure*}
    \includegraphics[width=\linewidth]{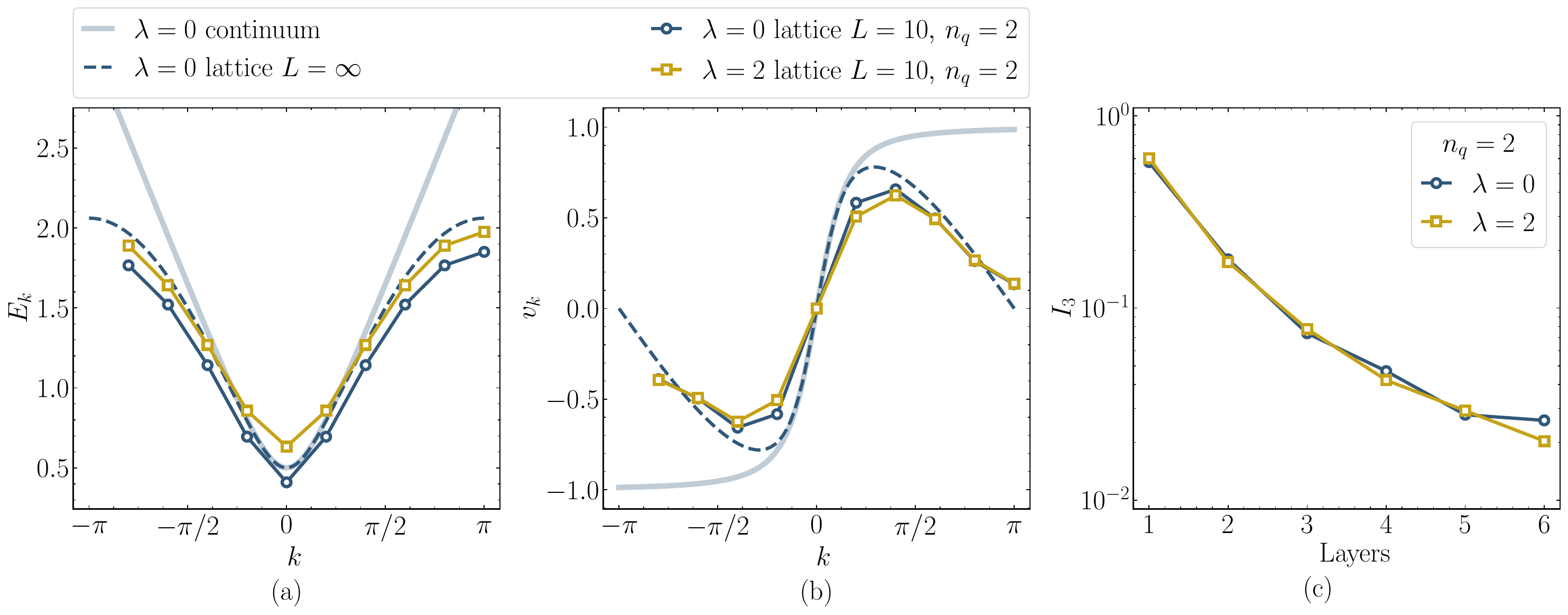}
    \caption{(a): The dispersion relation $E_k$ and (b): the group velocity $v_k$ for $m=1/2$ and $\lambda=0,2$. Exact computation is possible in the continuum (faint solid lines) and in the free lattice theory (dashed lines). The values of $E_k$ and $v_k$ are computed numerically by using ED on the Hamiltonian of Eq.~\eqref{eq:lattice_h} and projecting onto each momentum sectors (solid lines with circle and square markers). The numerical computations are done on a lattice of $L=10$ spatial sites with $n_q=2$ qubits per site. Discrete gradients are used for the numerical computations of $v_k$. (c): The convergence of the prepared particle wavepackets occupying three spatial sites with momentum $k=-\pi/3$, as a function of number of layers of the circuit from Fig.~\ref{fig:circuit_elements}c. The local infidelity spanning the width of the wavepacket, $I_3$, is used to measure the quality of the prepared wavefunction for $\lambda=0,2$.}
    \label{fig:e_k_v_k_wp_prep_convergence}
\end{figure*}

To reduce the impact of lattice artifacts, the momenta of the wavepackets must be carefully chosen. The infinite-volume lattice dispersion relation (Eq.~\eqref{eq:dispersion}) mimics that of the continuum, $E_{k,\text{cont.}} = \sqrt{m^2+k^2}$, for small $k$. These are shown by the faint solid and dashed lines respectively in Fig.~\ref{fig:e_k_v_k_wp_prep_convergence}a. The group velocity of a particle wavepacket with momentum $k$ is found in the infinite-volume lattice theory to be
\begin{align}
    v_k &= \frac{\partial E_k}{\partial k} = \frac{\sin{k}}{\sqrt{m^2 + 4 \sin^2{\frac{k}{2}}}}, \label{eq:group_velocity}
\end{align}
and in the continuum to be 
\begin{align}
    v_{k,\text{cont.}} &= \frac{\partial E_{k,\text{cont.}}}{\partial k} = \frac{k}{\sqrt{m^2 + k^2}}. \label{eq:group_velocity_cont}
\end{align}
These are plotted as the dashed and faint solid lines respectively in Fig.~\ref{fig:e_k_v_k_wp_prep_convergence}b. $E_k$ and $v_k$ can be computed numerically in the digitized theory for any value of $\lambda$ by using ED and projecting onto each momentum sector. These numerically-calculated values are shown in Figs.~\ref{fig:e_k_v_k_wp_prep_convergence}a and b by the solid lines with circle and square markers for $L=10$.

To simulate a high-energy collision, the highest momenta should be selected where the (digitized) lattice dispersion relation is still close to that of the continuum. With $L=10$ spatial sites, the resolution is poor and the digitized $v_k$ only match the continuum values at small $k$. The agreement can be improved with finer resolution by increasing the number of lattice sites. Finite-volume effects are reduced by using many lattice sites outside the interaction region. Based on this and on considerations of circuit depth and spread in $k$-space, wavepackets with $k=\pm\pi/3$ are chosen, with spread $\sigma_k=\pi/3$. The choice of $\sigma_k$ determines the support of the wavepacket in both position and momentum space. To limit the quantum resources required, the wavepacket width is truncated so that the circuits that create these states only act where the wavepackets have nonvanishing support. In this work, the spatial extent of the wavepackets is fixed to three lattice sites, and the circuits that prepare them act over $3n_q=6$ qubits.\footnote{Similar to the field digitization, choosing the extent of the wavepackets in position and momentum space according to the Nyquist-Shannon Sampling Theorem will ensure a double-exponential convergence in precision for future calculations.} A final choice in the setup of the simulation is the initial location of the wavepackets. The wavepackets should be well-separated at $t=0$ so they can be approximately treated as asymptotic ``in" states. While this is strictly true only for infinite separation, the wavepackets are local, and their extent is fixed by the width of the circuit preparing them. Taking advantage of the locality of the $\phi^4$ interaction, the wavepackets are initialized to be separated by one spatial site. This construction may need to be modified for theories where interactions have less locality.

Leveraging the analytic knowledge of the free theory, the wavefunction of the wavepacket $|\psi_\text{wp}\rangle$ defined in Eq.~\eqref{eq:wp_wavefunction} is used as $|\psi_\text{targ}\rangle$ for the variational optimization. The adiabatic turn-on of interactions given by Eq.~\eqref{eq:u_adiabatic} is simulated classically, producing interacting wavepacket wavefunctions that are then used for $|\psi_\text{targ}\rangle$ in SVC for the interacting theory. See App.~\ref{sec:wp_creation_details} for details on the classical determination of states used as $|\psi_\text{targ}\rangle$ for this step of state preparation. With this method, interacting wavepackets are prepared directly on top of the interacting vacuum. This way, in addition to reducing the circuit depth, SVC also removes the necessity for implementing the adiabatic evolution to turn on interactions. 

Because the region where the desired wavepacket has support does not grow with increasing system size $L$, the amount of qubits over which the variational circuit has to act is fixed by the spatial width of the wavepacket. Scaling these circuits is simple and straightforward since the number of parameters and structure do not change with system size. This is another example of using physical properties of the state, in this case locality of the wavepacket, to guide the design of the variational circuit. One layer of the brickwall ansatz used for this step of the state preparation is shown in Fig.~\ref{fig:circuit_elements}c. The building blocks of this circuit are the variational gates defined in Ref.~\cite{Madden:2021dax}. This block is particularly useful because any product of gates from the universal gate set $\{\text{CNOT},R_X,R_Y,R_Z\}$ can be written in terms of products of the block with different parameters. Each layer of the brickwall ansatz has a two-qubit gate depth of 2 and uses 20 variational parameters.\footnote{The number of parameters can be reduced by incorporating the physical structure of the wavepackets (e.g., the spatial symmetry about the peak) into the circuits.}

Since spatially-separated particles are required for the ``in" states of the scattering simulation, the wavepacket creation circuits are executed for both wavepackets in parallel. The same set of parameters is used to initialize both the left and right wavepackets to maintain symmetry. The circuit preparing the left $(k=+\pi/3)$ wavepacket is created in such a way that the resulting wavepacket is the mirror image of the right $(k=-\pi/3)$ wavepacket. This is indicated by the up (down) arrows for $k=-\pi/3$ $(k=+\pi/3)$ respectively, seen in the full circuit shown in Fig.~\ref{fig:full_circuit}. Site-wise SWAP gates are used to flip the qubit mapping in Eq.~\ref{eq:phi_digitized}, increasing the two-qubit gate depth of this part of the circuit by six.

Figure \ref{fig:e_k_v_k_wp_prep_convergence}c shows the convergence of the prepared wavepacket state as a function of the number of layers of the brickwall ansatz of Fig.~\ref{fig:circuit_elements}c. It is seen that the local infidelity spanning the wavepacket width, $I_3$, is systematically improved by adding more layers. Four layers of the circuit shown in Fig.~\ref{fig:circuit_elements}c are used to prepare the free and interacting wavepackets.\footnote{Note that with the parameters set to zero, inserting CNOTs between layers of this circuit will collapse the circuit to the identity. This is used in vacuum simulations to maintain the structure of the circuit.} These brickwall circuits have two-qubit gate depths of 14 (including extra site-wise SWAP gates), and 80 parameters in total. While it is possible to run an extrapolation of the wavepacket parameters, this is not necessary. The angles parameterizing the vacuum creation circuits are observed to change on the scale of $10^{-4}-10^{-3}$ for $L=10-12$ (see Table \ref{tab:gs_prep_angles}), which is at the precision tolerance of IBM devices. The extrapolated angles for the vacuum preparation are used to prepare the equivalent vacuum state on a smaller system size for purposes of training the wavepacket circuit. The vacuum preparation angles extrapolated to $L=60$ are used in an $L=12$ system to optimize the wavepacket preparation parameters. The resulting parameters $\vec{\theta}_\text{opt}$ are given in Table \ref{tab:wp_prep_angles}, and numerical $I_3$ values are given in Table \ref{tab:error_budget}.

\subsection{Time evolution}
\label{sec:circuits_time_evolution}
Scalable variational circuits can also be used to produce compressed time evolution circuits in place of non-variational methods such as Trotterization, QDrift \cite{Campbell:2019fez}, or Linear Combination of Unitaries \cite{Childs:2012gwh}. There has been much work in recent years in this direction. The structure of the ansatze that have been used varies widely, from circuits with parameterized Trotter terms \cite{Mansuroglu:2021azm,Tepaske:2022uad,Kotil:2022vmv,Tepaske:2023mfc}, to brickwall ansatze \cite{Mizuta:2022rrz,Keever:2022hfy,Miyakoshi:2023zzc,Causer:2023wpp,Kanasugi:2024ivt,Gacon:2024bsc,Gibbs:2024emw}. The related fast-forwarding techniques \cite{Commeau:2020aab,Gu:2021hyo} involve using variational circuits to approximately diagonalize a Hamiltonian to implement evolution in a number of gates that is constant with $t$ but scales with $L$ because of the diagonalization step.

\begin{figure*}
    \includegraphics[width=0.65\linewidth]{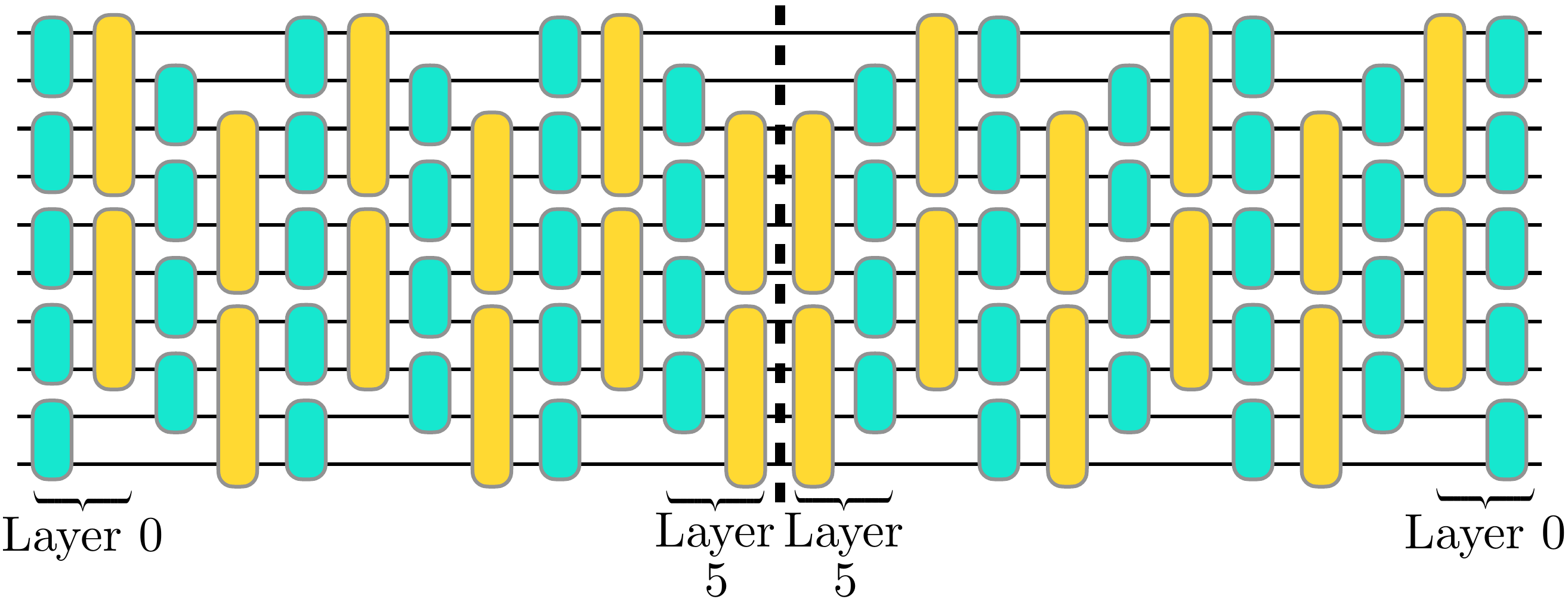}
    \caption{One variational step of the time evolution circuit. It consists of six forward layers followed by six reverse layers of the circuits of Fig.~\ref{fig:circuit_elements}d. Single- and two-body terms are used in an alternating brickwall fashion, mimicking the terms present in the system Hamiltonian.}
    \label{fig:time_evolution_circ_step}
\end{figure*}

In this work, a translationally-invariant brickwall ansatz is used to compress the time evolution operator. Note that while there is no guarantee that a shallow brickwall circuit can approximate the time evolution operator well, time evolution under any Hamiltonian with local interactions is subject to Lieb-Robinson bounds \cite{Lieb:1972wy}. Moreover, there is a maximum value for the group velocity of excitations from the lattice dispersion relation (Eq.~\eqref{eq:group_velocity}) \cite{Farrell:2024mgu} so it is reasonable to expect that low-depth brickwall circuits can reliably compress evolution under the Hamiltonian of Eq.~\eqref{eq:lattice_h} for sufficiently early times. The ansatz is made up of layers of the circuit shown in Fig.~\ref{fig:circuit_elements}e, arranged in a symmetric way, seen in Fig.~\ref{fig:time_evolution_circ_step}. Each variational step is made up of six forward layers and six reverse layers, with the reverse layers having the same parameters as the forward layers. This structure is chosen to mimic the structure of a second-order Trotter step. Each layer uses 12 parameters and has a two-qubit gate depth of two. The time evolution of the two-wavepacket state $e^{-itH}|\psi_\text{2wp}\rangle$ with 100 second-order Trotter steps on an $L=12$ system is used as $|\psi_\text{targ}\rangle$ to determine the variational circuits.

Similar to the wavepacket preparation step, the $L$-extrapolation is not necessary to implement time evolution for the times simulated in this work. However, in this case this simplification is only valid for early times $(t\sim L)$, because the time evolution circuits act on the whole lattice. Since the vacuum is not prepared exactly, and the time evolution is not implemented exactly, the vacuum regions will evolve in time \cite{Hayata:2024fnh}. This evolution will be captured by the variational optimization procedure, up to finite-size effects that scale as $O(t/L)$.\footnote{These effects are estimated and the vacuum evolution is used for error mitigation. See App.~\ref{sec:error_mitigation} for details.} To get around this limitation, the optimization can be run on larger systems for later times, or an $L$-extrapolation can be implemented. Propagating particles in a scattering simulation show more drastic effects from finite-size effects, such as wrapping around the lattice due to PBCs and re-scattering. Furthermore, larger wavepackets (needed for high-energy simulations or simulations approaching continuum) require more time to overlap, scatter, and fully separate. In these situations, it would be necessary to run the optimization for later simulation times on a larger lattice. While this is a limitation of the method in general, the scattering simulations are set up to require minimal time before scattering is observed (see Sec.~\ref{sec:circuits_wp_prep} for details). This places a limit both on the time required in the simulation, as well as the size of the interaction region that needs to be used for training the variational circuits. 

One promising way to use variational circuits for time evolution without running into these finite-size effects is to compress a single time evolution step, and use that step repeatedly in place of a Trotter step. Similar methods have been investigated in Refs.~\cite{Tepaske:2023mfc,Causer:2023wpp}. While the resulting circuits will be deeper than those produced by the current method, they will be more scalable with $L$. Another interesting direction is using MPS simulations of the time evolution operator for training \cite{Miyakoshi:2023zzc,Causer:2023wpp,Gibbs:2024emw}. This would remove the limitations placed on this method by finite-size effects. In this setup, the time evolution of one or many states may be used for training, as described in Sec.~\ref{sec:scalable_variational_circuits}. Methods that target time evolution operators in specific subspaces, such as restricted to the low-energy subspace, are currently being developed \cite{Kanasugi:2024ivt,Li:2024lrl}. The combination of subspace methods with variational time evolution has the potential to simplify the optimization task while also improving the quality of the produced circuits.

\begin{figure*}
    \includegraphics[width=\linewidth]{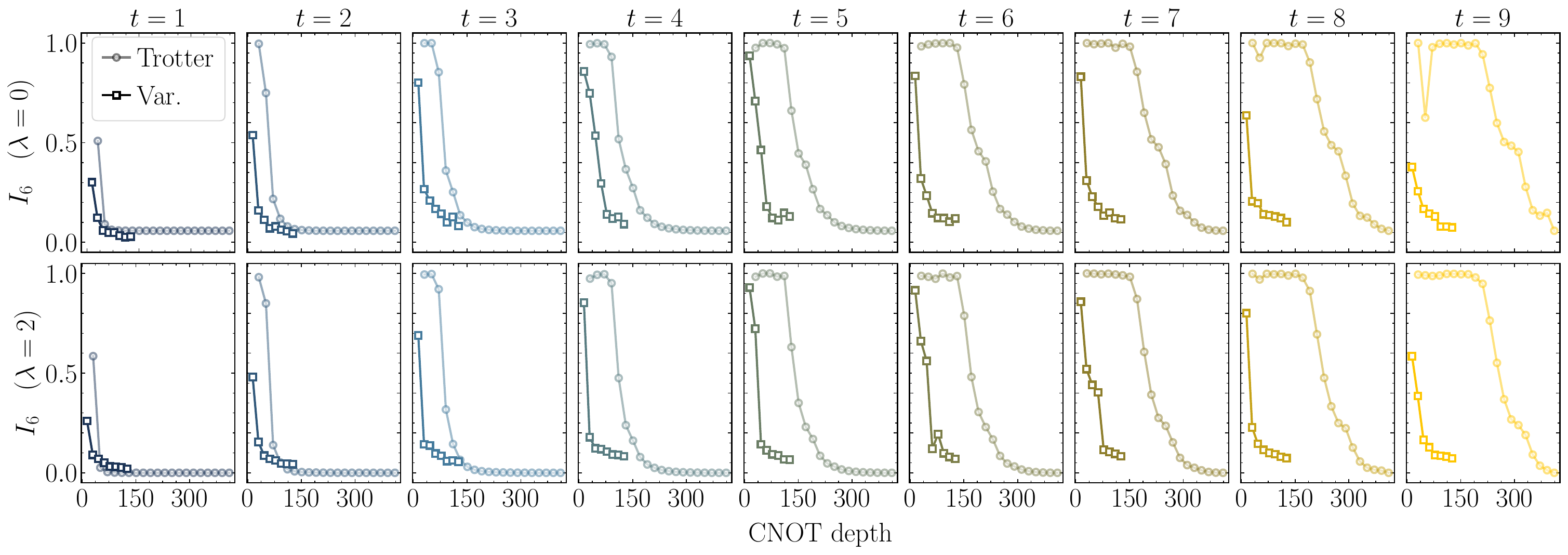}
    \caption{The convergence of the time-evolved single-wavepacket state on an $L=6$ system with $\lambda=0$ (top row) and $\lambda=2$ (bottom row) for $t=1-9$. The second-order Trotter time evolution (circles) is compared to the variational steps (squares) as a function of two-qubit gate depth. Two variational steps consisting of an increasing number of layers shown in Fig.~\ref{fig:circuit_elements}d are used. The infidelity of the whole $L=6$ system with the exact time-evolved state is used to measure the quality of the wavepacket state as it propagates through the lattice.}
    \label{fig:trot_vs_variational_fidelity}
\end{figure*}

Figure \ref{fig:trot_vs_variational_fidelity} shows the convergence of the time evolution variational circuits, compared to second-order Trotter circuits as a function of circuit depth. The plots show the infidelity as a function of two-qubit gate depth for the time evolution of a single wavepacket on an $L=6$ system. The Trotter time evolution circuits are constructed by adding more Trotter steps as described in Sec.~\ref{sec:qubit_representation}. Variational circuits fix the number of variational steps to two while increasing the depth of each step by adding layers of the circuit elements in Fig.~\ref{fig:circuit_elements}d following the structure of Fig.~\ref{fig:time_evolution_circ_step}. The results for $t=1-9$ show that the brickwall circuits are able to consistently compress the Trotter circuits. This can especially be seen at late times, where considerably deeper Trotter circuits are required to accurately implement the time evolution operator, whereas the variational circuits show a modest growth in depth. Sharp dips in the curves of both the variational and Trotter infidelities are results of finite-size effects, as well as of numerical optimization (for the variational lines). A difference in the infidelity between $\lambda=0$ (top row) and $\lambda=2$ (bottom row) can be seen, which is attributed to using fixed-depth circuits to describe states with different correlation lengths. These plots show that by increasing the number of layers (and so, the number of parameters), the variational ansatz is able to represent the time-evolved state with increasing accuracy. 

For the runs on the device, the number of variational steps is increased with the simulation time: $t=\{1,2,3\}$ use one variational step, $t=\{4,5,6\}$ use two, and $t=\{7,8,9\}$ use three. The number of layers in each step is kept fixed to six forward and six reverse layers, as in Fig.~\ref{fig:time_evolution_circ_step}. The steps are optimized separately for each simulation time to find better quality solutions for the variational circuit. They are applied in a Trotter-like fashion, so that the number of parameters is the same for all time steps. The time evolution portion of the variational circuit uses 72 parameters in total. The parameters $\vec{\theta}_\text{opt}$ that implement the time evolution are given in Tables \ref{tab:time_evolution_angles_1}-\ref{tab:time_evolution_angles_9} and the local infidelities $I_{10}$ for each time step are shown in Table \ref{tab:error_budget}.

\subsection{Summary of variational methods and error budget}
\label{sec:circuits_error_budget}
The choices in the design of the variational circuits and methods of determining parameters for a system of $L=60$ lattice sites are summarized below.
\begin{itemize}
    \item The vacuum $|\psi_\text{vac}\rangle$ is created using an input state preparation circuit and one step of SC-ADAPT-VQE, shown in Figs.~\ref{fig:circuit_elements}a and b. The circuits are trained on $|\psi_\text{targ}\rangle$ determined from ED with $L=2-12$, and parameters are extrapolated to $L=60$. This step of the state preparation has a two-qubit gate depth of 25 and three variational parameters.
    \item The two-wavepacket state $|\psi_\text{2wp}\rangle$ is initialized using two layered brickwall circuits in parallel (one layer is shown in Fig.~\ref{fig:circuit_elements}c). A single set of parameters is used for wavepackets with opposite momenta by spatially flipping the circuit and introducing site-wise SWAP gates. These circuits are determined using the $L=60$ vacuum parameters on an $L=12$ system, and the angles are not extrapolated. The prepared wavepackets span three lattice sites and have $k=\pm\pi/3$ and $\sigma_k=\pi/3$. The continuum expression for the $\lambda=0$ wavepacket (Eq.~\eqref{eq:wp_wavefunction}) and the resulting state after the adiabatic turn-on of $\lambda$ (Eq.~\eqref{eq:u_adiabatic}) are used as $|\psi_\text{targ}\rangle$ for the free and interacting theories, respectively. This step of the state preparation has a two-qubit gate depth of 14 and 80 parameters.
    \item Time evolution for $t=1-9$ is implemented using the translationally-invariant brickwall circuit shown in Fig.~\ref{fig:time_evolution_circ_step}. It has a second-order Trotter step-like structure, and the number of steps is increased with time. The circuits are optimized using time evolution of the two-wavepacket state $|\psi_\text{2wp}\rangle$ as $|\psi_\text{targ}\rangle$, determined using 100 second-order Trotter steps. They are trained on a lattice of size $L=12$ with the vacuum preparation parameters for $L=60$ and the wavepacket preparation parameters determined in the previous step. The parameters are not extrapolated. This part of the circuit has a two-qubit gate depth of $22\lceil\frac{t}{3}\rceil$ and 72 variational parameters.
\end{itemize}

The systematic errors introduced by the variational approximations used in this work can be quantified by running exact classical simulations using {\tt qiskit} \cite{qiskit2024} (see App.~\ref{sec:digitization_effects} for details on the digitization effects). The local infidelity $I_d$ (Eq.~\eqref{eq:local_infidelity}), which is observed to be independent of system size as a result of the SVC framework, is used to quantify the local quality of the prepared state compared to the state of interest. The width of the region, $d$, over which the reduced state infidelity is computed, is chosen to be representative of the given step of the algorithm. For vacuum preparation, a subsystem spanning lattice sites covering several correlation lengths is used. Because of the translational invariance of the vacuum, this region can be located anywhere on the lattice, and $I_d$ is observed to be uniform throughout.\footnote{Translational invariance is seen over pairs of sites. This is a result of the vacuum preparation scheme using only operators spanning two neighboring sites. See Sec.~\ref{sec:circuits_vac_prep} for a discussion.} For  wavepacket preparation, the wavepacket width is used. A subsystem spanning the area where the interaction takes place from $t=1-9$ is used for all steps of time evolution. The values of $I_d$ at each step of the simulation algorithm are shown in Table \ref{tab:error_budget}.

\begin{table}[t]
\centering
\renewcommand{\arraystretch}{1.4}
\begin{tabularx}{\linewidth}{|c|c||Y||Y||Y|Y|} \hline
\multicolumn{2}{|c||}{} &  \multicolumn{2}{c||}{$I_d$} & \# of parameters & Parameter values \\ \hline
\multicolumn{2}{|c||}{Simulation step} & $\lambda=0$ & $\lambda=2$ & &\\ \hline\hline
\multicolumn{2}{|c||}{Vacuum preparation $(d=4)$} & 0.0181 & 0.0071 & 3 & Table \ref{tab:gs_prep_angles}\\ \hline\hline
\multicolumn{2}{|c||}{Wavepacket preparation $(d=3)$} & 0.0555 & 0.0550 & 80 & Table \ref{tab:wp_prep_angles}\\ \hline\hline
\multirow{10}{*}{\makecell{Time\\evolution\\$(d=10)$}} & $t=1$ & 0.1680 & 0.1275 & 72 & Table \ref{tab:time_evolution_angles_1}\\ \cline{2-6}
& $t=2$ & 0.3307 & 0.2024 & 72 & Table \ref{tab:time_evolution_angles_2}\\ \cline{2-6}
& $t=3$ & 0.5412 & 0.3157 & 72 & Table \ref{tab:time_evolution_angles_3}\\ \cline{2-6}
& $t=4$ & 0.5692 & 0.3674 & 72 & Table \ref{tab:time_evolution_angles_4}\\ \cline{2-6}
& $t=5$ & 0.6930 & 0.3675 & 72 & Table \ref{tab:time_evolution_angles_5}\\ \cline{2-6}
& $t=6$ & 0.6447 & 0.4214 & 72 & Table \ref{tab:time_evolution_angles_6}\\ \cline{2-6}
& $t=7$ & 0.6106 & 0.4032 & 72 & Table \ref{tab:time_evolution_angles_7}\\ \cline{2-6}
& $t=8$ & 0.6087 & 0.3937 & 72 & Table \ref{tab:time_evolution_angles_8}\\ \cline{2-6}
& $t=9$ & 0.5932 & 0.4207 & 72 & Table \ref{tab:time_evolution_angles_9}\\ \cline{2-6}
\hline
\end{tabularx}
\renewcommand{\arraystretch}{1}
\caption{Values of the local infidelity $I_d$ for all steps of the simulation algorithm for both the free and interacting theories. The infidelity is computed using the target states $|\psi_\text{targ}\rangle$ specified at the beginning of this section. For vacuum preparation, a range of 4 lattice sites is used, which is roughly 2-$3\times$ the correlation length of the state. The local infidelity is uniform throughout the lattice because of translational invariance. For wavepacket preparation, the reduced state of the wavepacket (occupying three spatial sites) is used. For time evolution, the reduced state on 10 spatial sites is used, which contains the region where the interaction takes place. The number of variational parameters used for each step is shown in the third column. The table where the corresponding parameters are specified is given in the rightmost column.}
\label{tab:error_budget}
\end{table}

It can be seen that the infidelities for $\lambda=0$ are consistently larger than those for $\lambda=2$. This is a result of the interacting theory having a smaller correlation length than the free theory. Because circuits with equal numbers of parameters and equal two-qubit gate depths are used to initialize and time-evolve states for both values of $\lambda$, this difference in quality is expected. This is an indication that the circuit depth needs to be increased for states with larger $\xi$ in order to spread correlations over a greater number of lattice sites.

It should be noted that $I_d$, as well as other measures on the wavefunction, is a conservative estimate of the quality of observables calculated from these simulations. In fact, states with a modest overlap with the target state are seen to reproduce observables in the target state quite well. Appendix \ref{sec:exact_vs_variational} shows the similarity of observables computed using exact and variational methods.

\section{Results}
\label{sec:results}
\begin{figure*}[!ht]
\includegraphics[width=0.5\linewidth,scale=0.75]{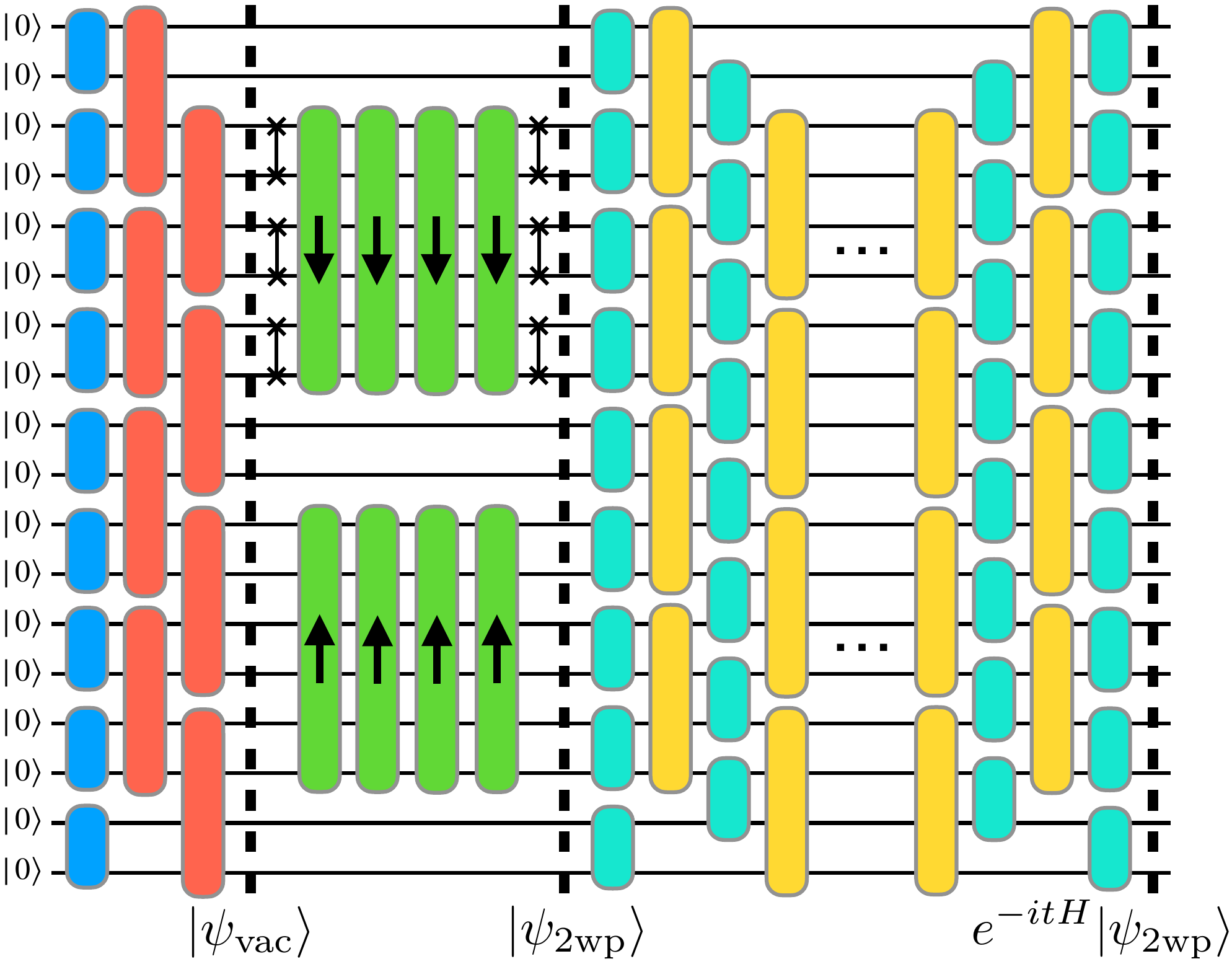}
\caption{The physics-informed SVC used to simulate wavepacket scattering in scalar field theory. The individual circuit elements are defined in Fig.~\ref{fig:circuit_elements}. First, the vacuum of the theory is created using SC-ADAPT-VQE. Wavepackets are initialized on top of this vacuum using a local brickwall ansatz. Then, a translationally-invariant brickwall ansatz is used to implement the time evolution. }
\label{fig:full_circuit}
\end{figure*} 

\begin{table}[h]
\centering
\renewcommand{\arraystretch}{1.4}
\begin{tabularx}{\linewidth}{|c||Y|Y|Y|Y|Y|Y|} \hline
\makecell{$t$} & \makecell{\# of variational\\steps\\(per $t$)} & \makecell{\# of two-qubit\\gates} & \makecell{Two-qubit\\gate depth} & \makecell{\# of\\PTs/layouts\\(per circuit)} & \makecell{\# of TREX twirls\\(per circuit)} & \makecell{\# of shots\\(per circuit)} \\\hline\hline
1-3 & 1 & 2284 & 59 & 80 & 2 & 8000\\\hline
4-6 & 2 & 3604 & 81 & 80 & 2 & 8000\\\hline
7-9 & 3 & 4924 & 103 & 80 & 2 & 8000\\\hline
\end{tabularx}
\renewcommand{\arraystretch}{1}
\caption{Details of the quantum simulations performed on 120 qubits of IBM's {\tt ibm\_fez} superconducting quantum computer. For a given simulation time $t$ (first column), the number of variational steps is given by $\lceil \frac{t}{3} \rceil$ (second column). The total number of two-qubit gates and corresponding two-qubit gate depths are given in columns three and four respectively, accounting for gate cancellations from different parts of the circuits and possible optimizations. Each circuit is executed using a number of Pauli-twirled circuits, with each one being mapped onto randomly-chosen rotation of the selected qubit layout; this number is given in column five. Measurement mitigation is applied using two TREX twirls for each Pauli-twirled circuit (sixth column). The number of shots per circuit is given in the rightmost column.}
\label{tab:device_run_params}
\end{table}

Using the SVC framework, the circuit shown in Fig.~\ref{fig:full_circuit} is scaled up to $L=60$ (120 qubits) and executed on {\tt ibm\_fez} to simulate the scattering of two wavepackets in the $\phi^4$ theory. The theory is digitized onto $n_q=2$ qubits with a cutoff of $\phi_\text{max}=1.5$ to represent the state of the field at each spatial site with a bare mass of $m=1/2$. Error mitigation is vital for the successful extraction of observables from noisy quantum simulations \cite{Kim:2023bwr}. An overview of the error mitigation methods used in this work is given here, and the implementation details and choices can be found in App.~\ref{sec:error_mitigation}. Once the bare circuits are created, several error mitigation layers are added. Dynamical decoupling (DD) \cite{Viola:1998jx,Ezzell_2023} is used to mitigate idle errors and crosstalk between qubits. Pauli twirling (PT) \cite{Wallman:2015uzh} is added to all two-qubit gates (CZ for {\tt ibm\_fez}) to convert coherent errors to incoherent stochastic noise, after which a depolarizing noise model is assumed for each observable of interest. Similar to PT for two-qubit gates, Twirled Readout Error eXtinction (TREX) \cite{Berg:2020ibi} is used to mitigate measurement errors. Observables are then estimated using Operator Decoherence Renormalization (ODR) \cite{Farrell:2023fgd,Farrell:2024fit,Urbanek:2021oej,ARahman:2022tkr}, which involves running a ``mitigation" circuit for every Pauli- and TREX-twirled ``physics" circuit. Circuits used for mitigation must be classically simulable, but must have similar error profiles to the physics circuits. In this work, circuits implementing the time evolution of the vacuum state $e^{-itH}|\psi_\text{vac}\rangle$ are used as mitigation circuits. Although the vacuum preparation and time evolution are implemented approximately and $e^{-itH}|\psi_\text{vac}\rangle \neq |\psi_\text{vac}\rangle$, the vacuum time evolution can be determined classically via an exponential extrapolation. See App.~\ref{sec:error_mitigation} for details on the classical calculation of this time evolution. Compared to methods that have been used in the past, it is found that the vacuum evolution circuits more accurately reflect the noise in the wavepacket circuits. This is especially noticeable for the case of brickwall circuits, where the gates are densely packed and noise spreads faster than in Trotter time evolution circuits. Using the mitigation circuits, the depolarization parameters are computed using the ratio of the measured outcome $\langle O_j\rangle_\text{meas}$ to the expected noiseless outcome $\langle O_j\rangle_\text{true}$ for each local observable $O_j$
\begin{align}
    p_j = \frac{\langle O_j\rangle_\text{meas}}{\langle O_j\rangle_\text{true}} = \frac{\langle \psi_\text{vac}|e^{itH} O_j e^{-itH}|\psi_\text{vac}\rangle_\text{meas}}{\langle \psi_\text{vac}|e^{itH} O_j e^{-itH}|\psi_\text{vac}\rangle_\text{true}}.
    \label{eq:odr_p_j}
\end{align}
The $p_j$ are then used to adjust for the noise in the physics circuits using the same equation in reverse. They are also used to filter out outlying measurements caused by local device imperfections. In addition to these methods, layout randomization is used to drive the noise closer to depolarizing noise. 

\begin{figure*}[t]
    \begin{subfigure}{\linewidth}
        \includegraphics[width=\linewidth]{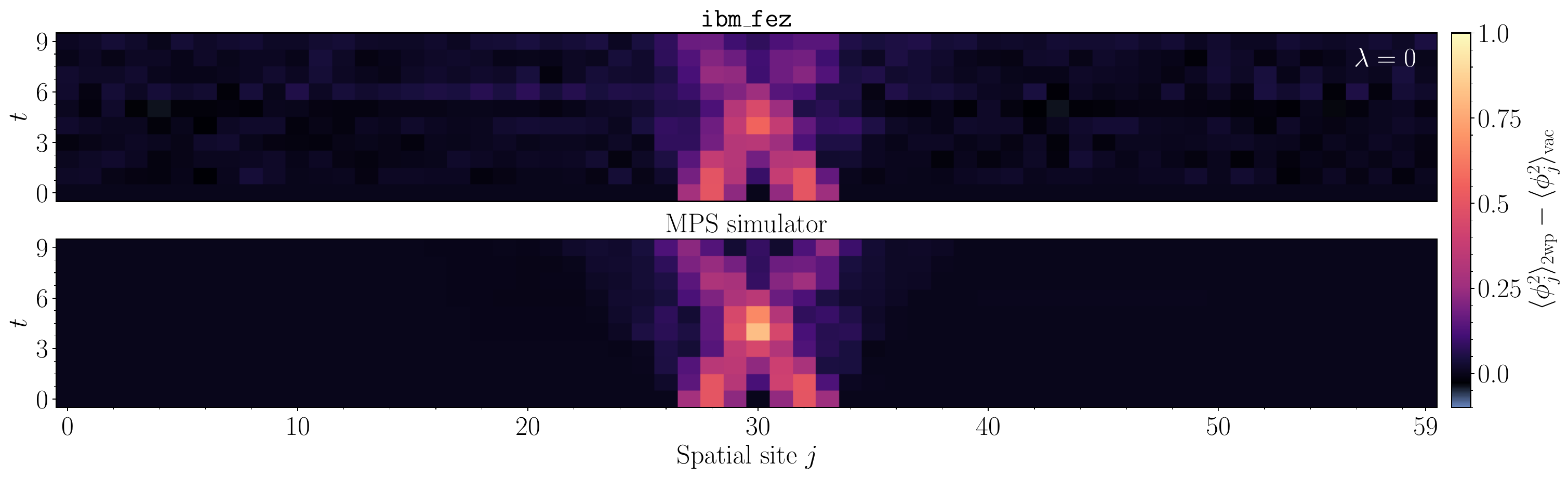}
        \caption{}
        \label{fig:scattering_heatmap_l_0}
    \end{subfigure}
    \begin{subfigure}{\linewidth}
        \includegraphics[width=\linewidth]{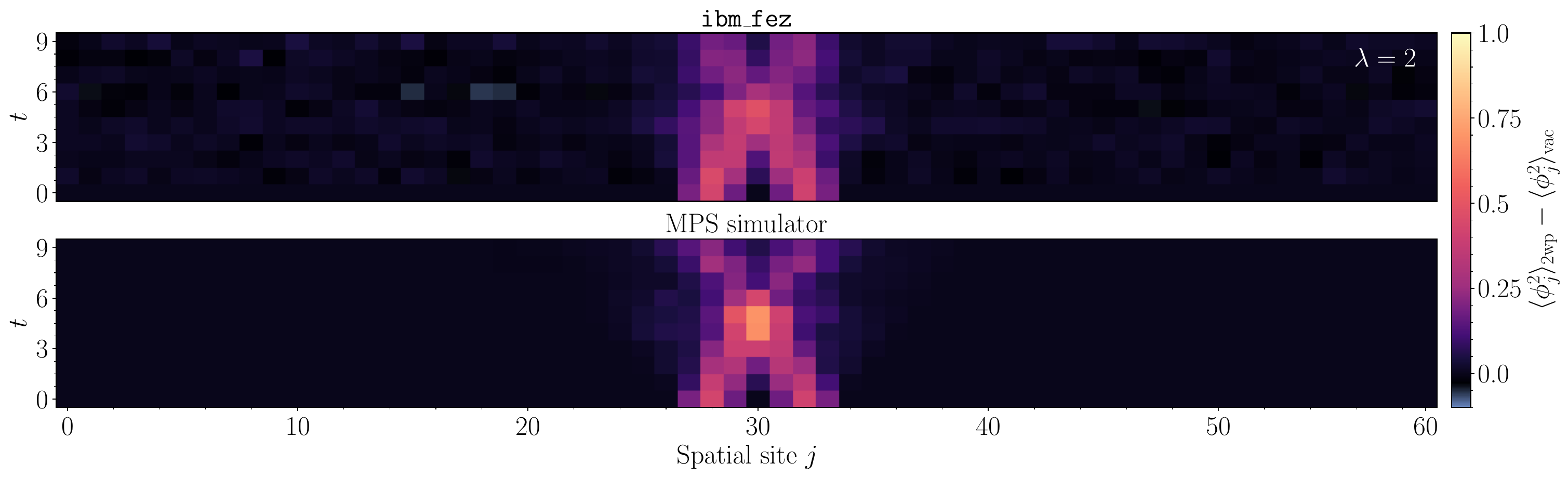}
        \caption{}
        \label{fig:scattering_heatmap_l_2}
    \end{subfigure}
\caption{The time evolution of the vacuum-subtracted $\langle\phi^2_j\rangle$ for a $L=60$ (120 qubits) lattice for (a) the free $(\lambda=0)$ and (b) interacting $(\lambda=2)$ theories. Two wavepackets are initialized with opposite momenta $(k=\pm\pi/3)$ at $t=0$ and centered at sites $j=28,\,32$. The particles travel toward each other, collide, and propagate as time goes on. Error-mitigated results from IBM's quantum computer {\tt ibm\_fez} (top panels) are compared to noiseless MPS circuit simulations (bottom panels). The vacuum of the theory is built using variational circuits determined by SC-ADAPT-VQE, while the wavepacket preparation and time evolution is implemented by variational brickwall circuits. In the device data, the vacuum evolution is evaluated via a classically-determined exponential extrapolation. A broken-down view for each time slice is shown in Fig.~\ref{fig:results_by_time}, and a zoomed-in comparison of the scattering region between the free and interacting theories is given in Fig.~\ref{fig:scattering_heatmap_small}. Details on the error mitigation strategies are given in the main text and in App.~\ref{sec:error_mitigation}.}
\label{fig:scattering_heatmap}
\end{figure*} 

\begin{figure*}[t]
\includegraphics[width=\linewidth]{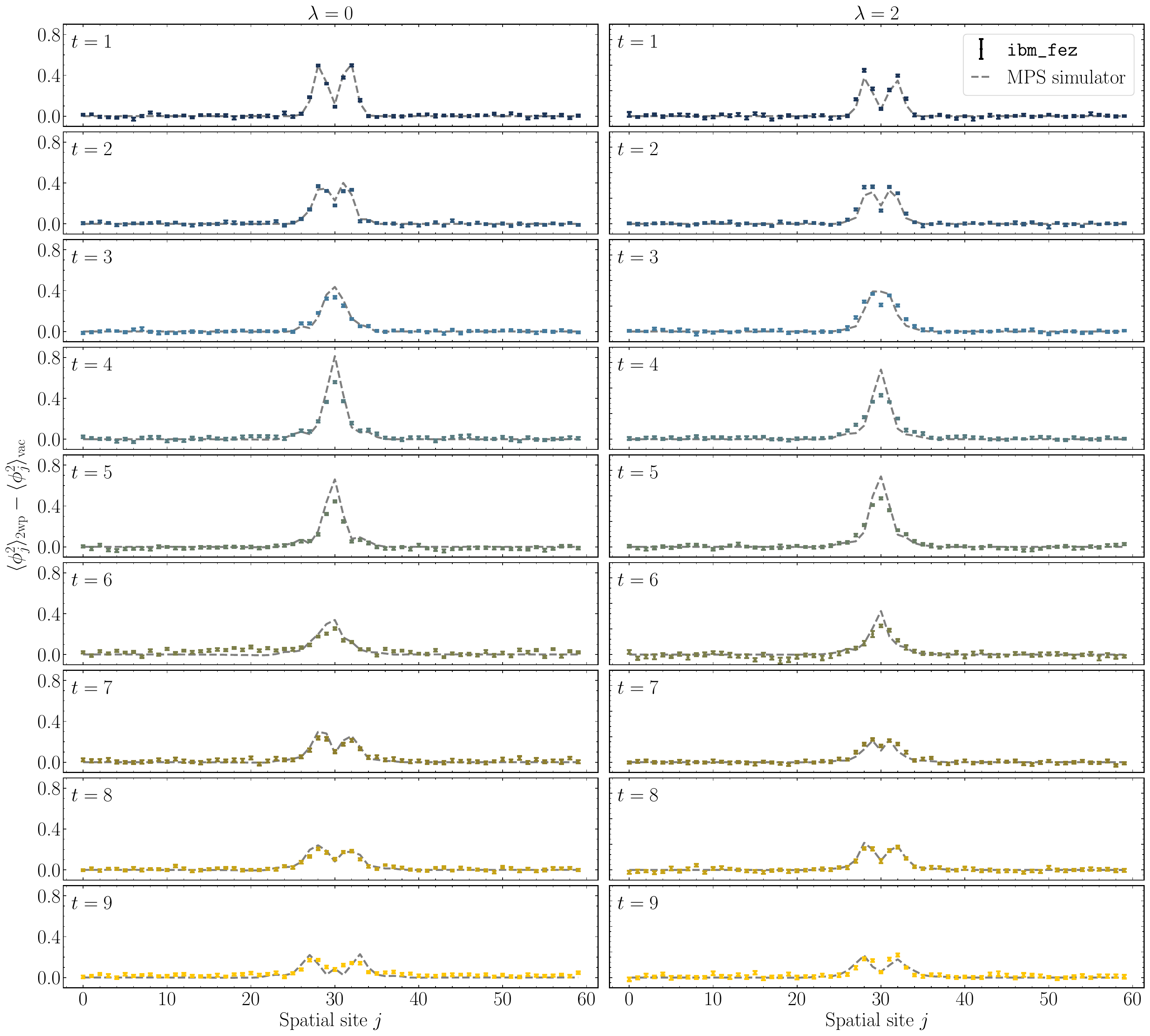}
\caption{A detailed view of the vacuum-subtracted $\langle\phi^2_j\rangle$ for the scattering of two wavepackets, as shown in Fig.~\ref{fig:scattering_heatmap}, by time step.  Noiseless MPS circuit simulations (grey dashed lines) are compared to data from IBM's quantum computer {\tt ibm\_fez} after error mitigation (points) for $\lambda=0$ (left column) and $\lambda=2$ (right column). In the device data, the vacuum evolution is evaluated via a classically-determined exponential extrapolation. The error bars represent one standard deviation of the bootstrap-resampled data.}
\label{fig:results_by_time}
\end{figure*} 

The two-qubit gate depths and counts, as well as other parameters of the runs on {\tt ibm\_fez} for times $t=1-9$, are shown in Table \ref{tab:device_run_params}. Circuits with a maximum two-qubit gate depth of 103 using 4924 two-qubit gates (for $t=7,8,9$) are used. For time $t$, $\lceil \frac{t}{3} \rceil$ variational steps are used (the variational step is defined in Sec.~\ref{sec:circuits_time_evolution} and shown in Fig.~\ref{fig:time_evolution_circ_step}). For each time $t$, four circuits are executed:  $e^{-itH}|\psi_\text{2wp}\rangle$ and $e^{-itH}|\psi_\text{vac}\rangle$ for $\lambda=0$ and $\lambda=2$. 80 Pauli-twirled circuits are run for each physics and mitigation circuit. For each Pauli-twirled circuit, two TREX twirled circuits are run for measurement mitigation. Each circuit executed on the device is evaluated using 8000 shots.

In these simulations, fluctuations of $\langle \phi^2_j \rangle$ above the vacuum are measured on the quantum device.\footnote{Note that it is possible to efficiently measure $\langle H_j\rangle$ using the same setup. Besides $\Pi^2_j$, all terms in the Hamiltonian of Eq.~\eqref{eq:lattice_h} are in the $\phi$ basis, so their expectation values can be computed using measurements in the $\phi$ basis. To measure in the $\Pi$ basis, a second set of circuits can be run with the $k_\phi$ local Fourier transform appended. The results can then be added together to compute $\langle H_j\rangle$. From this, it is straightforward to estimate quantities that are accessible to analytic computations, such as cross sections computed using perturbative quantum field theory.} The results from the device and from MPS circuit simulations for $\langle \phi^2_j\rangle_\text{2wp}-\langle \phi^2_j\rangle_\text{vac}$ for $\lambda=0$ and $\lambda=2$ are shown in Figs.~\ref{fig:scattering_heatmap_l_0} and \ref{fig:scattering_heatmap_l_2} respectively. A breakdown of the results by time step is shown in Fig.~\ref{fig:results_by_time}, and numerical values for these runs are given in Tables \ref{tab:results_t_1}-\ref{tab:results_t_9}. The uncertainties in the results from the quantum computer are estimated using bootstrap resampling. The expected results are determined using the {\tt qiskit} \cite{qiskit2024} MPS circuit simulator. This classical method works by converting the circuit to a tensor network acting on a MPS and applying the circuit to the MPS gate by gate while enforcing the cutoff and maximum bond dimension constraints at each step. See Ref.~\cite{qiskit_mps} for more information. In this work, a maximum bond dimension of 100 is used. For the circuits considered, it is seen that the results of these simulations are converged to $10^{-2}-10^{-3}$. 

The scattering process is seen by examining $\langle \phi^2_j\rangle_\text{2wp}-\langle \phi^2_j\rangle_\text{vac}$ in Fig.~\ref{fig:scattering_heatmap}. The propagation of particles is clearly identified as disturbances in the vacuum-subtracted $\langle\phi^2_j\rangle$ in the interaction region in the center (spatial sites 24-36). These are indicated as bright spots on the plots. The wavepackets, centered to the left and right of the center of the lattice at $t=0$, are initialized with opposite momenta. They travel toward each other and collide in the center. After the collision takes place, a light cone is seen to develop, indicating that the particles have propagated past each other and are traveling away from the point of the collision.

\begin{figure*}[t]
\includegraphics[width=\linewidth]{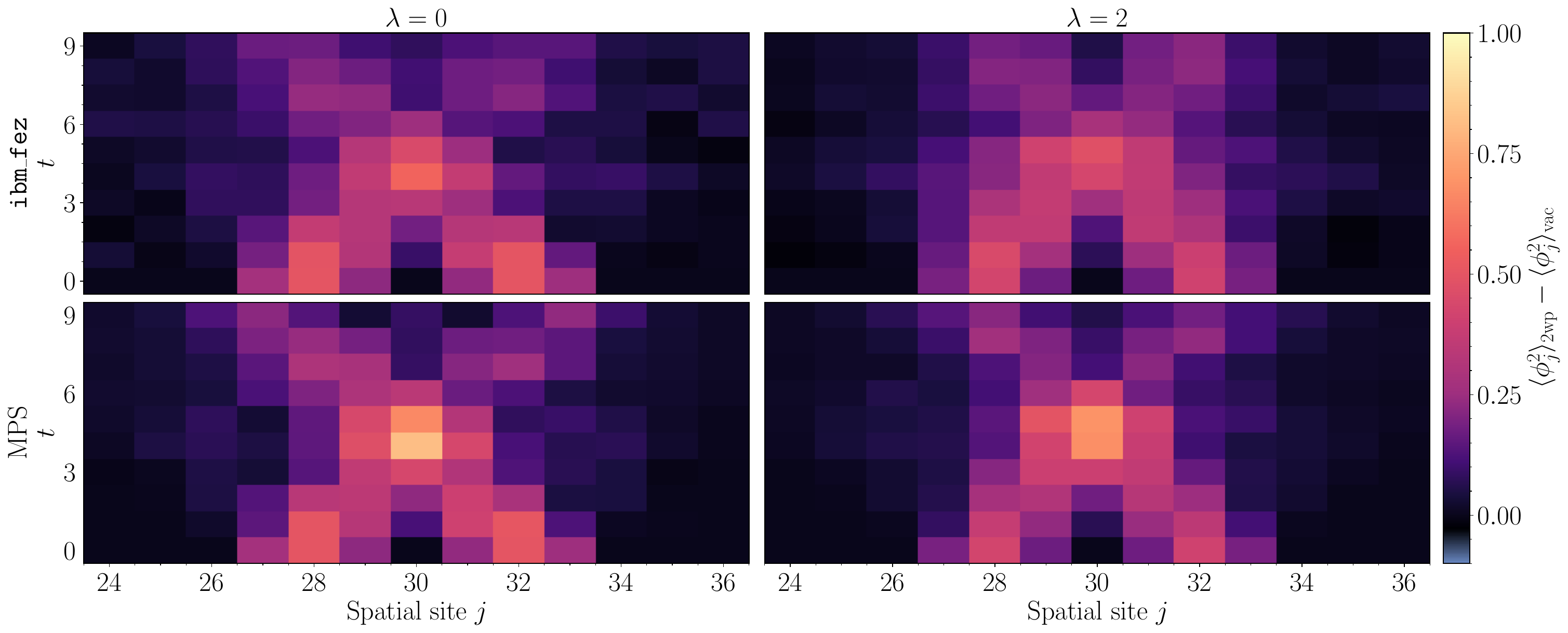}
\caption{A zoomed-in view of the interaction region of Fig.~\ref{fig:scattering_heatmap} highlighting the effect of interactions. The vacuum-subtracted $\langle \phi_j^2\rangle$ in the free theory (left panels) is compared to the interacting theory (right panels). The top panels show error-mitigated results from IBM's quantum computer {\tt ibm\_fez}, while the bottom panels show noiseless MPS circuit simulations.}
\label{fig:scattering_heatmap_small}
\end{figure*}

The effect of interactions is clearly seen when comparing results in the interaction region between the free and interacting theories as in Fig.~\ref{fig:scattering_heatmap_small}. In the free theory, the collision is seen to peak around $t=4$, after which the particles travel away from each other. When interactions are included in the theory, the peak of the collision is observed later (around $t=5$), and a time delay due to the interaction is visible. The difference between $\lambda=0$ and $\lambda=2$ can also be seen at late times $(t=9)$, where the free particles can be seen having traveled one more spatial site than the interacting particles. A time delay is expected because the $\phi^4$ potential is repulsive, particles with overlapping wavefunctions incur an energy penalty. In this sense, the time delay can be viewed as the interacting particles encountering a potential step and thus losing kinetic energy when their wavepackets overlap. As a result, the particles spend a longer time in the central region where they are able to interact. It is expected that this effect will be even more prominent at greater interaction strengths. In the case of the digitized simulations done in this work with $n_q=2$, the interactions and time delay are caused by digitization effects. See App.~\ref{sec:digitization_effects} for a detailed examination of these interactions.

The results in Figs.~\ref{fig:scattering_heatmap}, \ref{fig:results_by_time}, and \ref{fig:scattering_heatmap_small} show that by using a combination of variational circuit methods (SC-ADAPT-VQE and brickwall), together with error mitigation techniques, quantum computers are able to simulate scattering processes of interacting particles. The scalability of each step of this simulation algorithm enables the simulation of collision events on arbitrarily-large system sizes in a systematically-improvable manner. It should be noted that the observation of $\langle \phi^2_j\rangle_\text{2wp}-\langle \phi^2_j\rangle_\text{vac}=0$ in the vacuum regions is a nontrivial result, as it requires the wavepacket simulations from the quantum computer to agree closely with extrapolations from classical computations. In Figs.~\ref{fig:scattering_heatmap} and \ref{fig:results_by_time}, the data from the quantum computer is seen to qualitatively agree with the MPS circuit simulations. Increasing levels of noise (as well as uncertainties) can be seen in the results for later times, which is expected as a result of deeper circuits. There exist points where the error-mitigated results still disagree from the MPS data by several standard deviations (see Fig.~\ref{fig:results_by_time}). This is particularly noticeable in the interaction region during the collision of the particles, but occurs in vacuum regions as well. This highlights a weakness of the chosen error mitigation method: because the state chosen for error mitigation is $e^{-itH}|\psi_\text{vac}\rangle$, its evolution does not capture potential changes to wavepackets. Since the noise is state-dependent, these mitigation circuits cannot mimic the noise in the circuits simulating scattering perfectly. This reflects the simplicity of the assumed error model, and is an area where much improvement is expected in the future. Although errors persist even after the application of error mitigation, it is clear that quantum computers today are capable of probing scattering events, potentially shedding light on inelastic collisions in more complicated theories in the future.

\section{Summary and outlook}
\label{sec:discussion}
Quantum simulations hold the potential of reliably uncovering the dynamics of complex inelastic collision events, such as those that take place in extremely dense and hot environments. As a first step toward such computations, this work simulates elastic scattering in one-dimensional lattice scalar field theory. Scalable variational quantum algorithms are developed for both state preparation and time evolution, paving the path for future simulations at larger scales. The simulation begins by using a translationally-invariant SC-ADAPT-VQE ansatz to prepare the vacuum of the theory. Wavepackets representing particles are then created on top of the vacuum using a brickwall circuit, and a translationally-invariant brickwall ansatz is used to implement time evolution of the system. The scalability of all circuits in these simulations is a result of the physical properties of states in the system, namely their translational invariance, locality, and the presence of a mass gap. The circuits are determined classically on modest system sizes, and extrapolated to arbitrarily-large system sizes to be run on quantum devices. These methods are used to simulate the collision of two wavepackets using 120 qubits of IBM's 156 qubit superconducting quantum computer {\tt ibm\_fez}. The use of error mitigation and suppression techniques is crucial for the extraction of results that qualitatively agree with classical MPS simulations. Signatures of scattering and effects of the interaction strength are clearly identified in the vacuum-subtracted expectation value of the field determined from data from the quantum computer.

Overall, the utility of variational quantum algorithms in systematically compressing circuits is clear; the specialization of these methods to quantum simulations enables the use of physics to provide further performance improvements. These newly-developed scalable variational methods, together with existing techniques and new modifications to error mitigation strategies, enable the first simulation of wavepacket scattering in an interacting quantum field theory on a quantum computer. The resulting gate counts and resource requirements are many orders of magnitude below upper bounds for fault-tolerant algorithms. Furthermore, these results are the first demonstration of the extraction of qualitatively correct observables from quantum devices using variational brickwall circuits at scale. This work establishes variational algorithms as tools that are useful for the compression of all parts of a quantum simulation. In particular, it is demonstrated that the problem of barren plateaus that plagues many variational quantum computations is reduced through knowledge of the system being simulated (e.g., its symmetries and mass gap). Efficient variational circuits are determined by using physics to remove redundancies in the circuit parameterization. The development of variational circuit compression methods that are both scalable and amenable to error mitigation is useful for minimizing noise in the quantum device through reduced two-qubit gate depth. These methods and results constitute an important stepping stone toward the quantum simulation of inelastic scattering processes at high energies, where known classical methods are expected to fail. 

Improvements to the optimization step of the variational algorithm, as well as more efficient ansatze, are expected in the near future. This work uses the most simple iterative gradient-based optimization method. Recent work \cite{Stokes:2019pmg,Harrow:2021tta} has shown that it is possible to incorporate more sophisticated optimization protocols to improve the quality of the approximate quantum state. In parallel, ansatze for state preparation that take advantage of mid-circuit measurements \cite{Chen:2023tfg,Baumer:2023vrf,Malz:2023xve,Baumer:2024jng,Piroli:2024ckr} can be immediately used in simulations today; the combination of these methods with variational quantum algorithms is a promising direction. These modifications would allow for greater precision in variational calculations, while maintaining the same circuit depth, or even reducing it.

Highly-inelastic scattering events are believed to generate a large amount of entanglement, and as a result are prime candidates for demonstrations of quantum utility. Toward the simulation of inelastic collisions, several improvements to the current algorithm will be necessary. Initial states with energy higher than the two-particle threshold will need to be created, requiring more tightly peaked wavepackets in $k$-space. In turn, wavepackets that occupy a larger number of qubits will be required. The methods presented will need to be extended to the scalable creation of wavepackets of increasing sizes and their time evolution. The presence of a large number of lattice sites occupied by the vacuum will continue to provide the resolution in lattice momentum that will be necessary to observe inelastic effects. More qubits per site (i.e., increasing $n_q$) will be needed to achieve higher precision in these calculations, as well as to simulate genuine interactions. Similarly, larger system sizes and correlation lengths will be needed to approach continuum physics. All of these changes will require circuits with greater two-qubit gate depths, and improvements to the variational methods will be necessary. As the capabilities of state of the art quantum computers to realize large-scale system sizes improve, more sophisticated techniques for particle detection will be required. Further in the future, the extension of these methods to include more complex interactions, such as those described by non-Abelian gauge theories, will be required for the simulation of realistic particle collisions. 

\section*{Acknowledgments}
The author thanks Roland Farrell, Marc Illa, Zhiyao Li, Henry Froland, Anthony Ciavarella, and Martin Savage for helpful discussions and insightful comments. This work was supported in part by the U.S. Department of Energy, Office of Science, Office of Nuclear Physics, InQubator for Quantum Simulation (IQuS) \cite{iqus} under Award Number DOE (NP) Award DE-SC0020970 via the program on Quantum Horizons: QIS Research and Innovation for Nuclear Science. This work was also supported, in part, through the Department of Physics \cite{uw_phys} and the College of Arts and Sciences \cite{uw_artsci} at the University of Washington. This work has made extensive use of Wolfram {\tt Mathematica} \cite{Mathematica}, {\tt python} \cite{python}, {\tt julia} \cite{Julia-2017}, {\tt jupyter} \cite{4160251} notebooks in the {\tt conda} \cite{anaconda} environment, IBM's quantum programming environment {\tt qiskit} \cite{qiskit2024}, {\tt iTensor} \cite{ITensor,ITensor-r0.3}, and {\tt quimb} \cite{gray2018quimb} in this work. This work was enabled, in part, by the use of advanced computational, storage and networking infrastructure provided by the Hyak supercomputer system at the University of Washington \cite{uw_hyak}. This research was done using services provided by the OSG Consortium \cite{osg07, osg09, https://doi.org/10.21231/906p-4d78, https://doi.org/10.21231/0kvz-ve57}, which is supported by the National Science Foundation awards \#2030508 and \#1836650. The author acknowledges the use of IBM Quantum Credits for this work. The views expressed are those of the author, and do not reflect the official policy or position of IBM or the IBM Quantum team.

\clearpage
\appendix
\section{Digitization effects}
\label{sec:digitization_effects}
The Nyquist-Shannon Sampling Theorem \cite{shannon_collected_1993,10.5555/3179430.3179434,Macridin:2018oli,Macridin:2018gdw} guarantees that with a proper choice of $\phi_\text{max}$, the digitized field theory represents the continuous lattice field theory with errors that are exponentially suppressed with an increasing number of wavefunction sample points. In particular, the eigenfunctions of the continuous theory can be reconstructed from the digitized eigenfunctions up to errors that scale as $\epsilon \sim 2^{-2^{n_q}}$. The single-site wavefunction $\psi(\phi)$ can be approximately reconstructed from $n$ samples at points $\phi_n$ using the Whittaker–Shannon interpolation formula \cite{shannon_collected_1993}: 
\begin{align}
    \psi(\phi) &= \sum_{n=0}^{2^{n_q}-1} \psi(\phi_n)\, \text{sinc} \left(\frac{\phi+\phi_\text{max}-n\delta_\phi \phi_\text{max}}{\delta_\phi \phi_\text{max}}\right).\label{eq:shannon_interpolation}
\end{align}
Its multidimensional extensions can be used to reconstruct multi-site wavefunctions. 

A poorly chosen $\phi_\text{max}$ can introduce unwanted interactions due to the digitization and break the double-exponential convergence guaranteed by the Nyquist-Shannon Sampling Theorem. The optimal choice of $\phi_\text{max}$ for a given theory depends on $n_q$, as well as on the parameters $m$ and $\lambda$, since these control the strength of the potential. As a result, states with higher $m$ and $\lambda$ are more localized in $\phi$-space and less localized in $\Pi$-space. Previous approaches determined the optimal $\phi_\text{max}$ by minimizing the error in the commutator compared to the continuum value $||[\phi_i,\Pi_j]-i\delta_{ij}||$ \cite{10.5555/3179430.3179434,Macridin:2018oli,Macridin:2018gdw,Bauer:2021gek,Kane:2022ejm}, or minimizing the error in the eigenenergies by numerically solving the continuum theory for a small number of lattice sites \cite{Klco:2018zqz}. However, single-site considerations do not capture additions to the potential from neighboring sites, and numerical solutions become intractable for large systems due to the exponential resources required in the discretization of the system. In this work, $\phi_\text{max}$ is approximately chosen by including the single-site contributions from $H_\text{kin}$, so that the only term that is omitted from the Hamiltonian of Eq.~\eqref{eq:lattice_h} is $-\sum_{j=0}^{L-1}\phi_{j+1}\phi_j$. Similar to the previous methods, this is an approximation to the optimal $\phi_\text{max}$ for large system sizes, but it is closer to the true value.\footnote{In principle, the optimal $\phi_\text{max}$ can be determined following an approach similar to that of Sec.~\ref{sec:scalable_variational_circuits}, by finding $\phi_\text{max}$ as a function of the system size $L$ for small $L$, and extrapolating to $L$ of choice.} Following these considerations, the optimal $\phi_\text{max}$ is determined by using Eq.~\eqref{eq:shannon_interpolation} to maximize the overlap between the low-energy eigenstates in the digitized theory and the continuum theory solved numerically. With this method, the best choice for $\phi_\text{max}$ is 1.5 for $n_q=2$, $m=1/2$, $\lambda=0$, and 1.43 for $n_q=2$, $m=1/2$, $\lambda=2$. However, for $\lambda=2$ the difference between $\phi_\text{max}=1.5$ and the ideal value is minimal in terms of the effect on the representation of the continuum, and as a result $\phi_\text{max}=1.5$ is used for both the free and interacting theories throughout this work. In this case, the errors due to the suboptimal choice of $\phi_\text{max}$ are small compared to digitization errors stemming from a small $n_q$. 

\begin{figure*}
\begin{subfigure}{\linewidth}
    \includegraphics[width=\linewidth]{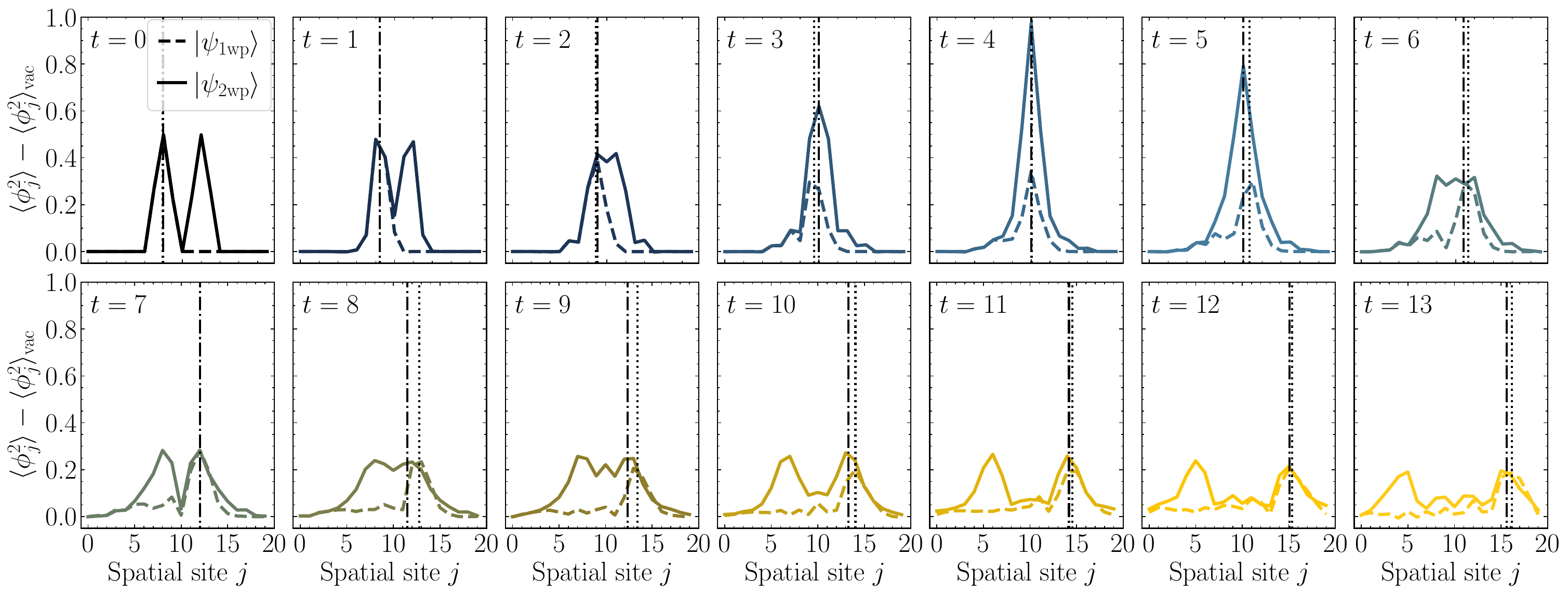}
    \caption{}
    \label{fig:single_vs_two_wp_free}
\end{subfigure}
\begin{subfigure}{\linewidth}
    \includegraphics[width=\linewidth]{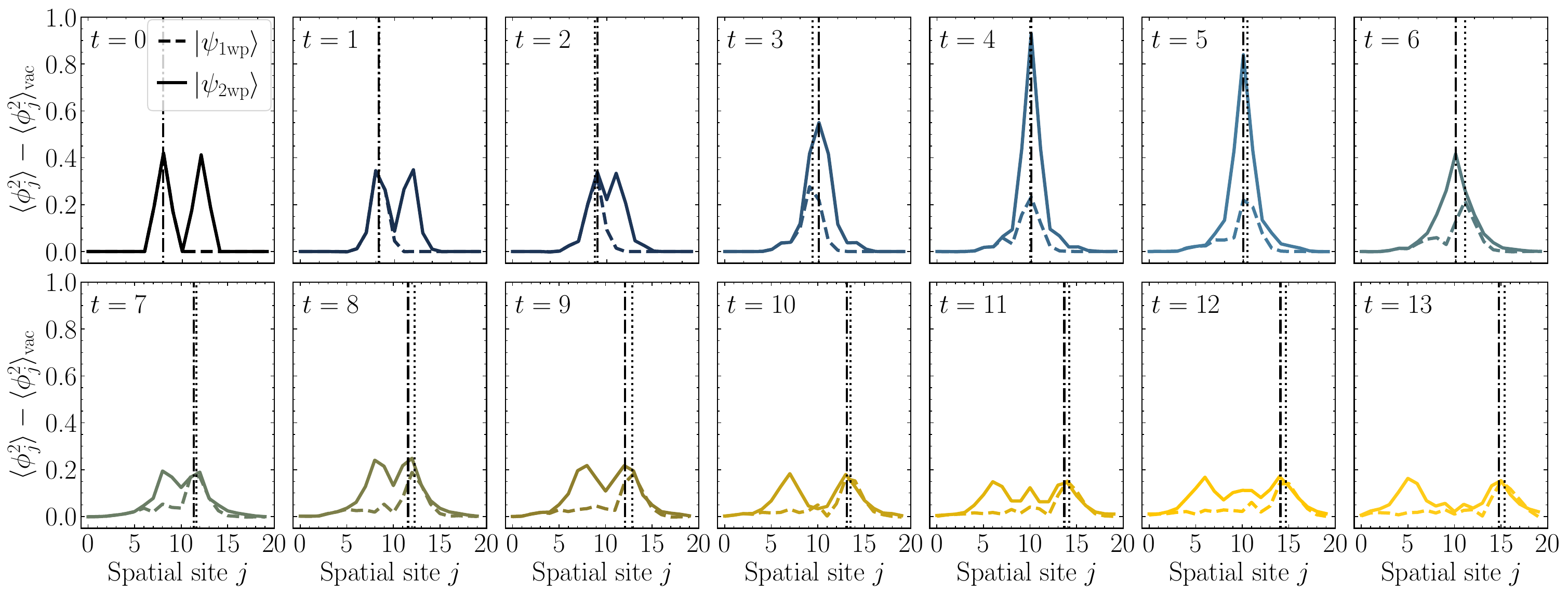}
    \caption{}
    \label{fig:single_vs_two_wp_int}
\end{subfigure}
\caption{Comparison of the time evolution of the single- and two-wavepacket states ($|\psi_\text{1wp}\rangle$ and $|\psi_\text{2wp}\rangle$ respectively), for (a) the free $(\lambda=0)$ and (b) the interacting $(\lambda=2)$ theories. The center of the left wavepacket is tracked for the single-wavepacket state with the dotted vertical line, and for the two-wavepacket state with the  dash-dotted vertical line. The centers are determined by fitting a Gaussian around the wavepacket region, and by taking the maximum value when the wavepackets are interacting. A system of $L=20$ spatial sites with $n_q=2,m=1/2$ is used. The simulations are done using a MPS circuit simulator with size $1/10$ Trotter steps and a maximum bond dimension of 100.}
\label{fig:single_vs_two_wp}
\end{figure*}

Two qubits per lattice site is the coarsest nontrivial digitization. As a result, the digitization effects are significant and must be considered. With $n_q=2$ and $\delta_\phi$ chosen as in Eq.~\eqref{eq:phi_digitized}, the $\phi^2_j$ and $\phi^4_j$ operators may be written as 
\begin{align}
    \phi^2_j = \frac{\phi_\text{max}^2}{9}\left(5 + 4Z_{2j}Z_{2j+1}\right),\\
    \phi^4_j = \frac{\phi_\text{max}^4}{81}\left(41 + 40Z_{2j}Z_{2j+1}\right).
\end{align}
From these expressions, it is seen that the non-identity portions of $H_\phi$ and $H_\text{int}$ in the Hamiltonian of Eq.~\eqref{eq:lattice_h} are proportional to each other. This implies that simulating a theory with $\lambda \neq 0$ is equivalent to simulating one with a larger mass, 
\begin{align}
    m' = \sqrt{m^2 + \frac{\lambda}{4!}\frac{20}{9}\phi_\text{max}^2}.
\end{align}
As a result of this, the interactions present in the simulations in this work are due to digitization effects. In principle, these interactions may be mapped to an expansion of the form $\sum_n c_{2n}\phi^{2n}$, but determining the coefficients $c_{2n}$ is difficult in practice. These interactions are examined in Fig.~\ref{fig:single_vs_two_wp}, where the time evolution of a single wavepacket is compared to that of a two-wavepacket state for both values of the coupling. By considering the single-wavepacket evolution, which is truly noninteracting, it can be seen that there is a time delay due to the interaction of two wavepackets for both couplings, with the time delay for the $\lambda=2$ theory being larger.

\begin{figure*}
    \includegraphics[width=\linewidth]{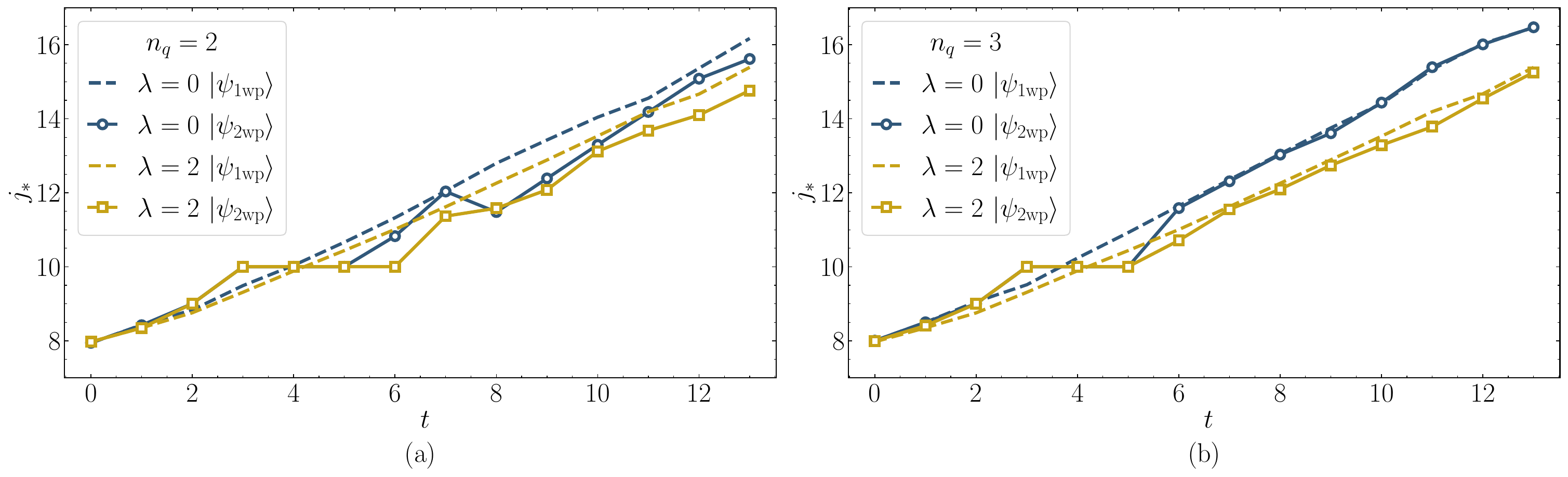}
    \caption{The center $j_*$ of the left wavepacket propagating in time with (a) $n_q=2$ and (b) $n_q=3$ digitization. The free and interacting theories are shown in blue and yellow, respectively. The single-wavepacket states $|\psi_\text{1wp}\rangle$ are marked with dashed lines, while the two-wavepacket states $|\psi_\text{2wp}\rangle$ are marked with solid lines. A system of $L=20$ spatial sites with $m=1/2$ is used. For $n_q=2$, $\phi_\text{max}=1.5$ is used for both $\lambda=0$ and $\lambda=2$ as in the simulations run on quantum computers in this work, while for $n_q=3$, the optimal $\phi_\text{max}$ values of 2.52 and 2.43 are used for the free and interacting theories, respectively. The simulations are done using a MPS circuit simulator with size $t=1/10$ Trotter steps and maximum bond dimension of 100.}
    \label{fig:center_tracking}
\end{figure*}

The centers $j_*$ of the wavepackets for all four cases are plotted in Fig.~\ref{fig:center_tracking}a. The dashed lines show the single-wavepacket time evolution. The slight difference in group velocity $v_k$, as expected from Fig.~\ref{fig:e_k_v_k_wp_prep_convergence}, is seen in the slopes of the $|\psi_\text{1wp}\rangle$ lines. By comparing the single-wavepacket and the two-wavepacket lines, the relative time delay between the $\lambda=0$ and $\lambda=2$ theories as a result of differences in $v_k$ is eliminated. The collision takes place between times $t=3-6$ indicated by the plateau, and the time delay is seen in the $\lambda=2$ plateau extending later in time than the $\lambda=0$ one. After an initial settling time, the ``asymptotic" time delay is also seen in the difference in the center points at a given time. The difference of $j_*$ between the $|\psi_\text{1wp}\rangle$ and $|\psi_\text{2wp}\rangle$ is larger for $\lambda=2$, indicating that the time delay due to interactions is greater in this theory.

To complete the discussion of digitization effects and interactions caused by them, simulations of the same process with $n_q=3$ are shown in Fig.~\ref{fig:center_tracking}b. These simulations are expected to more accurately capture the continuum physics because of the added precision from doubling the number of points where the continuum wavefunctions are sampled. In addition, the non-identity parts of $H_\phi$ and $H_\text{int}$ are not proportional with $n_q=3$, making the $\phi^4$ operator a genuine interaction term not present in the free Hamiltonian. Using the methods described above, the optimal $\phi_\text{max}$ is determined to be 2.52 for $\lambda=0$, and 2.43 for $\lambda=2$. All other parameters of the simulations are the same as for $n_q=2$. In the $\lambda=0$ simulations for $n_q=3$ there is no time delay, and a very small time delay for $\lambda=2$, which shows the strength of the digitization effects for $n_q=2$.

\section{Details on wavepacket creation}
\label{sec:wp_creation_details}

Section~\ref{sec:circuits_wp_prep} describes wavepacket preparation by learning an approximate shallow-depth variational circuit. This circuit is optimized to maximize the overlap with $|\psi_\text{targ}\rangle$ being the wavepacket produced by the non-digitized analytical wavefunction (Eq.~\eqref{eq:wp_wavefunction}) for $\lambda=0$, and after adiabatically turning on the interactions (Eq.~\eqref{eq:u_adiabatic}) for $\lambda=2$. The wavepackets represented by $|\psi_\text{targ}\rangle$ above are compared to those determined from ED by block-diagonalizing the Hamiltonian into momentum sectors in Table~\ref{tab:adiabatic_vs_ed}. The local infidelity over the three sites spanning the wavepacket, $I_3$, is seen to decrease with increasing $n_q$ indicating that the digitized wavefunction approximates the continuous-field case with increasing quality.\footnote{The wavepackets produced from ED can also be used directly as $|\psi_\text{targ}\rangle$, eliminating the need for the continuous-field approximation (Eq.~\eqref{eq:wp_wavefunction}) and adiabatic turn-on (Eq.~\eqref{eq:u_adiabatic}). This approach would encounter the same classical limitations in terms of scalability and is conceptually simpler. However, the direct ED method was not used in determining the parameters for circuits run on quantum hardware in this work.}

\begin{table}
\centering
\renewcommand{\arraystretch}{1.4}
\begin{tabularx}{\linewidth}{|c||Y|Y|} \hline
 & $I_3\,\, (n_q=2)$ & $I_3\,\, (n_q=3)$ \\\hline\hline
$\lambda=0$ & 0.03922 & 0.01471 \\\hline
$\lambda=2$ & 0.03973 & 0.01537 \\\hline
\end{tabularx}
\renewcommand{\arraystretch}{1}
\caption{Comparison of wavepackets produced from the continuous-field approximation (Eq.~\eqref{eq:wp_wavefunction}) with adiabatic turn-on using Eq.~\eqref{eq:u_adiabatic} for $\lambda=2$ to those determined from ED, as a function of the number of qubits used in the digitization $n_q$. The free and interacting theories are shown in the center and bottom rows, respectively. The wavepackets are created on an $L=6$ system for both methods. The local infidelity over the three sites spanning the wavepacket, $I_3$ is used. For $n_q=\{2,3\}$, the values of $\phi_\text{max}=\{1.5,3.1\}$ are used as determined in Ref.~\cite{Klco:2018zqz}.}
\label{tab:adiabatic_vs_ed}
\end{table}

The adiabatic turn-on is implemented as follows. The time evolution is implemented using Eq.~\eqref{eq:u_adiabatic} for a total time of $t_\text{ad}=100$, broken into $N_\text{ad}=10=O(\sqrt{\lambda t_\text{ad}})$ steps (as determined in Refs.~\cite{Jordan:2011ci,Jordan:2012xnu}). Each adiabatic step in Eq.~\eqref{eq:u_adiabatic} is Trotterized into 1000 ``forward" and 1000 ``backward" steps. The forward steps linearly increase the value of $\lambda$ once every 10 steps, while the backward steps evolve with fixed $\lambda$. As explained in Sec.~\ref{sec:lattice_scalar_field_theory}, the backward evolution steps serve to undo the unwanted time evolution to individual eigenstates composing the wavepacket.

The profile of $|\psi_\text{targ}\rangle$, and thus of the produced wavepackets is specified to be a Gaussian in $k$-space, as in Eqs.~\eqref{eq:single_particle_superposition} and \eqref{eq:wp_wavefunction}. This is slightly different from the approach JLP take in Refs.~\cite{Jordan:2011ci,Jordan:2012xnu}, where the form of the state is specified in position space. By specifying a Gaussian in $k$-space, the method used in this work also specifies a Gaussian in position space, where its size is then truncated to optimize device resources. The two approaches are equivalent, since JLP also truncate the spatial extent of the wavepackets.

\section{Comparison of exact and variational time evolution}
\label{sec:exact_vs_variational}
\begin{figure*}
\begin{subfigure}{\linewidth}
    \includegraphics[width=\linewidth]{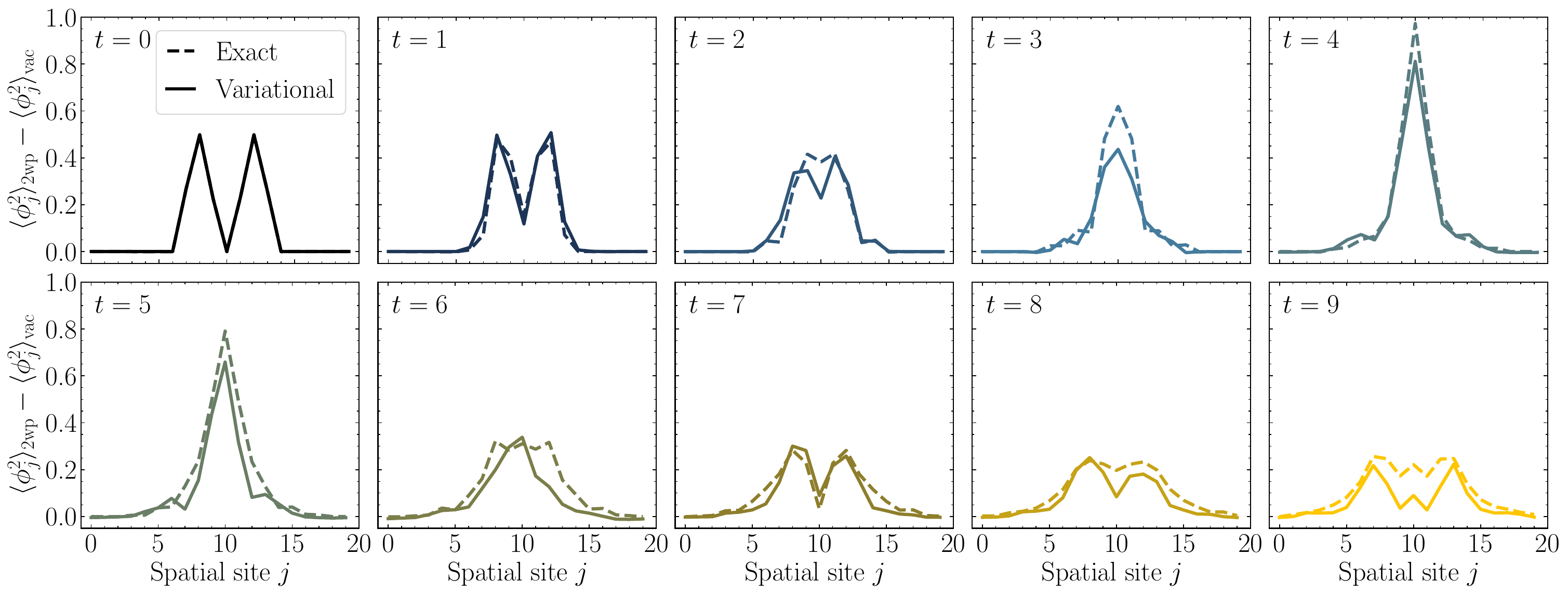}
    \caption{}
    \label{fig:exact_vs_variational_l_0}
\end{subfigure}
\begin{subfigure}{\linewidth}
    \includegraphics[width=\linewidth]{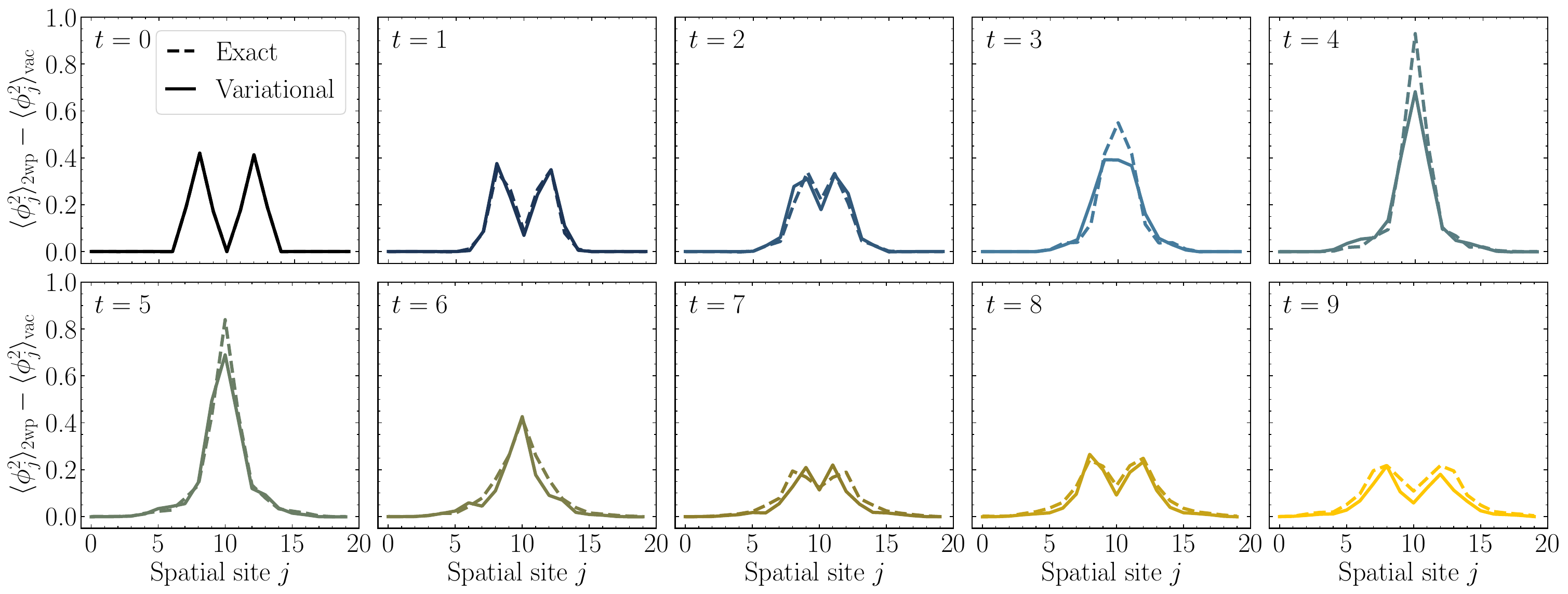}
    \caption{}
    \label{fig:exact_vs_variational_l_2}
\end{subfigure}
\caption{The vacuum-subtracted $\langle \phi^2_j \rangle$ for different times $t=0-9$ computed using a MPS circuit simulator for Trotterized time evolution (dashed lines) compared to its approximation using SVC (solid lines) for (a) the free $(\lambda=0)$ theory and (b) the interacting $(\lambda=2)$ theory. An $L=20$ (40 qubits) system is used with maximum bond dimension 100. The exact results use Trotterization with size $1/10$ Trotter steps.}
\label{fig:exact_vs_variational}
\end{figure*} 
In this appendix, noiseless MPS circuit simulations of an $L=20$ system are used to compare exact time evolution in the digitized theory with $n_q=2$ to time evolution approximated by SVC. Exact calculations are done using matrix exponentiation applied to the $t=0$ $|\psi_\text{2wp}\rangle$ state prepared via SVC methods. Figure \ref{fig:exact_vs_variational} compares $\lambda=0$ and $\lambda=2$. The SVC variational results can be seen to qualitatively track the exact time evolution, but degrade with time. The quality of the $\lambda=0$ variationally-determined observables is seen to be consistently worse than those for $\lambda=2$, which is again a consequence of the $\lambda=0$ theory having a longer correlation length. Although the local infidelity spanning the interaction region $I_{10}$ is high (see Table \ref{tab:error_budget}), the observables calculated from the SVC states still follow the exact values, suggesting that fidelity is a rough bound on the quality of prepared variational states. Since quantum devices compute observables, even states that have high infidelity may be used to approximately simulate the desired process.

In the last two $\lambda=0$ plots for $t=8,9$, the SVC observables are seen to deviate significantly from the exactly-computed observables. This is an indication that the circuit used to simulate these time steps is not deep enough and more variational layers should be used. The number of steps is incremented at $t=4$ and $t=7$, and it can be seen in Fig.~\ref{fig:exact_vs_variational} that the variational approximations get closer to the true values at these times, than for the preceding time steps.

\section{Optimization}
\label{sec:optimization}
Optimization of the vacuum and wavepacket preparation circuits proceeds in a straightforward way. The SC-ADAPT-VQE ansatz has only three parameters, and its symmetry-preserving structure results in fast convergence to the true vacuum. The locality of the wavepacket state enables the general brickwall circuit used in state preparation to function well. Because of its expressivity, it is able to initialize the wavepacket with high accuracy using a small number of layers. Both of these optimization steps are implemented in an iterative manner, where additional layers are added as needed until desired criteria are met, such as a combination of maximum circuit depth and desired value of $I_d$.

The time evolution circuit optimization is done in a sequential manner over the simulation times. Because of the Trotter step-like nature of the ansatz, results from optimizations of previous steps are used as initial guesses for future optimizations. For example, since $t=4$ uses two variational steps, and $t=2$ uses one step, the angles implementing the $t=2$ evolution are used as initial guesses for the $t=4$ optimization. Table \ref{tab:initial_guess_map} specifies how the initial guesses utilize parameters that have been optimized for past simulation times. The Broyden–Fletcher–Goldfarb–Shanno (BFGS) gradient-based optimization method \cite{Fletcher_1986} is used, with a tolerance of $10^{-5}$.

\begin{table}
\centering
\renewcommand{\arraystretch}{1.4}
\begin{tabularx}{\linewidth}{|c||Y|Y||c||Y|Y||c||Y|Y|} \hline
$t$ & \# of variational steps & Initial guess & $t$ & \# of variational steps & Initial guess & $t$ & \# of variational steps & Initial guess\\\hline\hline
1 & 1 & -     & 4 & 2 & $t=2$ & 7 & 3 & $t=5$\\\hline
2 & 1 & $t=1$ & 5 & 2 & $t=4$ & 8 & 3 & $t=7$\\\hline
3 & 1 & $t=2$ & 6 & 2 & $t=5$ & 9 & 3 & $t=8$\\\hline
\end{tabularx}
\renewcommand{\arraystretch}{1}
\caption{The map specifying the parameters from completed variational optimizations used as initial guesses for optimizations of circuits implementing later simulation times.}
\label{tab:initial_guess_map}
\end{table}

Gradient-based circuit optimization methods such as BFGS encounter challenges when the amount of parameters is increased due to the problem of barren plateaus. Barren plateaus occur as a result of circuit overparameterization, when gradients of the cost function with respect to the parameters become vanishingly small, preventing gradient-based local optimization algorithms from effectively making progress.\footnote{Interestingly, it was recently shown that the absence of barren plateaus indicates that there exists an efficient way to simulate the problem classically \cite{Cerezo:2023nqf}.} The barren plateau problem has been carefully studied (for a review, see Ref.~\cite{Larocca:2024plh}), as well as methods to overcome it in special cases \cite{Ostaszewski:2019vnn,Nakanishi:2020wok,Wierichs:2021nwf,Barison2021efficientquantum,Nadori:2024twv}. In the context of the optimizations run to approximate the time evolution operator using the ansatz circuit of Fig.~\ref{fig:time_evolution_circ_step}, examination of the optimization landscape for small systems shows many flat regions. Despite this, it can be seen (e.g., in Fig.~\ref{fig:trot_vs_variational_fidelity}), that by using knowledge of the physics of the system to design variational circuits, these circuits are able to capture the time evolution well.

Initial guesses are a crucial ingredient to tame barren plateaus when using gradient-based optimization methods. So-called ``warm starts" to standard optimization methods have been investigated in the context of parametrized circuit optimization \cite{Puig:2024rtm,Wang:2024pap}. In this light, it may be beneficial to use gradient-free methods, such as those taking advantage of the periodic dependence of the cost function on the parameters, to produce initial guesses for gradient-based optimizers. The barren plateau problem can be further reduced by introducing more structure into the circuit, such as through physical considerations as described in Sec.~\ref{sec:scalable_variational_circuits}. This can be viewed as a way to remove redundancy in the circuit.

The ansatz for time evolution for a specific simulation time is also built up in a greedy way, iteratively adding layers and using the parameters found for the previous layer as initial guesses. This way, with each layer of the variational ansatz, the algorithm creates a more precise approximation to the true time evolution. An important property of the circuit in Fig.~\ref{fig:time_evolution_circ_step} is that it implements the identity operator when all parameters are set to 0. Because of this, the layer-wise iterative optimization method ensures that additional layers increase the quality of the solution, since the optimizer can set all parameters to 0 if no other choice increases the overlap with the target state. This method helps to address the problem of barren plateaus by providing optimized initial guesses for future layers of the circuit. This is found to perform much better than optimizing all parameters from scratch, where most initial guesses land in a part of parameter space with vanishing gradients. This method is used as a heuristic when selecting the number of layers per variational step and the number of steps for each simulation time.

In addition to using variational parameters optimized for earlier simulation times as initial guesses, parameters optimized for Hamiltonians with different $\lambda$ may be used. It can be seen in Table \ref{tab:error_budget} that the $\lambda=0$ optimizations consistently perform worse than the $\lambda=2$ ones due to the larger correlation length in the free theory. Parameters implementing the time evolution in theories with smaller correlation lengths may be used as initial guesses for those with larger correlation lengths. For instance, the $\lambda=2$ parameters may help drive the $\lambda=0$ optimization to a better region of parameter space and produce a higher quality result. Overall, it is expected that there will be improvements to the optimization techniques, both based on the structure of the circuits, and on the hyperparameters and initial guesses.

\section{Error mitigation}
\label{sec:error_mitigation}
A suite of error mitigation techniques is used in this work, enabling extraction of meaningful results from noisy data from quantum devices. The central ingredient of the error mitigation strategy used in this work is ODR, which is a local version of decoherence renormalization \cite{Urbanek:2021oej,Farrell:2022wyt,Farrell:2022vyh}. In this method, known noise-free results of circuits that are easily simulated classically are compared to results of the same circuits from the quantum device. This ratio is then used in a spatially-local (operator-wise) fashion to measure the noise that has impacted each local expectation value. ODR assumes a Pauli noise model. The standard method of PT is used to convert the coherent noise in each circuit to Pauli noise, $\rho \rightarrow \sum_i p_i P_i \rho P_i$, where $\rho$ is a reduced state used to measure a local observable. The expectation value of an observable $O$ under this noise channel is given by $\langle O \rangle = \sum_i p_i\text{Tr}(P_i\rho P_i O)$. For Pauli observables, $P_i O P_i = \pm O$, and this channel rescales the measured values $\langle O\rangle_\text{meas}$ compared to their noise-free values $\langle O\rangle_\text{true}$, $\langle O\rangle_\text{meas} = p \langle O\rangle_\text{true}$. The error mitigation scheme works by estimating the parameter $p_j$ for this channel for each local observable $O_j$. 

In this work, ODR is run at the level of $\langle ZZ\rangle$ measurements, from which $\langle \phi^2_j \rangle$ is computed using Eq.~\eqref{eq:phi_qubits}. In the specific case of $n_q=2$ and $\phi_\text{max}=1.5$, $\langle \phi^2_j\rangle$ is given by
\begin{align}
    \langle \phi^2_j\rangle = 1.25 + \langle Z_{2j}Z_{2j+1}\rangle.
\end{align}
ODR estimates expectation values in the physics circuit $\overline{\langle ZZ \rangle}_\text{phys}$ twirl by twirl, using knowledge of the noiseless expectation values of the mitigation circuit $\langle ZZ\rangle_\text{mit,true}$ and its measurements from the device $\langle ZZ\rangle_\text{mit,meas}$ using Eq.~\eqref{eq:odr_p_j}:
\begin{align}
    \overline{\langle Z_{2j}Z_{2j+1}\rangle}_\text{phys} = \langle Z_{2j}Z_{2j+1}\rangle_\text{phys,meas}\frac{1}{p_j} = \langle Z_{2j}Z_{2j+1}\rangle_\text{phys,meas} \frac{\langle Z_{2j}Z_{2j+1}\rangle_\text{mit,true}}{\langle Z_{2j}Z_{2j+1}\rangle_\text{mit,meas}}.
\end{align}
ODR hinges on selecting mitigation circuits whose noise is as similar to the noise in the physics circuits as possible. The brickwall circuits used in this work are much more dense than those used in previous works where ODR was successfuly applied. This results in errors spreading much faster throughout the qubits. As a result, it is found in this work that previous choices of mitigation circuits, such as applying the time evolution operator $U(-t/2)U(t/2) = \mathbbm{1}$ (self-mitigation \cite{ARahman:2022tkr}) or $U(0) = \mathbbm{1}$ were not able to reflect the noise present in the physics circuit well. Similarly, using various ``Cliffordized" versions of the physics circuits as mitigation circuits did not capture the noise accurately for use in ODR.\footnote{Proposals to combine Zero Noise Extrapolation \cite{Temme:2016vkz} with ODR have seen fruitful results \cite{Ciavarella:2024fzw}.} This is particularly evident at late times, where the circuits are deeper and the mitigation circuits used previously deviate more from the physics circuit being implemented.

\begin{figure*}
\includegraphics[width=\linewidth]{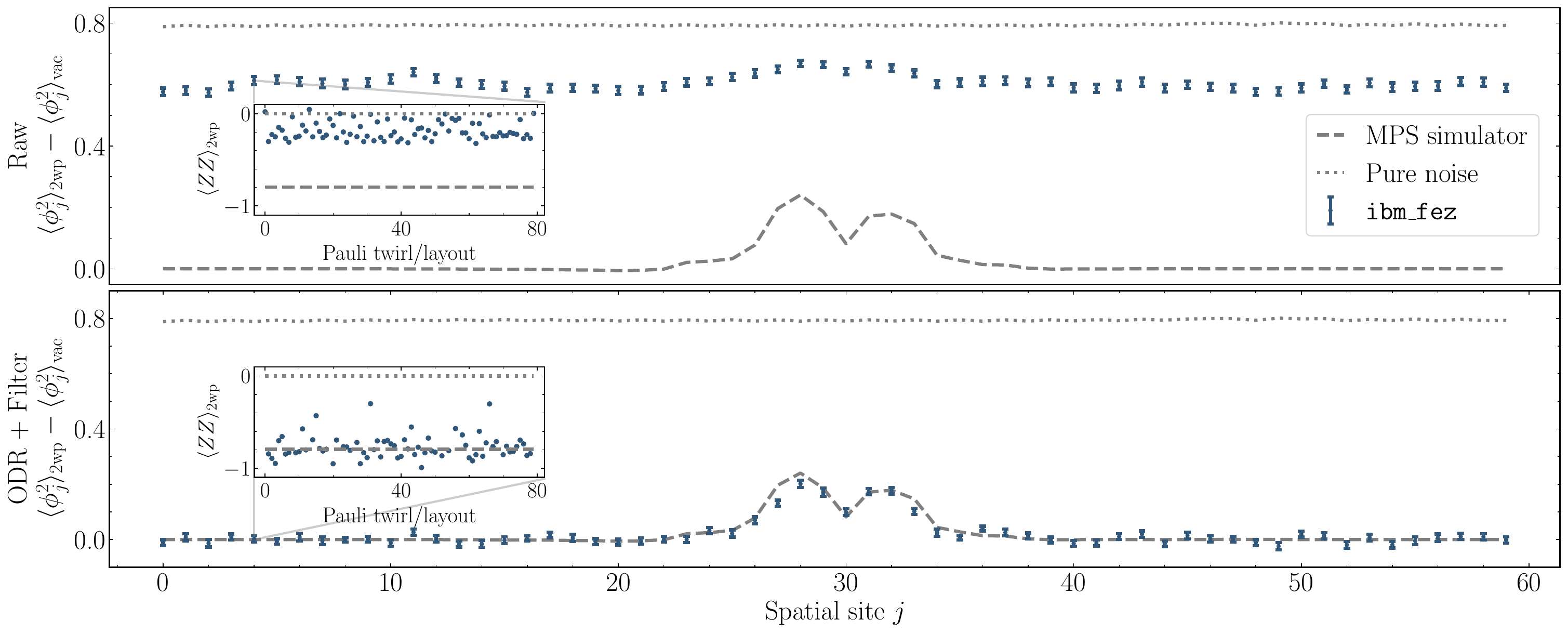}
\caption{The effect of ODR and filtering on the data from {\tt ibm\_fez} for $\lambda=0,\,t=8$. The points show data collected using 80 PTs, two TREX twirls, and 8000 shots for each circuit. The expectation value $\langle\phi^2_j\rangle_\text{2wp}-\langle\phi^2_j\rangle_\text{vac}$ is shown before (top panel) and after (bottom panel) ODR and filtering. The grey dashed line represents classical noiseless MPS circuit simulations, and the grey dotted line shows the results if the state had fully decohered to the completely mixed state. The error bars represent one standard deviation of the bootstrap-resampled data. ODR and filtering act on the $\langle ZZ\rangle_\text{2wp}$ expectation value twirl by twirl (insets).}
\label{fig:t_8_l_0_mitigation}
\end{figure*}

Instead, the circuits simulating the vacuum time evolution are used as mitigation circuits in this work. If the vacuum were prepared exactly, and if the time evolution operator were implemented exactly, time evolution would have no effect as the vacuum is an eigenstate of the Hamiltonian, $e^{-itH}|\psi_\text{vac}\rangle = |\psi_\text{vac}\rangle$. Since both the vacuum state preparation and the time evolution are implemented approximately using variational circuits, the prepared $|\psi_\text{vac}\rangle$ has some time evolution. However, this vacuum evolution can be computed classically, owing to the translation-invariant nature of $|\psi_\text{vac}\rangle$. Up to finite-size effects, the time evolution of translationally-invariant states does not change with system size. The vacuum time evolution for a given system size can be estimated following the same procedure as for the convergence of the variational parameters described in the SVC algorithm. Due to the fact that correlations in the vacuum decay exponentially in this theory, finite-size effects have an exponentially-small contribution. As a result, the infinite-volume vacuum expectation values $\langle \psi_\text{vac}|ZZ|\psi_\text{vac}\rangle_\infty$ (as well as those for any $L$) can be estimated by extrapolating from an exponential fit to the vacuum expectation values at a series of $L$. This provides a way to compute the $\langle ZZ\rangle_\text{mit,true}$. The values of these extrapolations for all simulation times in both the free and interacting theories, as well as the values at $L=60$ determined by MPS, are provided in Table \ref{tab:extrapolated_vacuum_evolution}. For the parameters and simulation times considered, this method is seen to give much better estimates of the noise, as a result of the same time evolution being applied in both the physics and mitigation circuits. The effect of ODR can be seen in Fig.~\ref{fig:t_8_l_0_mitigation}. The unmitigated data shown in the upper panel is much closer to the completely decohered result (grey dotted line), than the true result from MPS circuit simulations (grey dashed line). After the error mitigation scheme is applied (bottom panel) to measure the noise and adjust the results, measurements of $\langle\phi^2_j\rangle_\text{2wp}-\langle\phi^2_j\rangle_\text{vac}$ are much closer to the expected results from MPS. The action of ODR on $\langle ZZ\rangle_\text{2wp}$ measurements, from which $\langle \phi^2_j\rangle_\text{2wp}$ is calculated, is shown in the insets of Fig.~\ref{fig:t_8_l_0_mitigation} for lattice site $j=4$. 

\begin{table}
\centering
\renewcommand{\arraystretch}{1.4}
\begin{tabularx}{\linewidth}{|c||Y|Y|Y|Y||Y|Y|Y|Y|}
\hline
&  \multicolumn{4}{c||}{$\lambda=0$}  &  \multicolumn{4}{c|}{$\lambda=2$} \\\hline
\multirow{2}{*}{\makecell{$t$}} & \multicolumn{2}{c|}{Even sites} & \multicolumn{2}{c||}{Odd sites} & \multicolumn{2}{c|}{Even sites} & \multicolumn{2}{c|}{Odd sites} \\
\cline{2-9}
 & MPS & Extrapolation & MPS & Extrapolation & MPS & Extrapolation & MPS & Extrapolation \\
\hline\hline
 1 & 0.3949 & 0.3944 & 0.3882 & 0.3877 & 0.3580  & 0.3576 & 0.3469 & 0.3465 \\\hline
 2 & 0.4677 & 0.4670  & 0.4678 & 0.4671 & 0.3692 & 0.3688 & 0.3621 & 0.3617 \\\hline
 3 & 0.4970  & 0.4955 & 0.5073 & 0.5058 & 0.3477 & 0.3471 & 0.3551 & 0.3546 \\\hline
 4 & 0.4678 & 0.4627 & 0.4768 & 0.4717 & 0.3959 & 0.3941 & 0.3767 & 0.3748 \\\hline
 5 & 0.5339 & 0.5251 & 0.5621 & 0.5528 & 0.3620  & 0.3617 & 0.3672 & 0.3671 \\\hline
 6 & 0.5275 & 0.5331 & 0.5404 & 0.5201 & 0.4027 & 0.4017 & 0.3995 & 0.3985 \\\hline
 7 & 0.4349 & 0.4292 & 0.4227 & 0.4170  & 0.3693 & 0.3686 & 0.4010  & 0.4004 \\\hline
 8 & 0.4620  & 0.4596 & 0.4617 & 0.4551 & 0.3910  & 0.3902 & 0.3768 & 0.3757 \\\hline
 9 & 0.4780  & 0.4606 & 0.4987 & 0.4844 & 0.4143 & 0.4136 & 0.3946 & 0.3939 \\\hline
\end{tabularx}
\renewcommand{\arraystretch}{1}
\caption{ $\langle\psi_\text{vac}|e^{itH}\phi^2_je^{-itH}|\psi_\text{vac}\rangle$ that are used as ``true" values in the error mitigation scheme for the $L=60$ (120 qubits) simulations. Expectation values determined from classical statevector simulations by using an exponential extrapolation are compared to those determined by a MPS circuit simulation of the $L=60$ system. A maximum system size of $L=16$ (32 qubits) was used to compute the extrapolation. The even and odd sites have slightly different values because of the Trotterized state preparation, see Sec.~\ref{sec:circuits_vac_prep} for details.}
\label{tab:extrapolated_vacuum_evolution}
\end{table}

This method, similar to self-mitigation and other methods used earlier, is limited to early simulation times. In the case of the previous methods, the limiting factor was the growth of the difference in the noise between the physics and mitigation circuits with increasing circuit depth. Mitigation using the extrapolated vacuum evolution is limited by the quality of extrapolation. Finite-size effects grow as $O(t/L)$, so larger system sizes must be used to get reliable extrapolated values at later times. In previous choices of mitigation circuit, the physics circuit was modified to implement the identity and applied to the same initial state. In this work it is found that applying the same exact unitary mitigates errors in dense circuits (like the brickwall circuits of Fig.~\ref{fig:full_circuit}) much better, even though the starting state is different. This method is expected to work especially well in settings where perturbations on top of the vacuum are small, so that the state-dependent noise is similar in the physics and mitigation circuits. This is the case for wavepackets in scalar field theory (as seen in Fig.~\ref{fig:results_by_time} for example).

Note that it is possible to directly calculate an approximation to the vacuum evolution using MPS. In this work, results from classical MPS circuit simulations are used as ground truth to compare to results from {\tt ibm\_fez}. For the system parameters chosen and the times considered, it can be seen that the extrapolated vacuum evolution is quite close to the MPS predictions (see Table \ref{tab:extrapolated_vacuum_evolution}), with the differences only noticeable at late times. It is expected that highly-inelastic collisions will generate large quantities of entanglement and magic, thus being genuinely out of reach for simulations using classical methods such as MPS or Clifford+T. In this setting, the MPS values for the vacuum evolution may be used directly for mitigation. This is a reasonable expectation, as the ground states of one-dimensional gapped systems, such as the scalar field theory considered in this work, are known to obey area law entanglement \cite{Hastings:2007iok}, so their short-time evolution may be simulated with an MPS with a relatively small bond dimension.\footnote{The approximate variational methods may yield states that are more entangled than the true, non-variational states. This may be averted by using techniques similar to Refs.~\cite{Miyakoshi:2023zzc,Causer:2023wpp,Gibbs:2024emw} to train the variational circuits with MPS in the first place. Furthermore, MPS approximations to the $|\psi_\text{2wp}\rangle$ evolution circuit may be used as mitigation circuits in the future as well, provided they can be constructed in a way that preserves the structure of the noise.}

The ratio $p_j$ calculated from the results of the mitigation circuit is used to remove observations that have been particularly affected by noise. This procedure, referred to as filtering \cite{Farrell:2024fit}, discards measurements of a particular local observable if the ratio of the measured expectation value to the true expectation value is below a certain threshold; in this work the threshold is set to $p_j \geq 0.01$ for all runs. Once the ratio is applied to adjust the measurements of the physics circuit, a second round of filtering is done. This filter removes values of $\overline{\langle ZZ \rangle}_\text{phys} \geq 1$. Since the maximum possible value of $\langle ZZ\rangle=1$, any values above 1 are the result of the more decoherence being present for that observable in the mitigation circuits; these observations are discarded as well. These two filters work by removing outliers and extremely noisy data points, causing the size of the bootstrap samples for each local observable to be different. 

A loop of qubits connected by two-qubit gates is necessary because of the PBCs used in this work. As a result of the scalability of the simulation methods outlined in previous sections, it is possible to run circuits on the largest possible loop on {\tt ibm\_fez} (120 qubits). A specific loop is chosen based on a combination of the smallest errors and longest coherence time throughout the device. It is found in this work that the randomization of mapping from circuit qubits to device qubits (layout randomization) helps to tailor the noise closer to depolarizing noise, in addition to averaging out effects from noise due to localized imperfections on the device, such as specific noisy qubits, gates, and readout channels. In this work, a loop is fixed for all runs and the starting point of the loop is chosen at random for each circuit. For each run, the circuits are executed in an order that alternates between the mitigation (vacuum) and physics (wavepacket) circuits, for the same PT and layout. Since the noise on the device is expected to fluctuate with time, this is intended to more accurately measure the noise by running the mitigation circuits directly before the corresponding physics circuits. The same set of twirls and layouts is used for the physics and mitigation circuits.

To mitigate measurement errors, TREX is used. This method involves applying Pauli gates on all qubits at random prior to measurement, flipping the value of the classical result if the Pauli applied is $X$ or $Y$, and doing nothing in the case of $Z$ or $I$. Two TREX twirls are used for each Pauli-twirled circuit in this work. 

The IBM Heron processors have seen a 3-5$\times$ increase in performance over their Eagle predecessors, and have significantly decreased qubit crosstalk \cite{Gambetta_2023}. Despite this, it is still seen in this work that crosstalk remains a non-negligible source of error. DD is used to minimize this effect and remove erroneous evolution during qubit idle times. To mitigate these effects, the crosstalk-robust CRXY4 DD sequence \cite{PhysRevLett.131.210802} is used. This sequence consists of staggered XY4 sequences applied to neighboring qubits to cancel out collective evolution caused by crosstalk.\footnote{Recent work found that asymmetric implementations of $Y$ gates can have an adverse impact on the effectiveness of DD sequences \cite{vezvaee2024virtualzgatessymmetric}. As a result, all $Y$ gates used in this work, including the DD stage, are converted to the native gate set using the symmetric implementation $Y = R_Z(\pi/2)XR_Z(-\pi/2)$.}

In particular, the effects of crosstalk are noticeable in regions where qubits are idle for numerous circuit layers directly adjacent to ones that have gates acting on them. These types of errors are most prevalent in the wavepacket creation step of the simulation algorithm, where the wavepacket creation circuits are acting in limited regions of the quantum device (green blocks of Fig.~\ref{fig:full_circuit}). To mitigate these effects, barriers are inserted to isolate the times where both qubits are idle or single-qubit gates are being run on the wavepacket qubits, from the times where two-qubit gates are acting in the wavepacket region. The barriers instruct the circuit transpilation to add DD separately for these two cases because the noise is different during each process. 

\section{Variational parameters}
\label{sec:variational_params}
In this appendix, the variational parameters determined using the SVC framework presented in Sec.~\ref{sec:scalable_variational_circuits} are given for vacuum preparation, wavepacket excitation, and time evolution. Exponential extrapolation is run for the vacuum preparation parameters. The $L$-extrapolated vacuum parameters are used at $L=12$ to optimize the wavepacket preparation and time evolution circuits. 

\begin{table}[h]
\centering
\renewcommand{\arraystretch}{1.4}
\begin{tabularx}{\linewidth}{|c||Y|Y|Y||Y|Y|Y|}
\hline
&  \multicolumn{3}{c||}{$\lambda=0$}  &  \multicolumn{3}{c|}{$\lambda=2$} \\\hline
$L$ & $\theta_0$ & $\theta_1$ & $\theta_2$ & $\theta_0$ & $\theta_1$ & $\theta_2$ \\ 
\hline \hline
2 & 2.5517 & 1.3354 & 0.5659 & 2.6685 & 1.0626 & 0.4370 \\\hline
4 & 2.6346 & 1.0111 & 0.9131 & 2.7164 & 0.7923 & 0.7438 \\\hline
6 & 2.6462 & 0.9469 & 0.8703 & 2.7206 & 0.7547 & 0.7141 \\\hline
8 & 2.6490 & 0.9312 & 0.8580 & 2.7213 & 0.7491 & 0.7088 \\\hline
10 & 2.6497 & 0.9278 & 0.8536 & 2.7215 & 0.7478 & 0.7087 \\\hline
12 & 2.6499 & 0.9248 & 0.8536 & 2.7214 & 0.7466 & 0.7088 \\\hline\hline
$\infty$ &  2.6495 & 0.9273 & 0.8534 & 2.7214 & 0.7476 & 0.7085 \\\hline
\end{tabularx}
\renewcommand{\arraystretch}{1}
\caption{The variational parameters to prepare the vacuum as a function of system size, for both the free and interacting theory. They correspond to the angles $\theta_0, \theta_1, \theta_2$ defined in the circuit elements of Fig.~\ref{fig:circuit_elements}a. These parameters are determined using classical simulations and optimization. The infinite-volume values shown in the last row are calculated by an exponential extrapolation.}
\label{tab:gs_prep_angles}
\end{table}

\begin{table}
\centering
\renewcommand{\arraystretch}{1.4}
\begin{tabularx}{\linewidth}{|c||Y|Y|Y|Y||Y|Y|Y|Y|}
\hline
&  \multicolumn{4}{c||}{$\lambda=0$}  &  \multicolumn{4}{c|}{$\lambda=2$} \\\hline
$j$ & $i=0$ & $i=1$ & $i=2$ & $i=3$ & $i=0$ & $i=1$ & $i=2$ & $i=3$\\
\hline\hline
  0 & -0.1531 &  0.8023 &  0.5352 & -0.6361 &  3.0113 & -0.2089 &  1.8433 &  1.2508 \\\hline
  1 & -2.6753 &  1.7051 & -0.0746 &  1.5072 & -0.1898 &  1.9292 &  0.0173 & -1.3358 \\\hline
  2 &  0.9219 &  0.4939 &  0.0057 &  0.3665 &  2.0861 &  1.3400   &  0.1340  &  1.8122 \\\hline
  3 &  0.2726 & -0.3087 &  0.5203 & -0.2217 &  2.9159 &  1.3803 & -0.9289 &  0.8222 \\\hline
  4 &  1.6515 &  1.5371 &  1.3394 &  1.4023 &  1.4625 & -1.0540  &  1.5245 &  1.5592 \\\hline
  5 & -0.3017 & -0.2190  & -0.7053 & -0.0079 &  1.0102 &  1.5805 &  1.4121 & -0.0310  \\\hline
  6 &  0.4914 &  0.0186 & -0.0630  &  1.5343 & -0.1876 &  1.6883 &  0.8825 &  1.9420  \\\hline
  7 & -1.7130  & -0.2557 & -1.1006 & -0.7642 &  1.1893 & -0.0476 & -0.8028 &  1.2294 \\\hline
  8 & -1.5369 &  1.5928 &  1.5733 & -1.2954 &  1.3134 &  1.2069 &  1.5725 &  2.1823 \\\hline
  9 &  1.2792 & -0.0988 & -1.7023 & -0.0211 & -0.2964 & -1.7139 & -0.3014 &  1.7205 \\\hline
 10 &  1.5154 &  0.0565 & -1.0869 &  3.1620  & -0.6159 &  1.6691 &  1.4801 & -1.4323 \\\hline
 11 & -1.5202 &  0.8090  & -0.5202 & -1.5110  & -1.3265 &  1.2461 &  1.5431 &  2.1283 \\\hline
 12 &  0.5463 & -0.7434 & -0.8032 &  0.9535 &  0.6183 &  0.6455 & -1.4778 & -1.3675 \\\hline
 13 & -1.0278 & -0.5927 &  1.1085 &  1.7822 &  1.7852 &  1.7364 &  1.3890  &  0.0582 \\\hline
 14 &  1.7048 &  1.5187 &  1.3373 &  1.5012 & -0.9282 & -1.1880  &  2.7412 &  1.6371 \\\hline
 15 &  0.7122 & -0.5207 & -0.0962 &  0.0393 &  1.0079 & -0.7267 &  1.3845 &  0.2163 \\\hline
 16 &  1.7216 & -0.4310  &  0.7563 &  1.0700   &  1.2270  &  1.4292 &  0.9514 &  1.7011 \\\hline
 17 & -1.5542 & -0.0394 & -0.0974 &  1.5916 & -0.9657 & -0.2391 &  0.0401 &  1.7943 \\\hline
 18 &  1.5395 &  1.5849 &  1.1416 &  1.6014 &  2.9777 & -1.2030  &  1.0188 &  1.5363 \\\hline
 19 & -0.1785 &  1.1545 & -0.0179 & -1.6719 &  1.8511 &  0.5998 &  0.0393 & -0.5372 \\\hline
\end{tabularx}
\renewcommand{\arraystretch}{1}
\caption{The parameters $\theta_{ij}$ used to initialize Gaussian wavepackets of momenta $k=\pm\pi/3$ and spread in $k$-space $\sigma_k=\pi/3$ in the free and interacting theories. The $\theta_{ij}$ correspond to the angles defined in the brickwall layer shown in Fig.~\ref{fig:circuit_elements}c: $i$ labels the layer, and $j$ labels the parameter within layer $i$. These parameters are determined using classical simulations and optimization.}
\label{tab:wp_prep_angles}
\end{table}

\begin{table}
\centering
\renewcommand{\arraystretch}{1.4}
\begin{tabularx}{\linewidth}{|c||Y|Y|Y|Y|Y|Y||Y|Y|Y|Y|Y|Y|}
\hline
&  \multicolumn{6}{c||}{$\lambda=0$}  &  \multicolumn{6}{c|}{$\lambda=2$} \\\hline
$j$ & $i=0$ & $i=1$ & $i=2$ & $i=3$ & $i=4$ & $i=5$ & $i=0$ & $i=1$ & $i=2$ & $i=3$ & $i=4$ & $i=5$ \\
\hline\hline
  0 &  0.0843 & -0.0609 &  0.6955 &  0.2961 & -0.0480  & -0.0191 &  0.1727 &  0.0329 & -1.1632 &  0.0605 &  0.6813 &  0.1975 \\\hline
  1 & -0.2860  & -3.4066 & -2.9043 &  0.0167 &  2.2841 & -0.0292 &  0.8868 & -2.6254 &  1.6950  & -0.1890  & -3.8622 &  0.1428 \\\hline
  2 &  1.5618 &  0.1327 &  1.4740  & -0.1091 & -1.4579 & -1.3838 & -0.8353 &  0.1261 & -1.1718 & -0.1111 &  1.0946 & -0.7012 \\\hline
  3 &  0.2173 &  1.3044 & -1.1338 & -1.4324 & -1.3960  & -1.6492 & -0.2904 & -2.0721 &  1.6594 &  0.5606 &  1.0016 & -0.4645 \\\hline
  4 &  0.0121 & -1.1901 &  1.9423 & -1.4593 &  0.1350  & -1.1644 & -0.6027 &  3.2926 &  0.2786 & -0.8616 & -1.5139 &  0.5588 \\\hline
  5 & -1.5731 &  2.2241 &  1.3656 &  0.1695 & -0.6027 & -1.1536 &  1.4649 & -1.5636 &  0.7272 &  1.4701 &  2.0278 &  1.0915 \\\hline
  6 & -0.1173 &  0.7041 & -1.0110  &  1.5841 & -0.6495 & -0.0152 &  0.0666 & -1.5045 &  0.8148 &  0.6901 & -0.1011 &  0.0370  \\\hline
  7 &  1.8137 &  1.2280  &  1.5580  & -0.1221 &  0.3547 & -0.0793 &  1.7586 & -1.2382 &  0.9962 & -0.5063 &  0.1100   &  0.0887 \\\hline
  8 &  0.0636 & -1.0966 &  1.9644 & -1.4114 &  3.3265 &  1.0599 & -0.6402 &  2.2965 & -0.0883 & -1.1814 & -2.0575 &  0.8626 \\\hline
  9 & -1.6840  &  2.0991 &  1.5450  &  0.3095 &  0.5833 & -1.1253 &  1.3265 & -2.0403 &  0.9642 &  1.9077 &  1.3393 &  1.1265 \\\hline
 10 & -0.0975 &  0.7022 & -1.0872 &  1.6051 & -0.6503 & -0.0324 & -0.1070  & -1.1265 &  1.2493 & -0.1006 & -0.5110  & -0.0454 \\\hline
 11 &  1.8584 &  1.1814 &  1.5940  & -0.0322 &  0.3009 & -0.0683 &  1.6193 & -1.1086 &  0.9842 & -0.5982 & -0.0874 &  0.0748 \\\hline
\end{tabularx}
\renewcommand{\arraystretch}{1}
\caption{The parameters $\theta_{ij}$ used to implement time evolution for $t=1$ in the free and interacting theories. The $\theta_{ij}$ correspond to the angles defined in the brickwall layer shown in Fig.~\ref{fig:circuit_elements}d: $i$ labels the layer, and $j$ labels the parameter within layer $i$. These parameters are determined using classical simulations and optimization.}
\label{tab:time_evolution_angles_1}
\end{table}

\begin{table}
\centering
\renewcommand{\arraystretch}{1.4}
\begin{tabularx}{\linewidth}{|c||Y|Y|Y|Y|Y|Y||Y|Y|Y|Y|Y|Y|}
\hline
&  \multicolumn{6}{c||}{$\lambda=0$}  &  \multicolumn{6}{c|}{$\lambda=2$} \\\hline
$j$ & $i=0$ & $i=1$ & $i=2$ & $i=3$ & $i=4$ & $i=5$ & $i=0$ & $i=1$ & $i=2$ & $i=3$ & $i=4$ & $i=5$ \\
\hline\hline
  0 & -0.0127 & -0.0303 &  0.8574 &  0.2545 & -0.1172 &  0.0181 &  0.0398 &  0.0648 & -0.8169 &  0.3516 &  0.7947 &  0.1075 \\\hline
  1 & -0.7944 & -4.0568 & -3.6008 &  0.0228 &  3.4077 &  0.7981 &  1.1739 & -1.9846 &  1.7857 &  0.4659 & -4.8039 & -0.2786 \\\hline
  2 &  1.5237 &  0.3437 &  1.6710  & -0.1623 & -1.6623 & -1.6914 & -1.5506 &  0.1435 & -0.4705 & -0.0216 &  1.8417 & -1.0534 \\\hline
  3 &  0.0042 &  1.5494 & -1.0016 & -1.4038 & -1.9429 & -1.6069 & -0.0645 & -0.9712 & -1.0871 &  1.1405 &  1.3022 & -0.4437 \\\hline
  4 & -0.2237 & -0.8308 &  1.5671 & -1.4545 & -0.1409 & -0.5384 & -0.4929 &  2.1439 & -0.2443 & -1.7539 & -1.5463 & -0.2068 \\\hline
  5 & -1.5910  &  2.5192 &  1.4323 &  0.2501 & -0.8664 & -0.5929 &  1.2547 & -1.7316 & -0.1444 &  2.1785 &  1.9204 &  1.2651 \\\hline
  6 & -0.0115 &  0.6926 & -0.7056 &  2.0844 & -0.5686 &  0.0572 &  0.2956 & -1.9191 &  0.9858 & -0.5653 &  0.1543 &  0.0046 \\\hline
  7 &  2.6181 &  1.3091 &  1.4943 &  0.0006 & -0.3561 & -0.0597 &  0.4242 &  1.1723 &  1.2054 & -0.0798 &  0.1706 &  0.0134 \\\hline
  8 & -0.2804 & -0.9134 &  1.7664 & -1.2446 &  3.1131 &  0.5317 & -0.6196 &  2.0314 & -0.3388 & -1.9148 & -1.7189 & -0.1011 \\\hline
  9 & -1.5676 &  2.3869 &  1.5213 &  0.3573 &  0.8461 & -0.5576 &  1.1632 & -1.6832 &  0.0979 &  1.9924 &  2.1347 &  1.2702 \\\hline
 10 & -0.0285 &  0.6670  & -0.6825 &  2.0468 & -0.5351 &  0.0372 &  0.1732 & -2.0512 &  0.9512 & -0.7355 &  0.0777 &  0.0141 \\\hline
 11 &  2.5855 &  1.3686 &  1.4927 &  0.0083 & -0.3019 & -0.0333 &  0.3224 &  1.0980  &  1.3002 & -0.0688 &  0.3608 &  0.0228 \\\hline
\end{tabularx}
\renewcommand{\arraystretch}{1}
\caption{The parameters $\theta_{ij}$ used to implement time evolution for $t=2$ in the free and interacting theories. The $\theta_{ij}$ correspond to the angles defined in the brickwall layer shown in Fig.~\ref{fig:circuit_elements}d: $i$ labels the layer, and $j$ labels the parameter within layer $i$. These parameters are determined using classical simulations and optimization.}
\label{tab:time_evolution_angles_2}
\end{table}

\begin{table}
\centering
\renewcommand{\arraystretch}{1.4}
\begin{tabularx}{\linewidth}{|c||Y|Y|Y|Y|Y|Y||Y|Y|Y|Y|Y|Y|}
\hline
&  \multicolumn{6}{c||}{$\lambda=0$}  &  \multicolumn{6}{c|}{$\lambda=2$} \\\hline
$j$ & $i=0$ & $i=1$ & $i=2$ & $i=3$ & $i=4$ & $i=5$ & $i=0$ & $i=1$ & $i=2$ & $i=3$ & $i=4$ & $i=5$ \\
\hline\hline
  0 &  0.0021 & -0.0586 &  1.5776 &  0.0003 &  0.5290  &  0.4828 &  0.1081 &  0.0477 & -1.1605 &  0.5728 &  1.1499 &  0.2074 \\\hline
  1 & -1.5784 & -4.6994 & -3.1345 &  0.0540  &  3.4056 &  0.0368 &  0.1861 & -2.7631 &  2.0690  &  0.1181 & -3.8915 &  2.7622 \\\hline
  2 &  1.4004 &  0.6131 &  0.7863 & -0.1804 & -1.6037 & -0.9163 & -1.5665 &  0.1879 & -0.5834 &  0.0316 &  1.8911 & -1.1214 \\\hline
  3 &  0.1212 &  1.6162 & -1.5701 & -1.7525 & -1.5181 & -0.8125 & -0.0984 & -1.5692 & -0.2535 &  3.0750  &  1.6547 & -0.5287 \\\hline
  4 & -0.3352 & -1.5919 &  1.2529 & -0.4186 &  1.6014 & -1.5573 & -0.1697 &  1.5929 & -0.2217 & -1.5676 & -1.8083 & -0.4096 \\\hline
  5 & -1.5505 &  3.1467 &  1.4259 &  0.3900   & -1.4477 &  0.2438 &  1.6468 & -1.5093 & -0.3895 &  2.8569 &  1.8389 &  1.1812 \\\hline
  6 & -0.0018 &  1.1808 & -0.0031 &  1.5923 & -2.6329 & -0.9154 & -0.0441 & -1.8221 &  0.9875 &  1.0029 & -0.2364 &  0.0088 \\\hline
  7 &  3.1182 &  1.5498 &  1.5514 &  0.3567 & -0.0363 & -0.0068 &  1.0255 &  0.1220  &  1.6884 & -0.0175 & -1.8634 & -0.0213 \\\hline
  8 & -0.4566 & -1.4424 &  1.4231 & -0.5025 &  4.7225 &  1.5587 & -0.1532 &  1.6630  & -0.1710  & -1.5943 & -1.7456 & -0.4261 \\\hline
  9 & -1.5682 &  3.1396 &  1.4702 &  0.0032 &  1.4704 & -0.2056 &  1.6425 & -1.5363 & -0.3713 &  2.9048 &  1.9027 &  1.1827 \\\hline
 10 &  0.0028 &  0.9779 &  0.0008 &  1.5751 & -0.6327 &  0.9388 & -0.0570  & -1.7169 &  0.9733 &  0.9879 & -0.1879 & -0.0017 \\\hline
 11 &  3.1055 &  1.5655 &  1.6030  &  0.2539 & -0.0241 & -0.0084 &  1.0308 &  0.1463 &  1.6700   & -0.0398 & -1.8732 & -0.0112 \\\hline
\end{tabularx}
\renewcommand{\arraystretch}{1}
\caption{The parameters $\theta_{ij}$ used to implement time evolution for $t=3$ in the free and interacting theories. The $\theta_{ij}$ correspond to the angles defined in the brickwall layer shown in Fig.~\ref{fig:circuit_elements}d: $i$ labels the layer, and $j$ labels the parameter within layer $i$. These parameters are determined using classical simulations and optimization.}
\label{tab:time_evolution_angles_3}
\end{table}

\begin{table}
\centering
\renewcommand{\arraystretch}{1.4}
\begin{tabularx}{\linewidth}{|c||Y|Y|Y|Y|Y|Y||Y|Y|Y|Y|Y|Y|}
\hline
&  \multicolumn{6}{c||}{$\lambda=0$}  &  \multicolumn{6}{c|}{$\lambda=2$} \\\hline
$j$ & $i=0$ & $i=1$ & $i=2$ & $i=3$ & $i=4$ & $i=5$ & $i=0$ & $i=1$ & $i=2$ & $i=3$ & $i=4$ & $i=5$ \\
\hline\hline
  0 & -0.1073 &  0.0847 &  0.4879 &  0.3093 & -0.2566 & -0.0338 & -0.0025 & -0.0237 & -0.7062 &  0.2692 &  0.5673 & -0.2870  \\\hline
  1 & -0.5637 & -5.1000    & -2.6777 & -0.0863 &  3.3271 &  1.2604 &  1.0773 &  0.7424 &  2.4129 &  0.3060  & -4.2993 &  0.5121 \\\hline
  2 &  1.5109 &  0.4379 &  1.0296 & -0.1679 & -1.6439 & -0.8366 & -1.6468 &  0.3683 & -0.4062 &  0.0612 &  1.7473 & -1.9365 \\\hline
  3 &  0.0692 &  1.5616 & -1.0566 & -2.0557 & -2.3746 & -1.5466 & -0.0042 & -1.4611 &  0.0559 &  0.7209 &  0.9852 & -0.7805 \\\hline
  4 & -0.4893 & -1.0224 &  2.3117 & -1.6204 &  0.2698 & -0.5523 & -0.4652 &  2.3685 & -0.7806 & -1.5357 & -1.5061 &  0.0209 \\\hline
  5 & -1.6233 &  3.2011 &  1.8399 & -0.7357 & -0.8480  & -0.2008 &  1.5544 & -1.6221 &  0.9499 &  2.2305 &  2.2893 &  1.1904 \\\hline
  6 & -0.0555 &  1.5751 & -1.0267 &  1.3182 & -1.8387 &  0.1214 & -0.0479 & -1.2671 &  0.6778 & -0.0874 & -0.0155 & -0.0129 \\\hline
  7 &  3.6119 &  1.2176 &  1.5505 &  0.2732 & -0.9677 & -0.0461 & -2.3391 &  0.0142 &  1.3246 & -0.0462 & -0.7364 &  0.0004 \\\hline
  8 & -0.5405 & -0.9474 &  2.3937 & -1.5040  &  3.4139 &  0.5829 & -0.4848 &  2.2950  & -0.7543 & -1.5413 & -1.4418 &  0.0648 \\\hline
  9 & -1.4556 &  3.2751 &  1.6913 & -0.9322 &  0.7875 & -0.3782 &  1.4431 & -1.6935 &  0.9909 &  2.1455 &  2.0475 &  1.1984 \\\hline
 10 & -0.0087 &  1.5019 & -1.0804 &  1.3444 & -1.9556 &  0.0493 &  0.0243 & -1.2489 &  0.7149 &  0.0038 & -0.0335 & -0.0282 \\\hline
 11 &  3.4239 &  1.1735 &  1.5976 &  0.2317 & -1.0280  & -0.0450  & -2.4800   & -0.0288 &  1.4062 & -0.0953 & -0.7767 &  0.0393 \\\hline
\end{tabularx}
\renewcommand{\arraystretch}{1}
\caption{The parameters $\theta_{ij}$ used to implement time evolution for $t=4$ in the free and interacting theories. The $\theta_{ij}$ correspond to the angles defined in the brickwall layer shown in Fig.~\ref{fig:circuit_elements}d: $i$ labels the layer, and $j$ labels the parameter within layer $i$. These parameters are determined using classical simulations and optimization.}
\label{tab:time_evolution_angles_4}
\end{table}

\begin{table}
\centering
\renewcommand{\arraystretch}{1.4}
\begin{tabularx}{\linewidth}{|c||Y|Y|Y|Y|Y|Y||Y|Y|Y|Y|Y|Y|}
\hline
&  \multicolumn{6}{c||}{$\lambda=0$}  &  \multicolumn{6}{c|}{$\lambda=2$} \\\hline
$j$ & $i=0$ & $i=1$ & $i=2$ & $i=3$ & $i=4$ & $i=5$ & $i=0$ & $i=1$ & $i=2$ & $i=3$ & $i=4$ & $i=5$ \\
\hline\hline
  0 & -0.1326 &  0.0372 &  0.5814 &  0.3058 & -0.2071 & -0.1162 &  0.0000      & -0.0105 & -1.3237 &  0.6746 &  0.6885 &  1.3835 \\\hline
  1 & -0.7267 & -5.7599 & -3.2273 & -0.0727 &  3.6763 &  1.1732 & -0.0002 &  1.0968 &  1.5905 &  1.5804 & -5.7955 &  1.5715 \\\hline
  2 &  1.5629 &  0.4892 &  1.0416 & -0.1845 & -1.7186 & -1.1372 & -1.6402 &  0.3005 & -0.4687 & -0.0184 &  2.7857 & -1.6367 \\\hline
  3 & -0.0209 &  1.5764 & -1.1759 & -2.1123 & -2.2658 & -1.4966 &  0.0821 & -1.5130  &  0.0071 &  0.2494 &  1.5453 & -1.5708 \\\hline
  4 & -0.5286 & -1.0685 &  2.2987 & -1.7233 &  0.1708 & -0.5050  & -0.0255 &  1.5761 &  0.0117 & -3.1069 & -1.1159 &  0.2522 \\\hline
  5 & -1.6923 &  3.3742 &  2.0409 & -0.7240  & -0.6956 & -0.1600   &  1.4987 & -1.0560  & -0.2535 &  3.5926 &  1.5549 &  0.2101 \\\hline
  6 & -0.1063 &  1.4380  & -0.8177 &  1.2692 & -1.6112 &  0.0894 &  0.0001 & -0.9552 &  0.4211 & -1.2032 &  0.2553 &  0.4862 \\\hline
  7 &  4.3371 &  1.3833 &  1.5239 &  0.3667 & -0.7124 &  0.0020  & -2.6630  & -0.0029 &  1.5707 & -0.0319 & -1.5476 &  0.1618 \\\hline
  8 & -0.6487 & -0.9463 &  2.4319 & -1.6415 &  3.3190  &  0.5682 &  0.0047 &  1.6030  &  0.0016 & -3.1070  & -0.9698 &  0.2296 \\\hline
  9 & -1.3430  &  3.6764 &  1.7624 & -1.0195 &  0.8142 & -0.4189 &  1.3910  & -1.0438 & -0.2541 &  3.6141 &  1.5336 &  0.2097 \\\hline
 10 & -0.0081 &  1.1308 & -1.0171 &  1.4978 & -1.9940  &  0.0378 & -0.0011 & -0.9338 &  0.4016 & -1.1540  &  0.2346 &  0.6202 \\\hline
 11 &  3.9799 &  1.1996 &  1.6125 &  0.2875 & -1.0267 & -0.0509 & -2.6706 & -0.0246 &  1.5695 & -0.0200   & -1.5884 &  0.1631 \\\hline
\end{tabularx}
\renewcommand{\arraystretch}{1}
\caption{The parameters $\theta_{ij}$ used to implement time evolution for $t=5$ in the free and interacting theories. The $\theta_{ij}$ correspond to the angles defined in the brickwall layer shown in Fig.~\ref{fig:circuit_elements}d: $i$ labels the layer, and $j$ labels the parameter within layer $i$. These parameters are determined using classical simulations and optimization.}
\label{tab:time_evolution_angles_5}
\end{table}

\begin{table}
\centering
\renewcommand{\arraystretch}{1.4}
\begin{tabularx}{\linewidth}{|c||Y|Y|Y|Y|Y|Y||Y|Y|Y|Y|Y|Y|}
\hline
&  \multicolumn{6}{c||}{$\lambda=0$}  &  \multicolumn{6}{c|}{$\lambda=2$} \\\hline
$j$ & $i=0$ & $i=1$ & $i=2$ & $i=3$ & $i=4$ & $i=5$ & $i=0$ & $i=1$ & $i=2$ & $i=3$ & $i=4$ & $i=5$ \\
\hline\hline
  0 &  0.0411 & -0.1769 &  0.5015 &  0.8762 &  0.3464 & -0.0139 &  0.0030  & -0.0092 & -1.4508 &  0.6506 &  0.8001 &  1.1527 \\\hline
  1 & -0.0942 & -4.5314 & -4.1438 & -0.0091 &  2.9710  &  2.2259 & -0.0019 &  1.1037 &  1.8273 &  1.5725 & -4.2827 &  1.5692 \\\hline
  2 &  1.5761 &  0.5747 &  0.6118 & -0.5092 & -1.5826 & -1.5449 & -1.5997 &  0.2543 & -0.3165 & -0.0120  &  2.8019 & -1.7114 \\\hline
  3 & -0.0601 &  1.5681 & -1.4277 & -1.6061 & -1.5133 & -1.8533 &  0.0943 & -1.5557 &  0.0137 &  0.2642 &  1.6118 & -1.5714 \\\hline
  4 & -0.3745 & -0.7460  &  3.0021 & -2.1116 & -0.0530  & -1.3761 & -0.0057 &  1.5674 & -0.0024 & -3.1086 & -0.9774 &  0.1754 \\\hline
  5 & -1.5873 &  4.5537 &  1.6295 & -1.0644 & -0.4700   & -0.8731 &  1.5394 & -0.9324 & -0.2696 &  3.3930  &  1.5974 &  0.0565 \\\hline
  6 &  0.0100   &  1.0453 &  0.2529 &  1.4969 & -0.9874 &  0.0076 & -0.0109 & -0.6916 &  0.6663 & -1.5556 &  0.1951 &  0.6161 \\\hline
  7 &  3.0271 &  0.2209 &  1.6037 &  0.1586 & -0.9758 &  0.1598 & -2.6591 & -0.0200   &  1.5562 &  0.0239 & -1.5693 &  0.1799 \\\hline
  8 & -0.3817 & -0.7077 &  3.0607 & -2.039  &  3.1999 &  1.3527 &  0.0047 &  1.5856 &  0.0037 & -3.1055 & -1.0376 &  0.2258 \\\hline
  9 & -1.5157 &  4.5203 &  1.6534 & -1.0739 &  0.4792 & -0.8984 &  1.4434 & -0.9179 & -0.2720  &  3.3773 &  1.5878 &  0.0779 \\\hline
 10 &  0.0023 &  1.0685 &  0.2252 &  1.4997 & -0.9731 &  0.0103 & -0.0042 & -0.7239 &  0.5917 & -1.5104 &  0.2549 &  0.5399 \\\hline
 11 &  3.0216 &  0.1465 &  1.5569 &  0.2173 & -0.9865 &  0.1577 & -2.6715 & -0.0153 &  1.5556 &  0.0135 & -1.6080  &  0.1801 \\\hline
\end{tabularx}
\renewcommand{\arraystretch}{1}
\caption{The parameters $\theta_{ij}$ used to implement time evolution for $t=6$ in the free and interacting theories. The $\theta_{ij}$ correspond to the angles defined in the brickwall layer shown in Fig.~\ref{fig:circuit_elements}d: $i$ labels the layer, and $j$ labels the parameter within layer $i$. These parameters are determined using classical simulations and optimization.}
\label{tab:time_evolution_angles_6}
\end{table}

\begin{table}
\centering
\renewcommand{\arraystretch}{1.4}
\begin{tabularx}{\linewidth}{|c||Y|Y|Y|Y|Y|Y||Y|Y|Y|Y|Y|Y|}
\hline
&  \multicolumn{6}{c||}{$\lambda=0$}  &  \multicolumn{6}{c|}{$\lambda=2$} \\\hline
$j$ & $i=0$ & $i=1$ & $i=2$ & $i=3$ & $i=4$ & $i=5$ & $i=0$ & $i=1$ & $i=2$ & $i=3$ & $i=4$ & $i=5$ \\
\hline\hline
  0 & -0.0038 &  0.0528 & -0.0417 &  0.2083 &  0.0104 & -0.0076 &  0.3432 &  0.0404 & -1.0961 &  0.4267 &  1.1327 & -0.0044 \\\hline
  1 & -1.5722 & -4.7573 & -4.6618 & -0.0137 & -0.1573 & -0.2374 &  1.5457 & -0.2010  &  1.5569 & -0.0334 & -4.7501 & -0.5418 \\\hline
  2 &  1.5930  &  1.9531 &  0.8304 & -1.6783 & -1.5760  & -1.5841 & -1.5684 &  0.0604 & -0.4015 & -0.0202 &  1.5155 & -1.8278 \\\hline
  3 &  0.0228 &  1.5732 & -1.5732 & -1.5716 & -2.7976 & -1.5621 & -0.0520  &  0.0594 & -1.5664 &  1.7422 &  1.5781 & -0.0860  \\\hline
  4 & -1.9226 & -0.2589 &  2.7112 & -2.0992 &  0.0877 & -0.4603 & -1.2895 &  2.7726 & -0.1389 & -2.6832 & -2.0356 & -0.0042 \\\hline
  5 & -1.5758 &  4.6627 &  1.5745 &  3.3032 & -0.0037 & -0.7178 &  0.4848 & -1.4714 & -0.1673 &  1.6008 &  3.0597 &  1.2465 \\\hline
  6 &  0.0072 &  0.3859 & -0.2240  &  3.0196 & -0.4894 & -0.0041 &  0.0162 & -2.2549 &  0.8874 & -0.7982 &  0.3035 & -0.0108 \\\hline
  7 &  3.1959 &  1.5730  &  1.5746 & -0.0129 &  0.2335 &  0.0013 & -1.4587 &  1.5969 &  1.6607 & -0.0131 &  0.5083 &  0.0036 \\\hline
  8 & -1.8641 & -0.2125 &  2.7557 & -2.0107 &  3.1658 &  0.5005 & -1.4565 &  2.5604 &  0.0662 & -2.0192 & -1.5055 & -0.4916 \\\hline
  9 & -1.5729 &  4.6746 &  1.5732 &  3.2918 &  0.0080  & -0.7120  &  1.3067 & -1.5503 & -0.0798 &  1.5939 &  1.2361 &  1.2577 \\\hline
 10 &  0.0017 &  0.3691 & -0.2857 &  2.6193 & -0.1122 & -0.0086 &  0.4963 & -1.9117 &  0.8412 & -0.6049 &  0.8032 &  0.0027 \\\hline
 11 &  3.1826 &  1.5762 &  1.5861 & -0.0231 &  0.2348 &  0.0014 & -0.7771 &  1.5899 &  1.0078 &  0.0033 &  0.5246 & -0.0099 \\\hline
\end{tabularx}
\renewcommand{\arraystretch}{1}
\caption{The parameters $\theta_{ij}$ used to implement time evolution for $t=7$ in the free and interacting theories. The $\theta_{ij}$ correspond to the angles defined in the brickwall layer shown in Fig.~\ref{fig:circuit_elements}d: $i$ labels the layer, and $j$ labels the parameter within layer $i$. These parameters are determined using classical simulations and optimization.}
\label{tab:time_evolution_angles_7}
\end{table}

\begin{table}
\centering
\renewcommand{\arraystretch}{1.4}
\begin{tabularx}{\linewidth}{|c||Y|Y|Y|Y|Y|Y||Y|Y|Y|Y|Y|Y|}
\hline
&  \multicolumn{6}{c||}{$\lambda=0$}  &  \multicolumn{6}{c|}{$\lambda=2$} \\\hline
$j$ & $i=0$ & $i=1$ & $i=2$ & $i=3$ & $i=4$ & $i=5$ & $i=0$ & $i=1$ & $i=2$ & $i=3$ & $i=4$ & $i=5$ \\
\hline\hline
  0 &  0.0017 &  0.1202 & -0.2994 &  0.2670  &  0.008  & -0.0089 &  2.0762 &  0.0757 & -1.9467 &  0.5305 &  1.2240  & -0.0050  \\\hline
  1 & -1.5720  & -4.7144 & -4.7091 &  0.0071 & -0.1362 & -0.1595 &  1.5288 & -0.7801 &  1.5855 & -0.1109 & -4.7125 & -0.4436 \\\hline
  2 &  1.5624 &  2.4975 &  0.7148 & -2.4083 & -1.5697 & -1.5723 & -1.5883 & -0.0486 & -0.6387 &  0.0399 &  1.5001 & -1.5977 \\\hline
  3 &  0.0355 &  1.5617 & -1.5725 & -1.5728 & -3.3672 & -1.5360  & -0.0729 &  0.8390  & -1.5779 & -0.1046 &  1.6059 & -0.0781 \\\hline
  4 & -2.9469 & -0.5245 &  2.2046 & -2.2773 &  0.0940  & -0.5461 & -1.6336 &  1.2492 &  0.6364 & -1.0464 & -1.9926 & -0.3007 \\\hline
  5 & -1.5805 &  4.7094 &  1.5692 &  3.2779 & -0.0022 & -1.0045 & -0.2026 & -1.4905 &  1.9586 &  1.5602 &  3.1292 &  1.2552 \\\hline
  6 &  0.0011 &  0.7103 & -0.2456 &  2.9032 &  0.0792 & -0.0064 & -0.1564 & -2.5584 &  1.6904 & -1.5897 & -0.2588 &  0.0067 \\\hline
  7 &  3.1444 &  1.5738 &  1.5574 & -0.0258 &  0.1597 &  0.0009 & -1.3099 &  1.5913 &  2.0877 & -0.0052 &  0.4419 & -0.0027 \\\hline
  8 & -2.8916 & -0.5562 &  2.2471 & -2.2285 &  3.0843 &  0.5374 & -1.6450  &  1.2560  &  0.5729 & -1.2107 & -1.8533 & -0.3861 \\\hline
  9 & -1.5770  &  4.7101 &  1.5731 &  3.2801 &  0.0088 & -0.9555 & -0.1530  & -1.5376 &  1.9699 &  1.5432 &  2.9588 &  1.2599 \\\hline
 10 & -0.0034 &  0.5683 & -0.3675 &  3.2224 & -0.2362 & -0.0142 & -0.0932 & -2.7019 &  1.7268 & -1.6101 & -0.4026 &  0.0064 \\\hline
 11 &  3.1386 &  1.5736 &  1.5699 & -0.0124 &  0.1559 & -0.0004 & -1.2813 &  1.5986 &  2.0409 & -0.0089 &  0.4543 & -0.0007 \\\hline
\end{tabularx}
\renewcommand{\arraystretch}{1}
\caption{The parameters $\theta_{ij}$ used to implement time evolution for $t=8$ in the free and interacting theories. The $\theta_{ij}$ correspond to the angles defined in the brickwall layer shown in Fig.~\ref{fig:circuit_elements}d: $i$ labels the layer, and $j$ labels the parameter within layer $i$. These parameters are determined using classical simulations and optimization.}
\label{tab:time_evolution_angles_8}
\end{table}

\begin{table}
\centering
\renewcommand{\arraystretch}{1.4}
\begin{tabularx}{\linewidth}{|c||Y|Y|Y|Y|Y|Y||Y|Y|Y|Y|Y|Y|}
\hline
&  \multicolumn{6}{c||}{$\lambda=0$}  &  \multicolumn{6}{c|}{$\lambda=2$} \\\hline
$j$ & $i=0$ & $i=1$ & $i=2$ & $i=3$ & $i=4$ & $i=5$ & $i=0$ & $i=1$ & $i=2$ & $i=3$ & $i=4$ & $i=5$ \\
\hline\hline
  0 & -0.0008 &  0.2800   & -0.2194 &  0.3617 & -0.0039 &  0.1371 &  1.6507 &  0.0406 & -1.6259 &  0.6184 &  1.3403 & -0.0024 \\\hline
  1 & -1.5719 & -4.7144 & -4.7125 &  0.0032 &  0.1182 &  0.0038 &  1.2220  & -1.5253 &  1.5760  & -0.0568 & -4.7190  & -0.4047 \\\hline
  2 &  1.4799 &  2.7295 &  0.8456 & -2.5016 & -1.5347 & -1.5740  & -1.5872 & -0.0099 & -0.6879 &  0.0060  &  0.5950  & -1.6816 \\\hline
  3 &  0.0767 &  1.5663 & -1.5709 & -1.5705 & -3.1967 & -1.3495 & -0.0371 &  0.7480  & -1.5209 & -0.5640  &  1.5786 & -0.0048 \\\hline
  4 & -3.3481 & -1.0828 &  1.7347 & -2.9603 &  0.2967 &  0.5598 & -1.5525 &  1.5285 &  0.6970  & -1.5498 & -1.9947 & -0.3791 \\\hline
  5 & -1.5795 &  4.7124 &  1.5657 &  3.0253 & -0.0025 & -0.8377 & -0.2366 & -1.1993 &  2.1882 &  1.5553 &  3.1469 &  1.2310  \\\hline
  6 & -0.0013 &  0.7131 & -0.3294 &  3.1614 & -0.0275 &  0.0018 & -0.0657 & -2.5103 &  1.6019 & -1.6033 & -0.0107 & 0.0000      \\\hline
  7 &  3.1395 &  1.5733 &  1.5699 &  0.0879 & -0.0030  & -0.0001 & -0.6230  &  1.5356 &  2.0720  &  0.0043 &  0.4026 & -0.0003 \\\hline
  8 & -3.4245 & -0.9444 &  1.8114 & -2.9586 &  0.2792 &  0.5453 & -1.5796 &  1.4988 &  0.6103 & -1.5431 & -1.9636 & -0.3887 \\\hline
  9 & -1.5741 &  4.7115 &  1.5722 &  3.0286 & -0.0033 & -0.7946 & -0.1634 & -1.2214 &  2.1575 &  1.5569 &  3.1081 &  1.2298 \\\hline
 10 & -0.0042 &  0.6035 & -0.4230  &  3.1141 &  0.0180  & -0.0166 &  0.0585 & -2.6936 &  1.5923 & -1.5523 & -0.0084 & -0.0008 \\\hline
 11 &  3.1352 &  1.5725 &  1.5752 &  0.1049 & -0.0049 &  0.0013 & -0.5291 &  1.5413 &  1.9455 &  0.0013 &  0.4044 &  0.0008 \\\hline
\end{tabularx}
\renewcommand{\arraystretch}{1}
\caption{The parameters $\theta_{ij}$ used to implement time evolution for $t=9$ in the free and interacting theories. The $\theta_{ij}$ correspond to the angles defined in the brickwall layer shown in Fig.~\ref{fig:circuit_elements}d: $i$ labels the layer, and $j$ labels the parameter within layer $i$. These parameters are determined using classical simulations and optimization.}
\label{tab:time_evolution_angles_9}
\end{table}

\FloatBarrier

\section{Tables}
In this appendix, expectation values $\langle\phi_j^2\rangle_\text{2wp}-\langle\phi_j^2\rangle_\text{vac}$ are tabulated for $\lambda=0$ and $\lambda=2$. Error-mitigated results from {\tt ibm\_fez} are compared to classical data from a MPS circuit simulator. Larger uncertainties in the data from the quantum computer can be seen as a result of increasing circuit depth: the circuits for $t=1,2,3$ have a depth of 59 (Tables \ref{tab:results_t_1}, \ref{tab:results_t_2}, \ref{tab:results_t_3}), those for $t=4,5,6$ have a depth of 81 (Tables \ref{tab:results_t_4}, \ref{tab:results_t_5}, \ref{tab:results_t_6}), and those for $t=7,8,9$ have a depth of 103 (Tables \ref{tab:results_t_7}, \ref{tab:results_t_8}, \ref{tab:results_t_9}).

\begin{table}
\centering
\begin{tabularx}{\linewidth}{|c||Y|Y||Y|Y|}
\hline
\rule{0pt}{10pt} & \multicolumn{4}{c|}{\large$\langle\phi_j^2\rangle_\text{2wp}-\langle\phi_j^2\rangle_\text{vac}$}\\\hline\hline
\rule{0pt}{10pt} \multirow{2}{*}{\makecell{Spatial site\\$j$}} &  \multicolumn{2}{c||}{$\lambda=0$}  &  \multicolumn{2}{c|}{$\lambda=2$} \\\cline{2-5}
\rule{0pt}{10pt} & MPS & {\tt ibm\_fez} & MPS & {\tt ibm\_fez} \\
\hline\hline
 0 & 0.000 & 0.014(10)  & 0.000 & 0.018(23)  \\
  1 & 0.000 & 0.017(10)  & 0.000 & -0.008(09) \\
  2 & 0.000 & -0.005(11) & 0.000 & 0.009(10)  \\
  3 & 0.000 & -0.011(10) & 0.000 & 0.015(10)  \\
  4 & 0.000 & -0.015(10) & 0.000 & -0.005(12) \\
\hline
\end{tabularx}
\caption{Numerical values for the expectation value $\langle\phi_j^2\rangle_\text{2wp}-\langle\phi_j^2\rangle_\text{vac}$ at $t=1$ from MPS and {\tt ibm\_fez} for the free and interacting theories, as shown in Figs.~\ref{fig:scattering_heatmap} and \ref{fig:results_by_time}.}
\label{tab:results_t_1}
\end{table}

\begin{table}[h]
\centering
\begin{tabularx}{\linewidth}{|c||Y|Y||Y|Y|}
\hline
\rule{0pt}{10pt} & \multicolumn{4}{c|}{\large$\langle\phi_j^2\rangle_\text{2wp}-\langle\phi_j^2\rangle_\text{vac}$}\\\hline\hline
\rule{0pt}{10pt} \multirow{2}{*}{\makecell{Spatial site\\$j$}} &  \multicolumn{2}{c||}{$\lambda=0$}  &  \multicolumn{2}{c|}{$\lambda=2$} \\\cline{2-5}
\rule{0pt}{10pt} & MPS & {\tt ibm\_fez} & MPS & {\tt ibm\_fez} \\
\hline\hline
 0 &  0.000 & 0.007(10)  &  0.000 & 0.003(09)  \\
  1 &  0.000 & 0.013(08)  &  0.000 & -0.004(12) \\
  2 &  0.000 & 0.021(11)  &  0.000 & -0.003(11) \\
  3 &  0.000 & 0.008(08)  &  0.000 & 0.004(09)  \\
  4 &  0.000 & -0.013(11) &  0.000 & 0.007(11)  \\
\hline
\end{tabularx}
\caption{Numerical values for the expectation value $\langle\phi_j^2\rangle_\text{2wp}-\langle\phi_j^2\rangle_\text{vac}$ at $t=2$ from MPS and {\tt ibm\_fez} for the free and interacting theories, as shown in Figs.~\ref{fig:scattering_heatmap} and \ref{fig:results_by_time}.}
\label{tab:results_t_2}
\end{table}

\begin{table}[h]
\centering
\begin{tabularx}{\linewidth}{|c||Y|Y||Y|Y|}
\hline
\rule{0pt}{10pt} & \multicolumn{4}{c|}{\large$\langle\phi_j^2\rangle_\text{2wp}-\langle\phi_j^2\rangle_\text{vac}$}\\\hline\hline
\rule{0pt}{10pt} \multirow{2}{*}{\makecell{Spatial site\\$j$}} &  \multicolumn{2}{c||}{$\lambda=0$}  &  \multicolumn{2}{c|}{$\lambda=2$} \\\cline{2-5}
\rule{0pt}{10pt} & MPS & {\tt ibm\_fez} & MPS & {\tt ibm\_fez} \\
\hline\hline
  0 &  0.000 & -0.013(11) &  0.000 & 0.007(09)  \\
  1 &  0.000 & -0.007(09) &  0.000 & 0.007(09)  \\
  2 &  0.000 & 0.002(11)  &  0.000 & 0.002(08)  \\
  3 &  0.000 & 0.010(08)  &  0.000 & 0.028(12)  \\
  4 &  0.000 & 0.004(08)  &  0.000 & 0.018(07)  \\
\hline
\end{tabularx}
\caption{Numerical values for the expectation value $\langle\phi_j^2\rangle_\text{2wp}-\langle\phi_j^2\rangle_\text{vac}$ at $t=3$ from MPS and {\tt ibm\_fez} for the free and interacting theories, as shown in Figs.~\ref{fig:scattering_heatmap} and \ref{fig:results_by_time}.}
\label{tab:results_t_3}
\end{table}

\begin{table}[h]
\centering
\begin{tabularx}{\linewidth}{|c||Y|Y||Y|Y|}
\hline
\rule{0pt}{10pt} & \multicolumn{4}{c|}{\large$\langle\phi_j^2\rangle_\text{2wp}-\langle\phi_j^2\rangle_\text{vac}$}\\\hline\hline
\rule{0pt}{10pt} \multirow{2}{*}{\makecell{Spatial site\\$j$}} &  \multicolumn{2}{c||}{$\lambda=0$}  &  \multicolumn{2}{c|}{$\lambda=2$} \\\cline{2-5}
\rule{0pt}{10pt} & MPS & {\tt ibm\_fez} & MPS & {\tt ibm\_fez} \\
\hline\hline
 0 &  0.000 & 0.023(12)  &  0.000 & 0.008(15)  \\
  1 &  0.000 & 0.012(11)  &  0.000 & 0.001(10)  \\
  2 &  0.000 & 0.004(11)  &  0.000 & 0.014(10)  \\
  3 &  0.000 & 0.005(11)  &  0.000 & 0.019(11)  \\
  4 &  0.000 & -0.020(13) &  0.000 & 0.003(12)  \\
\hline
\end{tabularx}
\caption{Numerical values for the expectation value $\langle\phi_j^2\rangle_\text{2wp}-\langle\phi_j^2\rangle_\text{vac}$ at $t=4$ from MPS and {\tt ibm\_fez} for the free and interacting theories, as shown in Figs.~\ref{fig:scattering_heatmap} and \ref{fig:results_by_time}.}
\label{tab:results_t_4}
\end{table}

\begin{table}[h]
\centering
\begin{tabularx}{\linewidth}{|c||Y|Y||Y|Y|}
\hline
\rule{0pt}{10pt} & \multicolumn{4}{c|}{\large$\langle\phi_j^2\rangle_\text{2wp}-\langle\phi_j^2\rangle_\text{vac}$}\\\hline\hline
\rule{0pt}{10pt} \multirow{2}{*}{\makecell{Spatial site\\$j$}} &  \multicolumn{2}{c||}{$\lambda=0$}  &  \multicolumn{2}{c|}{$\lambda=2$} \\\cline{2-5}
\rule{0pt}{10pt} & MPS & {\tt ibm\_fez} & MPS & {\tt ibm\_fez} \\
\hline\hline
 0 &  0.001 & 0.005(10)  &  0.000 & 0.008(13)  \\
  1 &  0.000 & -0.016(15) &  0.000 & -0.011(13) \\
  2 &  0.001 & 0.019(12)  &  0.000 & -0.021(14) \\
  3 &  0.000 & -0.030(14) &  0.000 & -0.010(14) \\
  4 &  0.001 & -0.038(15) &  0.000 & 0.003(11)  \\
\hline
\end{tabularx}
\caption{Numerical values for the expectation value $\langle\phi_j^2\rangle_\text{2wp}-\langle\phi_j^2\rangle_\text{vac}$ at $t=5$ from MPS and {\tt ibm\_fez} for the free and interacting theories, as shown in Figs.~\ref{fig:scattering_heatmap} and \ref{fig:results_by_time}.}
\label{tab:results_t_5}
\end{table}

\begin{table}[h]
\centering
\begin{tabularx}{\linewidth}{|c||Y|Y||Y|Y|}
\hline
\rule{0pt}{10pt} & \multicolumn{4}{c|}{\large$\langle\phi_j^2\rangle_\text{2wp}-\langle\phi_j^2\rangle_\text{vac}$}\\\hline\hline
\rule{0pt}{10pt} \multirow{2}{*}{\makecell{Spatial site\\$j$}} &  \multicolumn{2}{c||}{$\lambda=0$}  &  \multicolumn{2}{c|}{$\lambda=2$} \\\cline{2-5}
\rule{0pt}{10pt} & MPS & {\tt ibm\_fez} & MPS & {\tt ibm\_fez} \\
\hline\hline
 0 &  0.001 & 0.022(10)  &  0.000 & 0.025(18)  \\
  1 &  0.001 & -0.010(11) &  0.000 & -0.036(16) \\
  2 &  0.001 & 0.033(10)  &  0.000 & -0.020(17) \\
  3 &  0.001 & 0.010(10)  &  0.000 & -0.021(25) \\
  4 &  0.001 & 0.036(11)  &  0.000 & -0.003(17) \\
\hline
\end{tabularx}
\caption{Numerical values for the expectation value $\langle\phi_j^2\rangle_\text{2wp}-\langle\phi_j^2\rangle_\text{vac}$ at $t=6$ from MPS and {\tt ibm\_fez} for the free and interacting theories, as shown in Figs.~\ref{fig:scattering_heatmap} and \ref{fig:results_by_time}.}
\label{tab:results_t_6}
\end{table}

\begin{table}[h]
\centering
\begin{tabularx}{\linewidth}{|c||Y|Y||Y|Y|}
\hline
\rule{0pt}{10pt} & \multicolumn{4}{c|}{\large$\langle\phi_j^2\rangle_\text{2wp}-\langle\phi_j^2\rangle_\text{vac}$}\\\hline\hline
\rule{0pt}{10pt} \multirow{2}{*}{\makecell{Spatial site\\$j$}} &  \multicolumn{2}{c||}{$\lambda=0$}  &  \multicolumn{2}{c|}{$\lambda=2$} \\\cline{2-5}
\rule{0pt}{10pt} & MPS & {\tt ibm\_fez} & MPS & {\tt ibm\_fez} \\
\hline\hline
  0 &  0.000 & 0.023(14)  &  0.000 & -0.003(10) \\
  1 &  0.000 & 0.012(16)  &  0.000 & 0.008(15)  \\
  2 &  0.000 & 0.015(17)  &  0.000 & 0.014(12)  \\
  3 &  0.000 & 0.025(16)  &  0.000 & -0.001(13) \\
  4 &  0.000 & 0.002(16)  &  0.000 & -0.003(11) \\
\hline
\end{tabularx}
\caption{Numerical values for the expectation value $\langle\phi_j^2\rangle_\text{2wp}-\langle\phi_j^2\rangle_\text{vac}$ at $t=7$ from MPS and {\tt ibm\_fez} for the free and interacting theories, as shown in Figs.~\ref{fig:scattering_heatmap} and \ref{fig:results_by_time}.}
\label{tab:results_t_7}
\end{table}

\begin{table}[h]
\centering
\begin{tabularx}{\linewidth}{|c||Y|Y||Y|Y|}
\hline
\rule{0pt}{10pt} & \multicolumn{4}{c|}{\large$\langle\phi_j^2\rangle_\text{2wp}-\langle\phi_j^2\rangle_\text{vac}$}\\\hline\hline
\rule{0pt}{10pt} \multirow{2}{*}{\makecell{Spatial site\\$j$}} &  \multicolumn{2}{c||}{$\lambda=0$}  &  \multicolumn{2}{c|}{$\lambda=2$} \\\cline{2-5}
\rule{0pt}{10pt} & MPS & {\tt ibm\_fez} & MPS & {\tt ibm\_fez} \\
\hline\hline
 0 &  0.000 & -0.004(10) &  0.000 & -0.024(16) \\
  1 &  0.000 & 0.011(12)  &  0.000 & -0.014(15) \\
  2 &  0.000 & -0.007(13) &  0.000 & -0.013(10) \\
  3 &  0.000 & 0.012(10)  &  0.000 & -0.028(16) \\
  4 &  0.000 & 0.010(12)  &  0.000 & -0.019(13) \\
\hline
\end{tabularx}
\caption{Numerical values for the expectation value $\langle\phi_j^2\rangle_\text{2wp}-\langle\phi_j^2\rangle_\text{vac}$ at $t=8$ from MPS and {\tt ibm\_fez} for the free and interacting theories, as shown in Figs.~\ref{fig:scattering_heatmap} and \ref{fig:results_by_time}.}
\label{tab:results_t_8}
\end{table}

\begin{table}[h]
\centering
\begin{tabularx}{\linewidth}{|c||Y|Y||Y|Y|}
\hline
\rule{0pt}{10pt} & \multicolumn{4}{c|}{\large$\langle\phi_j^2\rangle_\text{2wp}-\langle\phi_j^2\rangle_\text{vac}$}\\\hline\hline
\rule{0pt}{10pt} \multirow{2}{*}{\makecell{Spatial site\\$j$}} &  \multicolumn{2}{c||}{$\lambda=0$}  &  \multicolumn{2}{c|}{$\lambda=2$} \\\cline{2-5}
\rule{0pt}{10pt} & MPS & {\tt ibm\_fez} & MPS & {\tt ibm\_fez} \\
\hline\hline
  0 &  0.000 & 0.007(12)  &  0.000 & -0.017(18) \\
  1 &  0.000 & 0.017(11)  &  0.000 & -0.003(15) \\
  2 &  0.000 & 0.033(11)  &  0.000 & 0.022(15)  \\
  3 &  0.000 & 0.020(14)  &  0.000 & 0.007(13)  \\
  4 &  0.000 & -0.001(13) &  0.000 & 0.036(14)  \\
\hline
\end{tabularx}
\caption{Numerical values for the expectation value $\langle\phi_j^2\rangle_\text{2wp}-\langle\phi_j^2\rangle_\text{vac}$ at $t=9$ from MPS and {\tt ibm\_fez} for the free and interacting theories, as shown in Figs.~\ref{fig:scattering_heatmap} and \ref{fig:results_by_time}.}
\label{tab:results_t_9}
\end{table}

\clearpage
\bibliography{refs}
\end{document}